\documentclass[twocolumn]{aa}
\usepackage{graphicx}
\usepackage{natbib}
\newcommand{\cpt}[1]{\caption{\small{#1}}}

\begin{document}
  \title{The History of the Baryon Budget}
  \subtitle{Cosmic Logistics in a Hierarchical Universe}
  
  \author{Yann Rasera \and  Romain Teyssier}
  
  \offprints{Yann Rasera}
   
  \institute{DAPNIA, Service d'Astrophysique, Batiment 709, 
    L'Orme des Merisiers,
    91191 Gif--sur--Yvette Cedex, FRANCE\\
    \email{romain.teyssier@cea.fr}}
   
  \date{Received March 22, 2005; accepted May 3, 2005}
   
  \abstract{
       
    Using a  series of high-resolution N-body hydrodynamical numerical
    simulations, we investigate several scenarios for the evolution of
    the baryon budget in galactic  halos.   We derive individual  halo
    star formation history (SFH), as well as the global star formation
    rate  in the universe.  We  develop a simple analytical model that
    allows   us to compute   surprisingly  accurate predictions,  when
    compared  to  our   simulations, but  also   to  other simulations
    presented  in \cite{Springel03b}.  The  model depends  on two main
    parameters: the    star formation time scale    $t_*$ and the wind
    efficiency $\eta_{\rm w}$. We also compute,  for halos of a given  mass,
    the baryon fraction  in each of the  following phases:  cold discs
    gas, hot halo  gas and  stars.   Here again, our analytical  model
    predictions are in good agreement  with simulation results, if one
    correctly takes into account finite resolution effect.  We compare
    predictions of  our analytical   model  to  several  observational
    constraints, and conclude  that a very narrow  range of the  model
    parameters is allowed. The important role played by galactic winds
    is outlined, as well as a possible  `superwind' scenario in groups
    and clusters.  The `anti-hierarchical' behavior of observed SFH is
    well  reproduced   by   our best   model   with  $t_*=3$~Gyr   and
    $\eta_{\rm w}=1.5$.  We obtain in  this case a present-day cosmic baryon
    budget of $\Omega_{*}\simeq 0.004$, $\Omega_{\rm cold}\simeq  0.0004$,
    $\Omega_{\rm hot}\simeq   0.01$    and   $\Omega_{\rm back}  \simeq  0.02$
    (diffuse background).
  
     }
  
  \maketitle
  
  \section{Introduction}
  
  Thanks  to  recent advances  in  the  observation of  high-redshift
  star-forming   galaxies,   and  to   the   improved  statistics   in
  low-redshift  galaxy   surveys,  it  is  now  possible   to  have  a
  quantitative view of the star  formation history in the universe, in
  a   global,   volume-averaged   sense  \citep{Hughes98,   Steidel99,
  Flores99, Glazebrook99, Yan99, Massarotti01, Giavalisco03}, but also
  for  each  galactic  halo,  in a  statistical,  mass-averaged  sense
  \citep{Heavens04}.

  In  the  standard picture of galaxy   formation,  small mass objects
  collapse first, when gravitational instability enters the non-linear
  regime.  This  so-called `hierarchical  scenario' relies heavily  on
  the observation of  primordial density fluctuations imprinted on the
  Cosmic  Microwave   Background    \citep{Spergel03}.   These   first
  generation  objects eventually form  stars, the so-called Population
  III  stars, whose properties  are   likely to differ  strongly  from
  present-day galactic stars. The contribution of Population III stars
  to the global star formation history  in the universe is currently a
  matter   of debate:  they   likely  contribute to  a prompt  initial
  enrichment of the intergalactic medium,  as well as a possible early
  reionization epoch,   but  they may  also   cause a strong  negative
  feedback  on  the star  formation   efficiency at   this very  early
  epoch. In this paper, we only adress star formation occurring inside
  galactic discs, concentrating on  quiescent modes of star formation.
  This means  that  we  only consider  atomic  processes to  drive the
  cooling and  subsequent   catastrophic  collapse of   halo gas  that
  ultimately leads to the    formation of a  centrifugally   supported
  disc. Molecular processes that     are  relevant for  first    stars
  formation, or  for molecular clouds  formation inside galactic discs
  are neglected in this study.
  
  Our main  concern is therefore to follow  the formation of the first
  generation of gas rich, rotating  discs, and the subsequent  merging
  hierarchy that leads to larger  and larger discs,  up to present day
  spiral galaxies, as  well    as galaxy groups  and  clusters.    The
  standard approach   in  current  galaxy  formation scenario   is  to
  consider  that  hot gas virializes  first  into extended dark matter
  halo potential well. Depending on the Virial temperature of this hot
  gas, it cools rather rapidly,  looses pressure support and collapses
  up to  a point where centrifugal  equilibrium sets in.  It is inside
  these  cold, rotating discs that  star formation takes  place at the
  end of a  complex  cascade of turbulent fragmentation  and molecular
  cloud formation.  On cosmological scale,   models of star  formation
  still    rely on     heuristic  recipes, partially    based  on
  observational  constraints,  and   partially based   on  theoretical
  arguments.
  
  This rather simple  picture is altered by  feedback processes due to
  these stars.   First  of all, massive  stars create  collectively  a
  strong  UV background  that  photo-ionizes the intergalactic medium,
  preventing small mass halos from building  their virialized, hot gas
  component, and altering their  cooling efficiency.  This means that,
  after  the universe   is re-ionized,  we  have to   introduce a mass
  threshold (or equivalently  a  Virial  temperature  threshold) below
  which  star formation is   suppressed.  This threshold allows us  to
  distinguish between a diffuse  component, defined by small mass dark
  matter halos  with  a very  low  gas fraction,  and  a  star forming
  component, defined by   high mass dark  matter halos  where disc and
  star  formation is possible.  The diffuse  component is often called
  `the Lyman alpha forest' or  `the smooth  baryon background' in  the
  literature.  In the present paper, we name  the other component `the
  star forming  halos'.    Following  standard definitions  in  galaxy
  formation studies, we further  divide even more the baryons  sitting
  in these  `galactic halos' into 3   different phases: hot  gas, cold
  discs  and  stars.  The  purpose of   this paper  is  to  compute as
  precisely  as  possible the evolution of  these  4  phases,  for the
  universe  as a whole, as well  as for individual  dark matter halos,
  thus studying the baryon logistics.

  A lot of  other   different feedback processes are   currently under
  close examination in the  literature, like  supernovae-driven winds,
  young  protostellar   jets, AGN-driven jets   and associated buoyant
  bubbles, and so on and so forth. Such outflows are actually observed
  in high redshift galaxies  \citep{Martin99}.  They consist  of
  high-velocity fountains of  ionized   gas, expelled from   a parent,
  star-forming disc. These winds are likely to be caused by collective
  outbreaks of  supernovae bubbles \citep{DeAvillez01}.   Other energy
  sources are however possible, like  a central massive black hole, or
  collective jet ram pressure from star  forming clouds. Regardless of
  their physical origin,  these outflows are fundamental in explaining
  the metal  enrichment in the  intergalactic  medium, as well  as the
  high-metallicity  observed in rich galaxy  cluster.  The modeling of
  such  winds will also be  addressed  in this paper,  in a simplified
  way, and their impact on the star  formation history in the universe
  as a   whole will be  computed.  We  will investigate in more
  detail the    impact of winds   on  the star  formation  history of
  individual star forming halos in a follow-up paper.

  Several approaches have been   used to investigate the  evolution of
  the baryons in the hierarchical model of galaxy formation. Numerical
  simulations \citep{Cen92,  Navarro93, Mihos94,   Katz96,   Gnedin97,
  Thacker00,   Springel03a}   and semi-analytic models \citep{White91,
  Somerville99, Kaufmann99, Cole00,   Hatton03}  have been  the   most
  popular techniques.  In a  recent paper, an analytical approach  has
  been proposed  by \cite{Springel03c} to   compute the star formation
  history of  the universe.   Here, we   present also a  simple
  self-consistent   analytical  model.    This  model,  validated   on
  simulations, allows us  to quickly  compute  the evolution  of the 4
  baryons components, for the  Universe as a whole,   and also for  an
  individual average halo.

  The paper  is organized as follows: in  the next section, we present
  numerical simulations of  star formation in a $\Lambda$CDM universe,
  based on  our Adaptive Mesh Refinement  code named RAMSES.   We then
  describe a  simple  analytical model  that allows  us to compute the
  star formation    history inside individual halos   and in the
  universe as a whole.   In  the section~4,  we compare our  model to
  simulation  results,  showing   that,    once  calibrated  to    our
  simulations, the  model works very well  in predicting the evolution
  of the 4  different baryon phases.  We  finally compare our model to
  several observational  constraints.  Our analytical  approach allows
  us to explore efficiently the parameter space,  which is in our case
  limited  to a 2-dimensional space: star  formation time $t_*$ versus
  wind efficiency $\eta_{\rm w}$.    We finally confront  our predictions to
  current  observational  constraints,  and  found   a  rather  narrow
  parameter range.
  
  \section{Simulations}
  
  Our  simulations were  performed using the  Adaptive Mesh Refinement
  code   called   RAMSES    and    described in   detail  by
  \cite{Teyssier02}.  The N-body solver  is  very similar to  the  ART
  code \citep{Kravtsov97} and the  hydrodynamical solver is based on a
  second-order Godunov-type method,   called the MUSCL-Hancock  scheme
  \citep{Toro97}.  We   evolve the collisionless dark  matter particle
  distribution  by   solving  the  Vlasov-Poisson   equation, and  the
  baryonic component by solving the Euler equation with gravity source
  terms. More detail  about  our    hydrodynamical solver and     our
  refinement strategy are given in appendix \ref{method_num}.  We also
  solve for a heating and cooling source terms in the energy equation,
  assuming    primordial H  and   He   plasma  photo-ionized by    the
  \cite{Haardt96}  UV  background.  In dense  and cold  regions of the
  flow, we  turn  a  fraction of  the  gas  into collisionless  `star'
  particles.  This numerical approach is widely used in current galaxy
  formation studies    \citep{Cen92,   Navarro93,  Mihos94,    Katz96,
  Gnedin97, Thacker00}.  We will briefly recall here the properties of
  our own implementation.   We then  describe cosmological parameters,
  box sizes  and mass resolution we use  in  our three main simulation
  series.
  
  \subsection{Star formation recipe}

  \begin{table*}
  \begin{tabular}{|c|c|c|c|c|} 
    \hline
    Name & L10N64S30 & L10N128S30 & L10N256S30 & L10N512S30 \\
    \hline
    $L(h^{-1}\textrm{Mpc})$&10&10&10&10\\
    \hline
    $n_{\rm 0}$(cm$^{-3}$) &0.036&0.036&0.036&0.036\\
    \hline
    $t_{\rm 0}$(Gyr) &30&30&30&30\\
    \hline
    $\alpha_{\rm 0}$ &3&3&3&3\\
    \hline
    $N_{\rm cell}$& $64^3$&$128^3$&$256^3$&$512^3$\\
    \hline
    $\ell_{\rm max}$&5&5&5&5\\
    \hline
    $z_{\rm end}$&3&3&3&3\\
    \hline
    $m_{\rm DM}(h^{-1}\textrm{M}_{\odot})$&$2.8\times 10^8$&$3.4\times 10^7$
    &$4.3\times 10^6$&$5.4\times 10^5$\\
    \hline
    $\Delta x(h^{-1}\textrm{kpc})$&4.9&2.4&1.2&0.6\\
    \hline
  \end{tabular}
  \cpt{Runtime parameters for the `convergence study' simulation
  suite. $L$ is the box length, $n_{\rm 0}$ is the density threshold at $z=0$, $t_{\rm 0}$ is the star formation time scale for density $n_{\rm 0}$, $\alpha_{\rm 0}$ gives the evolution of the star formation time scale with redshift, $N_{\rm cell}$ is the number of cells at the coarse level, $\ell_{\rm max}$ is the level of refinement, $z_{\rm end}$ is the final redshift, $m_{\rm DM}$ is the dark matter particle mass and $\Delta_{\rm x}$ is the spatial resolution.}
  \label{table1}
  \end{table*}

  We now  describe our method  for implementing star formation  in the
  RAMSES  cosmological code.   It is  based on  an  heuristic approach
  adopted in  many, if not  all, cosmological studies. For  a complete
  review of various implementations, see \cite{Kay02}.  Basically, one
  considers  that  star formation  proceeds  at  a  given time  scale,
  written here $t_*$, in region  where one or several physical criteria
  are  fullfiled. In this  paper, as  well as  in many  previous other
  papers in the literature, we adopt  a simple scheme to turn gas mass
  into star particles, adding a source term in the continuity equation

  \begin{eqnarray}
    \left( \frac{D\rho}{Dt} \right)_* = & -\frac{\rho}{t_*}
    & \mbox{~if~} \rho > \rho_{\rm 0} (1+z)^{\alpha_{\rm 0}},
    \label{starrateeq} \\
    \nonumber
    \left( \frac{D\rho}{Dt} \right)_* = & 0 & \mbox{~otherwise},
  \end{eqnarray}

  \noindent where the threshold density may depend on redshift through
  index $\alpha_{\rm 0}$. The star  formation time scale $t_*$ is proportional
  to the local free-fall time

  \begin{equation}
    t_* = t_{\rm 0 }\left( \frac{\rho}{\rho_{\rm 0}} \right)^{-1/2}.
  \end{equation}

  \noindent  In the  literature, one  can find  basically  two different
  approaches:   a  density  threshold   constant  in   physical  units,
  corresponding here  to $\alpha_{\rm 0}=0$, and a  density threshold constant
  in comoving units, corresponding here to $\alpha_{\rm 0}=3$.

  This  simple  model  of   star  formation  has  been  discussed  and
  criticized  extensively in the  literature \citep{Kay02}:  we recall
  here  briefly  its  possible  physical  and  observational  origins.
  Spiral galaxies in  our nearby universe are seen to  form stars at a
  rate  given  by the  Kennicutt  law  \citep{Kennicutt98}, similar  to
  equation~\ref{starrateeq}, with volume densities $\rho$ and $\rho_{\rm 0}$
  replaced by average disc surface densities.

  \begin{equation}
    \Sigma_{\rm 0} = 10 \mbox{~M}_{\odot} /\mbox{pc}^2 \mbox{~and~} t_{\rm 0}=2 ~\mbox{Gyr}.
  \end{equation}

  The physical  origin of such  a behavior is not  clearly identified
  yet.    One   good  candidate   is   the  sustained,   interstellar,
  self-gravitating  turbulent cascade,  which controls  the  mass flux
  between  large scale  filaments in  the  disc and  small scale  star
  forming molecular  clouds, at  a rate given  by the local  (on large
  scale, though)  free-fall time \citep{Elmegreen02}.   Such a precise
  description  of  star  formation  is completely  irrelevant  for  our
  present cosmological study.   Using a `sub-cell approach', similar
  in spirit to the one developed by the fluid mechanics community and
  often named  the $k-\epsilon$ method, our current  modeling of star
  formation try to mimic in  a statistical sense the complex behavior
  of the interstellar medium.

  Along   the   same   lines,   \cite{Yepes97}   and   more   recently
  \cite{Springel03a}  have  developed   a  multiphase  model  of  the
  interstellar  medium,   based  on  \citet{McKee77}   early  work  on
  molecular clouds evaporation within supernovae remnants hot bubbles.
  This multiphase model offers an interesting alternative to
  the  previous turbulent  model.  It  also gives  the  possibility to
  computes self-consistently the star formation parameters, which are

  \begin{equation}
    n_{\rm 0} = 0.1 \mbox{~cm}^{-3} \mbox{~and~} \alpha_{\rm 0}=0 \mbox{~and~} t_{\rm 0}=2.1 \mbox{~Gyr}.
  \end{equation}

  Recall that in this case,  star formation is allowed in region whose
  gas  density  lies above  a  {\it  physical}  density threshold.   In
  cosmological  simulations, however,  the {\it  comoving}  density is
  usually  preferred to  define collapsed,  high-density  clumps where
  star  formation  is  likely  to occur.   In  \cite{Springel03b}  for
  example, equation~\ref{starrateeq} is augmented with the requirement
  that the  local {\it overdensity} should exceed  200. This guarantees
  that  star   formation  cannot  occur  in  smooth   regions  of  the
  cosmological flow, but only within collapsed, virialized halos.

  In the  present paper, we would  like to explore  a different option
  than the one  used in \cite{Springel03b}. We use  the following star
  formation density threshold

  \begin{equation}
    n_{\rm 0} = 0.036 \mbox{~cm}^{-3} \mbox{~and~} \alpha_{\rm 0}=3. 
  \end{equation}

  \noindent This corresponds to  a baryon overdensity threshold of $1.6
  \times 10^5$. This approach was also adopted in earlier works on the
  star  formation  history   in  the  universe  \citep{Cen92,  Kay02,
  Nagamine04},  with  threshold  overdensities  ranging  from  $5$  to
  $10^5$, depending on the authors.  In our case, this is motivated by
  our AMR  refinement scheme: $n_{\rm 0}(1+z)^3$ corresponds  exactly to the
  density  threshold  triggering   the  maximum  level  of  refinement
  $\ell_{\rm max}=5$.  When baryons cool and  settle down at the centre of
  their   host  dark   matter   halos,  the   density  does   increase
  dramatically.  Our  AMR code describes this  collapse accurately, by
  adding  recursively new cells  at the  centre of  the halo.   At the
  point when  the maximum  level of refinement  has been  reached, the
  numerical description of the collapse is not valid anymore.  We then
  turn on star formation, as our sub-cell modelling of this uncomplete
  collapse, that would have lead ultimately to the formation of one or
  several star forming molecular clouds.

  As soon  as star formation  is active,  we create collisionless star
  particles   of constant  mass  $m_*$.   As  we  explain in
  appendix~\ref{sectrefine}, this constant   mass is chosen to be
  equal to the initial  mass resolution in  the gas  distribution, in
  order to prevent spurious  refinement or de-refinement  triggered by
  star  formation.  Using a constant mass   for our star particles has
  also the advantage  of controlling the {\it  maximum number} of star
  particles  created at the end of  the simulation.  In order to solve
  for  the star  formation   source term (Eq.~\ref{starrateeq})   with
  constant  star    particles mass, we need    to  adopt  a stochastic
  approach, similar  to  the one  proposed in \cite{katz92}.   In star
  forming cells, we generate $N$ equal mass  star particles, where $N$
  is drawn from a Poisson process, with probability

  \begin{equation}
    P(N) = \frac{\lambda^N}{N!}\exp{(-\lambda)},
  \end{equation}

  \noindent and with parameter (or mean value)

  \begin{equation}
    \lambda = \left( \frac{\rho \Delta x^3}{m_*}\right) \frac{\Delta t}{t_*}.
  \end{equation}

  \noindent These star particles are created at each time step, at the
  very centre of their parent cell. They are given a velocity equal to
  the local fluid velocity, plus a random component that we take equal
  to  the  local sound  speed  in  the  gas. The  corresponding  mass,
  momentum and  internal energy is  of course removed from  the parent
  cell's conservatively.

  In the  simulation presented  here, the only  free parameter  is the
  star formation time scale at {\it comoving} threshold density, $t_{\rm 0}$.

  \subsection{Parameters}

 \begin{table*}
  \begin{tabular}{|c|c|c|c|} 
    \hline
    Name &L100N256S3 &L10N256S3 &L1N256S3\\ 
    \hline
    $L(h^{-1}\textrm{Mpc})$&100&10&1\\
    \hline
    $n_{\rm 0}$(cm$^{-3}$)&0.036&0.036&0.036\\
    \hline
    $t_{\rm 0}$(Gyr)&3&3&3\\
    \hline
    $\alpha_{\rm 0}$ &3&3&3\\
    \hline
    $N_{\rm cell}$&$256^3$&$256^3$&$256^3$\\
    \hline
    $\ell_{\rm max}$&5&5&5\\
    \hline
    $z_{\rm end}$&0&2.5&5.5\\
    \hline
    $m_{\rm DM}(h^{-1}\textrm{M}_{\odot})$&$4.3\times10^9$&$4.3\times10^6$&$4.3\times10^3$\\
    \hline
    $\Delta x(h^{-1}\textrm{kpc})$&12&1.2&0.12\\
    \hline
  \end{tabular}
  \cpt{ Runtime  parameters for  the `high efficiency' simulation
  suite. The meaning of the symbols is the same as table \ref{table1}.}
  \label{table2}
  \end{table*}

 \begin{table*}
  \begin{tabular}{|c|c|c|c|} 
    \hline
    Name &L100N256S30&L10N256S30&L1N256S30\\ 
    \hline
    $L(h^{-1}\textrm{Mpc})$&100&10&1\\
    \hline
    $n_{\rm 0}$(cm$^{-3}$)&0.036&0.036&0.036\\
    \hline
    $t_{\rm 0}$(Gyr)&30&30&30\\
    \hline
    $\alpha_{\rm 0}$&3&3&3\\
    \hline
    $N_{\rm cell}$&$256^3$&$256^3$&$256^3$\\
    \hline
    $\ell_{\rm max}$&5&5&5\\
    \hline
    $z_{\rm end}$&0&2.5&5.5\\
    \hline
    $m_{\rm DM}(h^{-1}\textrm{M}_{\odot})$&$4.3\times10^9$&$4.3\times10^6$&$4.3\times10^3$\\
    \hline
    $\Delta x(h^{-1}\textrm{kpc})$&12&1.2&0.12\\
    \hline
  \end{tabular}
  \cpt{Runtime  parameters for  the  `low efficiency' simulation
  suite. The meaning of the symbols is the same as table \ref{table1}.}
  \label{table3}
  \end{table*}

   We assume throughout this paper a single cosmological model, the so
  called  `concordance model', with  a cold dark matter component with
  $\Omega_{\rm m}=0.3$, a baryon  component with $\Omega_{\rm b}=0.04$  and a dark
  energy  component with  $\Omega_{\Lambda}=0.7$. The Hubble  constant
  was  set to  $h=0.7$.  The initial  power spectrum  is assumed to be
  Harrisson-Zeldovich with $n=1$ and a normalization constrained
  by  $\sigma_{\rm 8}=0.93$.   The exact functional   form   we use for  the
  transfer function is given in \cite{sugiyama95}.

  We  performed several AMR  simulations, varying  the  box size,  the
  number of  particles and the  star formation time scale  $t_{\rm 0}$.  The
  density  threshold  for  star   formation  was  set  to  $n_{\rm 0}=0.036$
  cm$^{-3}$, and  was held  constant in {\it  comoving units}.  In our
  notation,  this  translates into  $\alpha_{\rm 0}=3$.   This star  formation
  threshold  corresponds to the  overdensity threshold  triggering our
  last level of refinement $\ell_{\rm max}=5$.

  Our simulation suite can be organized in 3 different sets. The first
  set was designed to perform a  convergence study. We used a box size
  of  10$h^{-1}$   Mpc,  with  identical   physical  and  cosmological
  parameters.  We vary  only the initial number of  particles and grid
  cells,  from  $64^3$  to   $512^3$.   Runtime  parameters  for  this
  `convergence     study'    simulations    are     summarized    in
  Table~\ref{table1}.

  The second   set  of simulations was  designed   to explore a  `high
  efficiency'  SFR scenario with the   maximum dynamical range we  can
  afford (Table~\ref{table2}).  We used  for that purpose 3  different
  box  size: 1, 10 and  100 h$^{-1}$ Mpc.   All these simulations were
  performed using $256^3$ particles and the same number of grid cells.
  This   translates into  a dark   matter particle  mass of $4.3\times
  10^3$,  $4.3\times 10^6$ and   $4.3\times 10^9 $ h$^{-1}$M$_\odot$ .
  The third set of simulations is similar to  the second one but for a
  `low efficiency' SFR scenario (Table~\ref{table3}).

  The 2 last simulation sets span  a huge dynamical range in mass, but
  each simulation is valid for a limited range of redshifts.  When the
  typical scale of non linearity  reaches the box size, the simulation
  has  to  be  stopped:  the  simulated volume  is  not  statistically
  representative of the universe as  a whole, and spurious effects due
  to  periodic  boundary  conditions  become  visible.   The  stopping
  redshift was chosen to be 5, 2.5 and  0, for a box size of 1, 10 and
  100 h$^{-1}$ Mpc respectively.

  We have not implemented feedback processes in  the RAMSES code: this
  is currently under  development. As explained  in the introduction,
  galactic winds are a key  ingredient in computing the star formation
  history in  the universe.  In  this paper,  we use  SPH  simulations
  results obtained by  \cite{Springel03b} to estimate the influence of
  winds in the overall baryon budget.

  Three set of  AMR simulations performed by us  using the RAMSES code
  (`convergence  study', `high  efficiency' and  `low efficiency')
  and one set of  SPH simulations \citep{Springel03b} compose the data
  we analyze and discuss intensively in this paper.

  \subsection{Results}
  \label{results}
  We present  now general results obtained  by the RAMSES  code, for a
  typical cooling  and star  formation run. We  would like  to outline
  that   current  high-resolution   numerical   simulations  reproduce
  qualitatively the  global picture of galaxy  formation: fast cooling
  gas builds  up centrifugally supported  discs at the center  of dark
  matter  haloes, in which  star formation  quietly proceeds.  We will
  analyze our results more quantitatively in the next sections.
 
  \begin{figure*}
  \centering \includegraphics[width=\hsize]{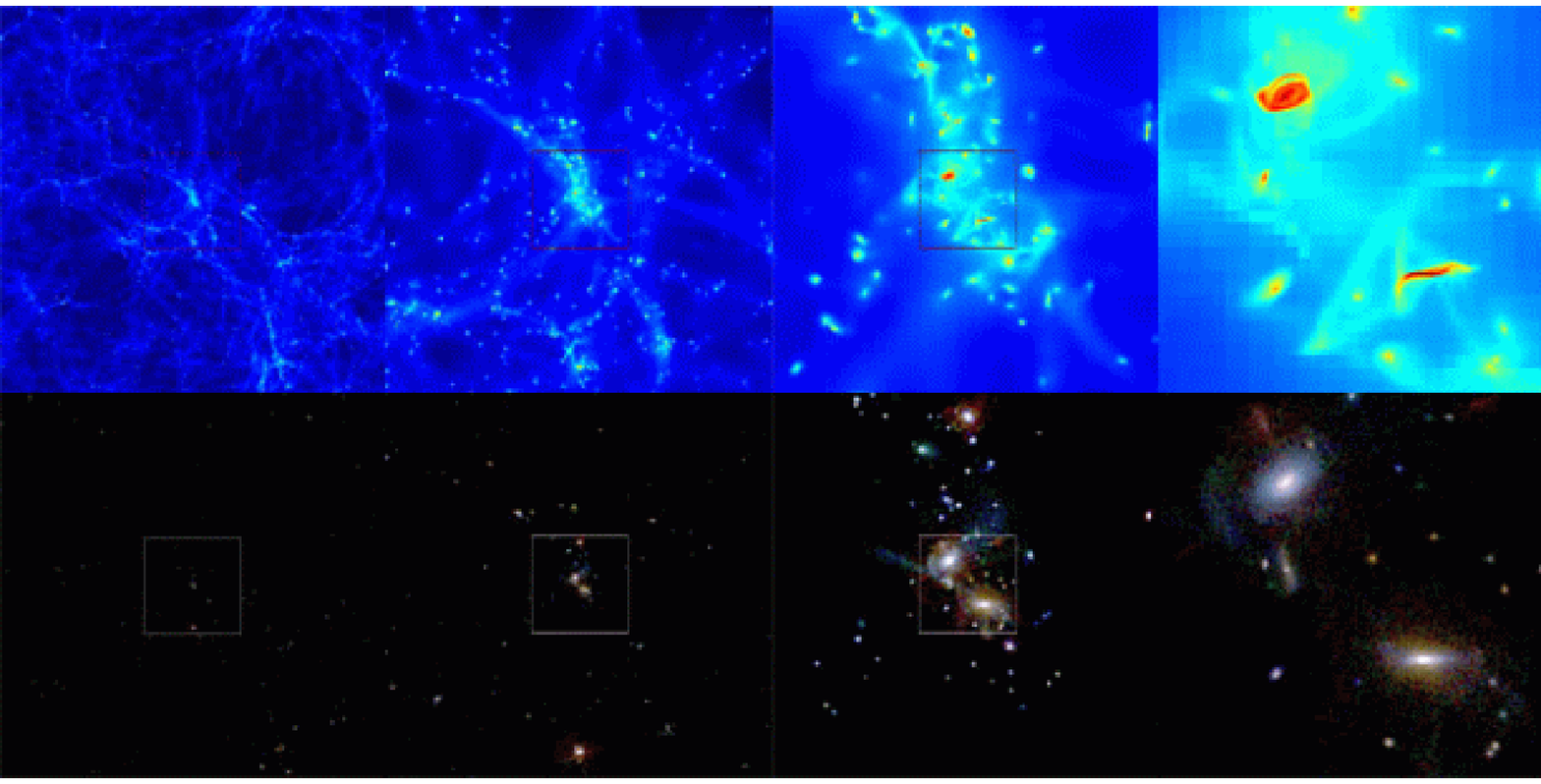}
    \cpt{Gas  density  (up)  and  stellar density  (down)  in  our
      highest resolution run L10N512S30.  The first map on the left of
      each  row shows a projection  of the  full periodic  volume. The
      squares  in each  map delimits  the zoom  region where  the next
      image in  the row  is defined.  For  color map  definitions, see
      text.}
    \label{zoom512}
  \end{figure*}
  
  Figure~\ref{zoom512} presents the  simulated density field projected
  along  one  principal  axis   of  the  comoving  periodic  box.  Run
  L10N512S30 is shown, at  redshift $z=2.8$. The projected gas density
  is shown  with a logarithmically  scaled color table.  The  4 images
  show a zooming  sequence, starting from the whole  periodic box with
  the typical large scale  filamentary structure, down to a particular
  region where 2 spiral discs are clearly visible.

  The  second set  of images  in Figure~\ref{zoom512}  shows  the same
  zooming sequence  for the  projected stellar density.   These images
  were computed using a `true color' color scheme: stars are divided
  into 3  families according  to the formation  redshift of  each star
  particle ($3<z_{\rm f}  < 4$, $4<z_{\rm f}<5$ and $5<z_{\rm f}$).  3 projected density
  maps are computed  for each family.  The 3  images are then combined
  using  the RGB  color scheme:  `Red'  stands here  for old  stars,
  `Blue' for  young stars  and `Green' for  intermediate formation
  redshift stars.
  
  These  rather spectacular  images compare  favorably  with observed
  high redshift  galaxies observed for  example with the  Hubble Space
  Telescope.  We have to be careful in making definitive conclusions:
  the  spiral   discs  we  observe  in  our   simulations  are  highly
  underresolved. Figure~\ref{zoom512} illustrates  this point: the AMR
  grid structure  clearly shows up  in the highest  resolution figure,
  demonstrating that these runs are  not meant to resolve the internal
  structure  of  galactic discs.   We  are  however  confident in  the
  computed baryon fraction inside each  halo, as soon as the halo mass
  is greater than a few hundred particles.

\begin{figure}                                     
    \centering
    \includegraphics[width=\hsize]{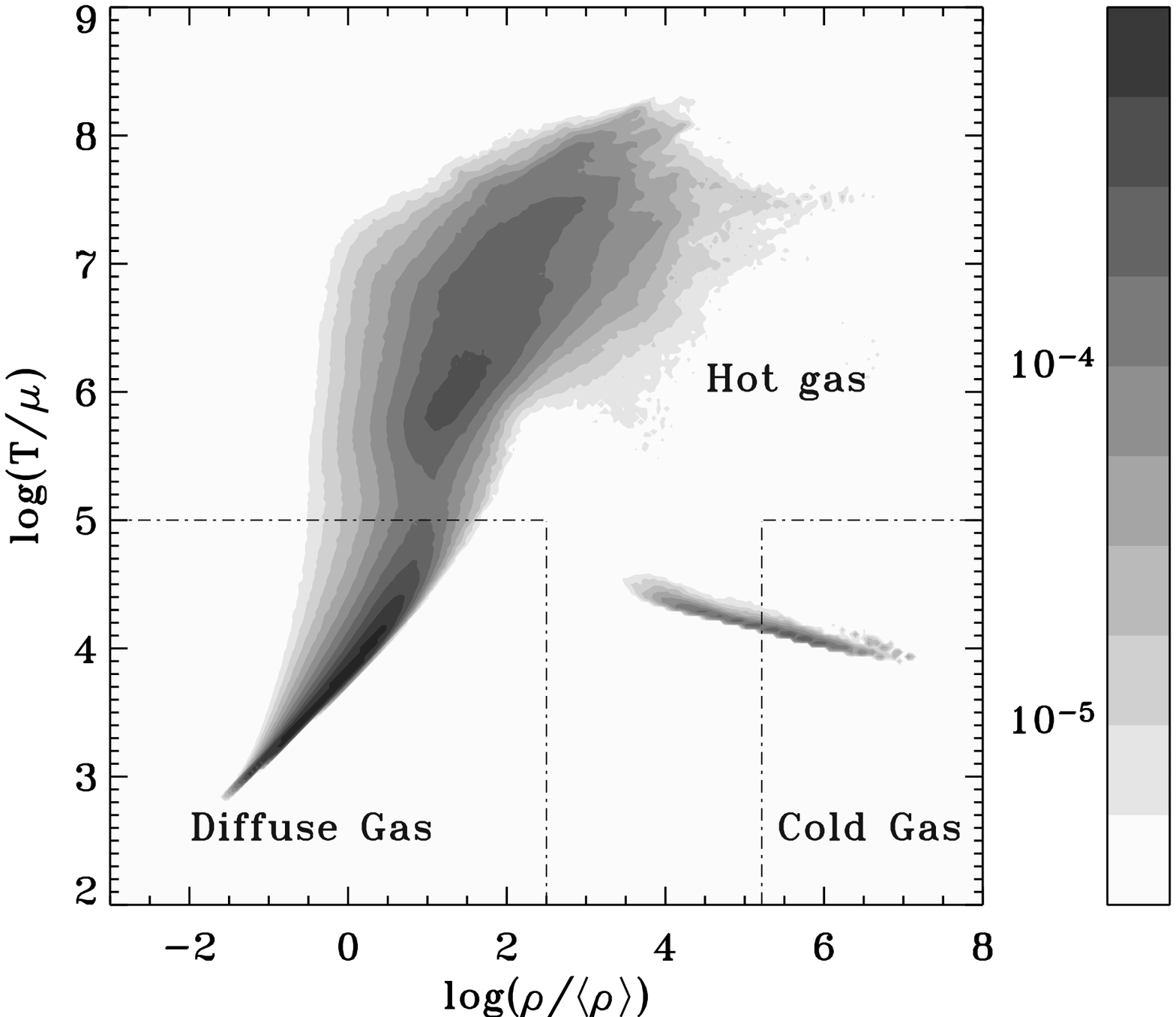} 
    \cpt{The different baryon phases in the $\rho-T$ diagram. Gray
    contours show a mass-weighted  histogram: the baryon mass fraction
    at a given  density and temperature. Each region  corresponds to a
    given phase (diffuse  background, hot or  cold gas), as defined in
    the text.}
    \label{phases}
    \end{figure}

  The traditional  way of describing  the baryon density field  is to
  divide  it into  4 phases.   This  phase separation  is apparent  in
  Figure~\ref{phases},  which shows the  density-temperature histogram
  at $z=0$ for  our large box run (100  h$^{-1}$ Mpc) L100N256S30. The
  first regime occurs at low density  ($\rho < 200 \bar \rho$) and low
  temperature ($T <  10^5$ K) in the phase space  diagram. This is the
  diffuse intergalactic medium, also  known as the Lyman alpha forest.
  The diffuse intergalactic UV flux is responsible for preventing this
  warm gas from collapsing into their parent dark matter halo.

  The second regime  occurs at high density ($\rho  > 10^5 \bar \rho$)
  and  low  temperature  ($T  <  10^5$  K),  and  corresponds  in  our
  simulations  to cold,  centrifugally supported  discs.  This  gas is
  termed here `cold gas'.  It is shown in Figure~\ref{phases} as the
  rightermost box.   In this region,  the quasi-neutral gas  follows a
  very tight $\rho$-T relation, typical of high-density HI discs.

  The remaining  gas corresponds to  shock heated gas  into virialized
  halos.  We call it `hot gas' although, when cooling is efficient,
  its  temperature never exceed  a few  $10^4$ K.  This gas,  in rough
  hydrostatic  equilibrium, spans  a large  density range,  from $\rho
  \simeq \bar \rho$ in the outskirts of dark matter halos up to $\rho
  \simeq 10^5  \bar \rho$ in the  X ray emitting  cores.  This ionized
  gas  is also rapidly  cooling from  the inside  out, leading  to the
  formation of the cold phase.

  The last component in the  baryon budget is the stellar phase. Stars
  originate from the highest density  tail in the phase space diagram.
  In our model, star  formation occurs above the overdensity threshold
  $\rho  > 10^5  \bar  \rho$,  which corresponds  to  our cold  phase.

  Molecular cooling  as well as  supernovae feedback are  not modeled
  here.  The  cold neutral gas  is in reality decomposed  into several
  components, typical of the interstellar medium: molecular, very cold
  clouds embedded  into hot,  supernovae driven bubbles.   The correct
  description of this multiphase interstellar medium is far beyond the
  possibilities   of  current   cosmological   simulations.   The   only
  possibility  is  to rely  on  sub-cell  modeling,  along the  lines
  described  for  example  in \cite{Yepes97}  and  \cite{Springel03a}.
  These  authors  noticed that  their  multiphase  modeling does  not
  affect the overall baryon budget  between the hot halo gas, the cold
  disc component  and the  stellar fraction. The  main effect  of this
  multiphase  approach was  to modify  the internal  structure  of the
  gaseous discs, altering the effective equation of state of the dense
  gaseous component.
  
\begin{figure*}
    \centering
	\includegraphics[width=\hsize]{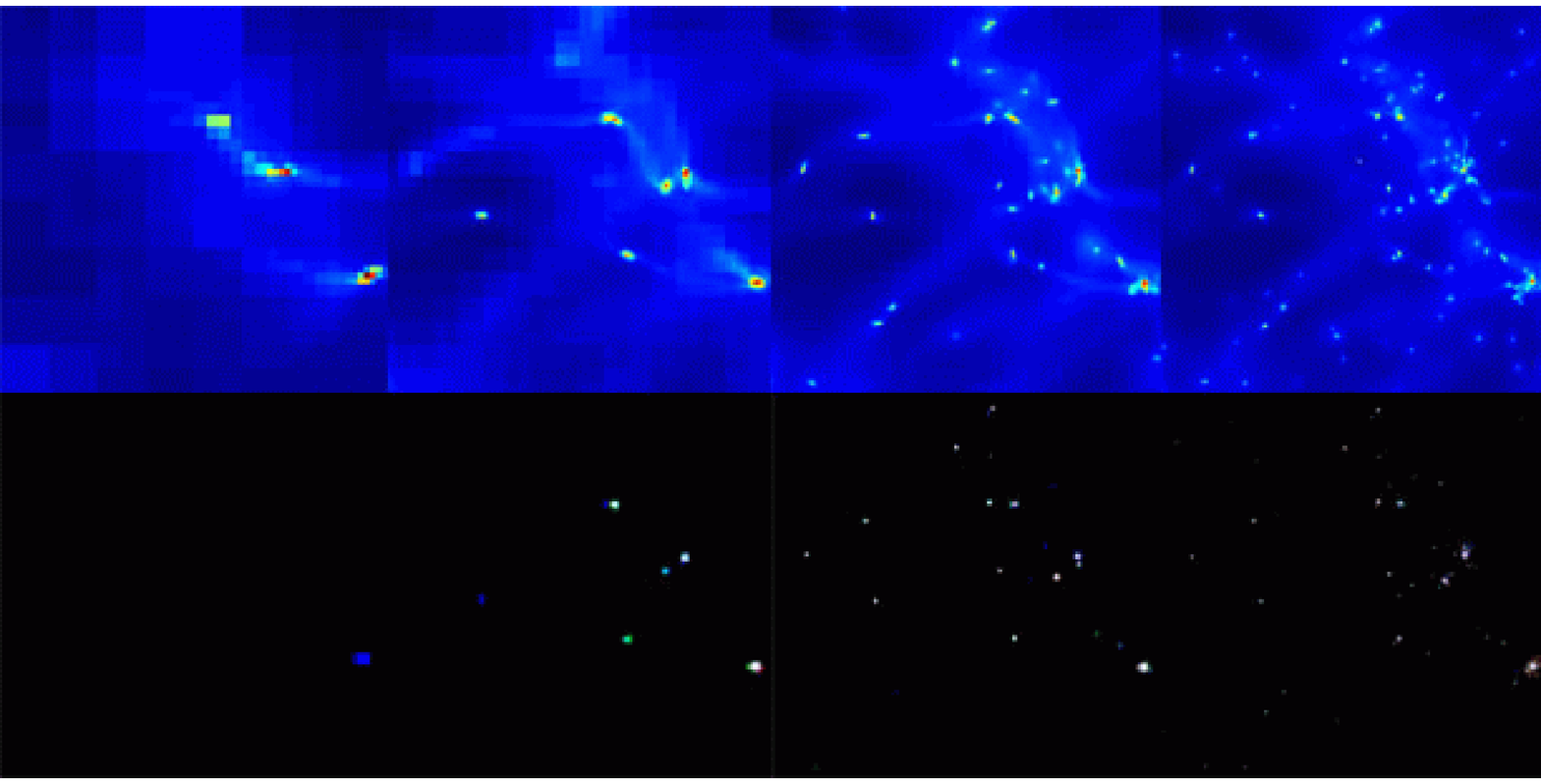}
    \cpt{Projected  density  maps   (top)  and  projected  stellar
      density (bottom) of  our `convergence study' simulations series
      ($L =  10$ h$^{-1}$ Mpc).  The  image size is  1.25 h$^{-1}$ Mpc
      wide on  a side. From left  to right, the number  of dark matter
      particles  is increased  from  $64^3$, to  $128^3$, $256^3$  and
      $512^3$. }
    \label{resolmap}
  \end{figure*}

  The most important feature  that  is likely  to affect the  computed
  baryon  budget  is the  mass resolution  of  the  code.  We have  to
  carefully assess its  effect on our  results.   Let us first
  examine Figure~\ref{resolmap}, which shows  a sequence of images  of
  the projected gas   density fields   in  a  chosen region  of    the
  `convergence   study' suite.  Initial   conditions  were   generated
  self-consistently  for these 4  simulations, in order to recover the
  same large  scale distribution. From  the lowest resolution run with
  $64^3$  particles  to start with,  up to  the highest resolution run
  with $512^3$ dark matter particles, one clearly sees the spectacular
  increase in small  mass haloes along the  filaments and in the voids
  in between.

  If we now examine in Figure~\ref{resolmap} the corresponding stellar
  density  distributions,  we see that  small  haloes appear devoid of
  stars.  Small haloes   are indeed unsufficently resolved   to
  reach the  high density contrast required   to allow star formation.
  Cooling is also very likely affected by the poor resolution in these
  small halos.   We  will see  in the   next  sections that   for  our
  simulations the minimum mass  for a halo  to host  stellar particles
  lies around 400 dark matter particles.

  Another  effect of  mass  resolution  is also  visible  in the  same
  figure. Large galactic discs obtained  at a given resolution tend to
  fragment at higher resolution, leading to smaller discs with several
  orbiting satellites. These satellites are the remnants of progenitor
  halos, which form earlier  at smaller mass. Insufficient resolution
  also affects  the history of  mass assembly inside  galactic halos.
  Figure~\ref{resolmap}  shows that  the color  of a  given  galaxy is
  affected  by the  finite mass  resolution. At  low  resolution, star
  formation  is artificially  delayed  to late  times  and the  galaxy
  appears blue, while at higher resolution, the correct star formation
  history is recovered  and the same galaxy appears  red.  The purpose
  of this  paper is  to carefully estimate  the effect of  finite mass
  resolution on our predictions.

  \section{Model}
  
  In this section, we present a simple analytical model to compute the
  evolution of the baryon budget in the universe. The purpose of this
  analytical model is to shed light on the complex behavior of our
  numerical simulations.  Our analytical treatment of  the cosmic star
  formation  history  is, of  course,  unaffected  by finite  resolution
  effect. It is however a  crude model, and a careful comparison
  with numerical simulations is required to validate our approach.
  
  This  model differs  with   the `semi-analytic' modeling  of  galaxy
  formation,  an approach   pursued by  several  teams \citep{White91,
  Somerville99, Kaufmann99,  Cole00, Hatton03}. These models are based
  on    a quite sophisticated    treatment of  the  physics  of galaxy
  formation: cooling, star formation   and spiral discs  evolution are
  few   examples  among  the  numerous ingredients  in semi-analytical
  modeling.

  In  this paper, our  goal is  to compute the  evolution in  the mass
  fraction of the 4  different components in the baryon  distribution:
  diffuse background, hot  virialized  plasma, cold neutral  discs and
  stars. We will make  predictions for the   universe as a whole,  but
  also for individual halos of  a given mass.  Semi-analytical models,
  when coupled  to N body  simulations, can predict the baryon history
  for individual  halo, based on the  specific merging history  at the
  origin of  the halo hierarchical   mass assembly.  Our  simple model
  allows us to compute only the average  baryon history for halos of a
  given mass range.

  In  a recent  paper, an  analytical  approach has  been proposed  to
  compute    the   star    formation   history    in    the   universe
  \citep{Springel03c}.    The   proposed  method   was   to  use   the
  Press-Schechter  theory  for   the  dark  matter  halos  statistics
  together with the star formation rate as a function of halo mass, as
  measured  in SPH numerical  simulations.  In  the present  paper, we
  develop  a  fully self-consistent  analytical  model, slightly  more
  complex than the one proposed  by \cite{Springel03c}, but based on a
  similar  approach. This  self-consistency allows  us to  compute the
  mass  fraction  evolution  of  the  4 baryon  components,  for  the
  universe as a whole, and also for an individual {\it average} halo.

  \subsection{Halo Model}

  The   method we use   in this  paper  to compute  the star formation
  history  is   based on   what is  usually    called the  halo  model
  \citep{Cooray02}.   The idea is to  decompose dark matter and baryon
  density   fields   into a  collection   of   virialized halos, whose
  distribution is described by the \cite{Press74} mass function.  This
  approach    was introduced  to    describe  galaxy  and dark  matter
  clustering, using  two  additional ingredients:  linear halo biasing
  theory   and  the Navarro, Frenk  \&   White (1995)  density profile
  \citep{Ma00,  Seljak00}.   Later on,  several  authors used  similar
  tools  to compute the Sunyaev-Zeldovich  power  spectrum, as well as
  the mean Comptonization parameter \citep{Cooray00, Refregier02}.

  A feature common to most of these earlier works is the rather static
  picture they give to dark  matter halos evolution.  We propose here
  to use a  similar approach, in order to compute  the history of star
  formation  within each  dark matter  halo.  Our  methodology differs
  somewhat  from  earlier works,  since  it is  based  on  a two  step
  approach. 

  In  the  first step,  using static  halo  model,   we compute  the mass
  transfer rates between each phase.  In the second step, using
  these computed mass transfer  rates, we solve  for the mass fraction
  evolution equations, a  system of first  order Ordinary Differential
  Equations (ODE).    Our analytical model is   therefore based on the
  computation of the baryon logistics.  In order to compute the baryon
  mass fraction locked  up in stars for  example, we need to solve the
  complete set of ODE from  a large redshift ($z$ =  200 say) down  to
  $z$=0.

  In this paper,  we define the halo mass $M$  as $M_{\rm 200}$, the total
  mass  enclosed  in  radius  $R_{\rm 200}$, where  the  mean  overdensity
  (relative to the background density) is $\Delta = 200$.
  \begin{eqnarray}
    M=M_{\rm 200}=\frac{4\pi}{3} \bar \rho(z) \Delta R_{\rm 200}^3.
  \end{eqnarray}
  This  choice differs  from earlier  definitions, where  $\Delta$ was
  either defined relative  to the critical density, or  $\Delta$ was a
  function of redshift, as  suggested by the spherical collapse model.
  As  shown  by  \cite{Jenkins01}  and \cite{White02},  these  earlier
  definitions  are not  suited  to the  Press  \& Schechter  approach.
  Although  they are based  on physical  principles, they  destroy the
  self-similarity of the Press \& Schechter mass function.
  
  In this  paper, we therefore adopt  $\Delta = 200$  (relative to the
  mean background  density) independent of redshift. We  also use very
  often  the halo  circular  velocity $V_{\rm 200}$  and  the halo  Virial
  temperature  $T_{\rm 200}$,  instead  of  the  halo  mass.   The  Virial
  temperature must  not be considered as a  true physical temperature,
  but rather as  yet another mass parameterization. In  this paper, the
  Virial temperature is defined as
  \begin{eqnarray}
    k_{\rm B}T_{\rm 200}=\frac{\mu m_{\rm H}}{2}\frac{GM_{\rm 200}}{R_{\rm 200}},
    \label{virial}
  \end{eqnarray}
  (with $\mu$ the mean molecular weight) and the circular velocity as
  \begin{eqnarray}
     V_{\rm 200}= \sqrt{\frac{GM_{\rm 200}}{R_{\rm 200}}}.
    \label{circular}
  \end{eqnarray}  

  In the 5 following sections, we are going to compute the cosmic rates between the 4 phases. If necessary the reader can go directly to section \ref{chain}. 

  \subsection{Minimal Mass}

  The first component in the baryon budget  is the diffuse background.
  This may be the  most important one, since   it is the  reservoir of
  fresh gas that   will eventually  feed   star forming halos at   all
  epochs. It is usually called the  Inter Galactic Medium (IGM) or the
  Lyman Alpha Forest. We need to give  a precise definition of what we
  call `diffuse background' in this paper.

  The diffuse background  is the baryon  component associated to  dark
  matter halo   with masses lower  than  the  minimal mass $M_{\rm min}$,
  below  which  cooling is  inefficient  and  pressure  forces prevent
  baryons to  collapse in their potential wells.  This minimal mass is
  therefore fundamental because  it  is the transition  between `star
  forming halos' and `diffuse' one.

  This Minimal Mass is  taken to be the  maximum between the Filtering
  Mass and the  Minimal Cooling Mass. The filtering  mass $M_{\rm F}$ is the
  minimal halo mass  above which baryons  are able to  fall into their
  dark matter halo potential wells (see Appendix~\ref{Filtering}). The
  minimal  cooling mass $M_{\rm cool}$ is the  mass above which the gas is
  able     to     cool   and  therefore      to    form    stars  (see
  Appendix~\ref{coolingmodel}).

  We implement this using a smooth  function of both Virial temperatures
  \begin{eqnarray} T_{\rm min} = T_{\rm F} + T_{\rm cool}. \end{eqnarray}

  \begin{figure}
    \centering \includegraphics[width=\hsize]{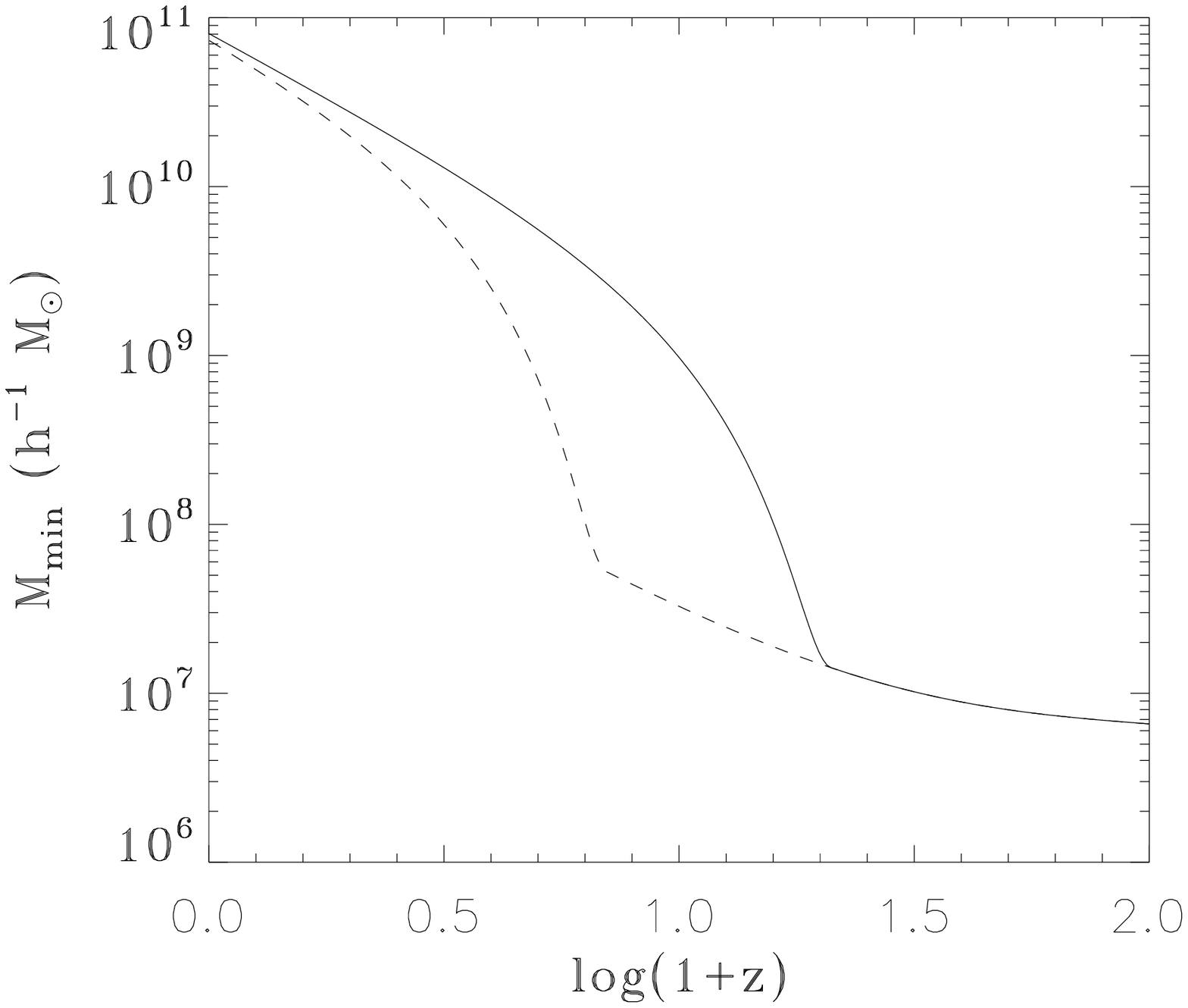}
    \cpt{Time  evolution of  the  Minimal Mass  $M_{\rm min}$ for  two
      reionization  scenarios:  $z_{\rm r}$=20   (solid  line)  and  $z_{\rm r}$=6
      (dashed line).   In both case, the  background temperature after
      reionization  was  set  to  $T_{\rm r}=6\times 10^3$  K.  The  Minimal
      Cooling temperature was also set to $T_{\rm cool}=6\times 10^3$ K.}
    \label{figtf}
  \end{figure}

  \noindent  The Minimal  Mass is  finally computed  using  the Virial
  relation   (Eq.~\ref{virial}).    We   plot  in   Figure~\ref{figtf}
  $M_{\rm min}$  as   a  function  of   redshift,  for  our   two  extreme
  reionization scenarios.   At early times, this  mass remains roughly
  constant, independent  of redshift.  As  reionization proceeds, this
  Minimal Mass increases steadily, up  to a rather large value $M_{\rm min}
  \simeq 10^{11}$ h$^{-1}$M$_{\odot}$ today.

  \subsection{Cosmic Accretion Rate}
  
  Using the Press \& Schechter formalism, we now compute the mass fraction in
  the diffuse background and the mass fraction in star forming halos.
  Since  the Minimal  Mass is  considered here  as the  mass threshold
  between these two components,  and assuming that the baryon fraction
  in each halo is equal to the universal one, we have
  \begin{eqnarray}
    f_{\rm hot} = f(M>M_{\rm min}) = f_{\rm b}
    \int^\infty_{\nu_{\rm min}} \sqrt{\frac{2}{\pi}}\exp(-\nu^2/2)d\nu,
  \end{eqnarray}
  with
  \begin{eqnarray}
      \nu_{\rm min} = \frac{\delta_{\rm c}(t)}{\sigma(M_{\rm min})}
      \mbox{~~~and~~~}
      \delta_{\rm c}(t) = \frac{1.686}{D^+(t)},
  \end{eqnarray}
  where $D^+$  is the linear  growth factor and  $\sigma(M_{\rm min})$ the
  variance  of  the  density   field  smoothed  at  the  Minimal  Mass
  scale. The  rate at which  baryons are transferred from  the diffuse
  background to  star forming  halos is computed  by taking  the time
  derivative of the previous equation
  \begin{eqnarray}
    \dot{f}_{\rm acc} = \frac{df_{\rm hot}}{dt} = -  \frac{df_{\rm back}}{dt}
    = - f_{\rm b}\dot{\nu}_{\rm min}\sqrt{\frac{2}{\pi}}
    \exp(-\nu_{\rm min}^2/2),
  \end{eqnarray}
  We then define  the Cosmic Accretion Rate of  fresh diffuse gas into
  star forming halos by
  \begin{eqnarray}
    \dot{f}_{\rm acc} = \omega_{\rm acc}f_{\rm back},
    \label{accmass}
  \end{eqnarray}
  where the accretion rate, in units of Gyr$^{-1}$, is given by
  \begin{eqnarray}
    \omega_{\rm acc} = - \dot{\nu}_{\rm min}\sqrt{\frac{2}{\pi}}
    \frac{\exp(-\nu_{\rm min}^2/2)}{\mbox{erfc}(\nu_{\rm min}/\sqrt 2)}.
    \label{accrate}
  \end{eqnarray}
  This last equations give the mass accretion rate of diffuse gas into
  star forming  halos {\it in the  general case}, for  which the mass
  fraction  in the  diffuse  component  is allowed  to  vary from  its
  canonical value. Note that this accretion rate has nothing in common
  to   the  traditional   mass  accretion   rate  on   a   given  halo
  \citep{Lacey93}. This  rate gives the fraction of  fresh diffuse gas
  dispatched among all star forming  halos. This fresh gas in assumed
  to be transferred  exclusively to the hot plasma  component. The two
  variables $f_{\rm back}$ and $f_{\rm hot}$ refer therefore to the total mass fraction
  in the background and the  total mass fraction in the hot component,
  both integrated over the PS distribution.
  
  It is also  possible to compute the Cosmic Accretion  Rate on a halo
  by   halo  basis,   using  the   Extended  Press   Schechter  theory
  \citep{Bond91,   Lacey93}.  This  theory   allows  to   compute  the
  progenitors mass  distribution as  a function of  time, for  a given
  parent halo  mass $M_{\rm 0}$,  up to the  `halo formation  time' $t_{\rm 0}$.
  The individual  Cosmic Accretion Rates  are very similar to  the one
  computed  for the  whole universe.   We follow  the  same procedure,
  computing first  the mass fraction in star  forming halos, assuming
  that each progenitor hosts a  baryon fraction equal to the universal
  one.
  \begin{eqnarray}
    f_{\rm hot}(M_{\rm 0},t_{\rm 0}) = f_{\rm b}
    \int^\infty_{\nu_{\rm min}} \sqrt{\frac{2}{\pi}}\exp(-\nu^2/2)d\nu,
  \end{eqnarray}
  with this time
  \begin{eqnarray}
      \nu_{\rm min}(M_{\rm 0},t_{\rm 0}) = \frac{\delta_{\rm c}(t) -\delta_{\rm c}(t_{\rm 0})}
	 {\sqrt{\sigma(M_{\rm min})^2-\sigma(M_{\rm 0})^2}}.
	 \label{extnuf}
  \end{eqnarray}
  The  accretion rate  is then  computed exactly  as for  the previous
  case, using Equations~\ref{accmass}  and \ref{accrate}, with however
  a different value for $\nu_{\rm min}$ given by Equation~\ref{extnuf}. We
  want to stress again that  we do not consider accretion of satellite
  halos on the  most massive progenitor, which is  the traditional way
  of  computing the accretion  rate.  Here,  we consider  accretion of
  diffuse gas on all star forming progenitors of the final halo.

  \begin{figure}
    \centering \includegraphics[width=\hsize]{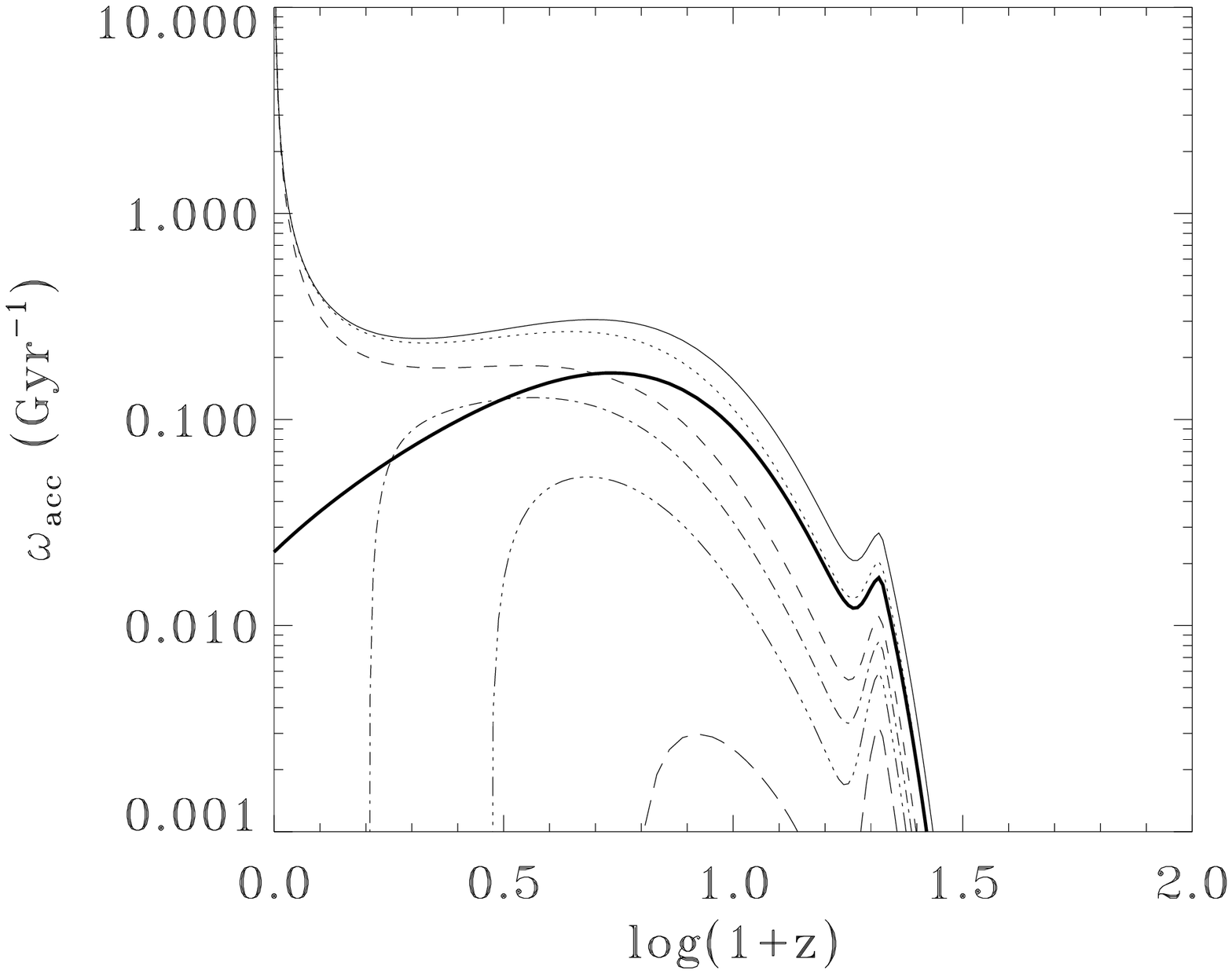}
    \cpt{Cosmic Accretion Rate for the $\Lambda$CDM cosmology with
      $z_{\rm r}=20$ for the universe as  a whole (thick solid line) and for
      various halo  masses (thin lines).   Halo mass are, from  top to
      bottom,  $M_{\rm 0}  =   10^{13}$,  $10^{12}$,  $10^{11}$,  $5  \times
      10^{10}$,  $2.5 \times 10^{10}$  and $10^{10}$ h$^{-1}$M$_{\odot}$. The
      halo formation redshift is set to $z_{\rm 0}=0$.}
    \label{figwacc}
  \end{figure}

  This fresh gas  contributes to fill up dark  matter halos with hot,
  virialized   gas.   Hot   gas  coming   from  satellite   halos  is
  automatically  accounted for in  our formalism.   If we  assume very
  short cooling  rates and  instantaneous star formation,  this Cosmic
  Accretion  Rate is  nothing but  the Star  Formation History  in the
  universe.  In a realistic case, star formation and cooling introduce
  a delay in the curve.

  This cosmic  accretion rate  depends on the  thermal history  of the
  background gas,  on the density power  spectrum through $\sigma(M)$,
  and     on    the     cosmological    model     through    $D^+(t)$.
  Figure~\ref{figwacc} shows the accretion  rates (in Gyr$^{-1}$) as a
  function of redshift, for a $\Lambda$CDM universe, and for different
  halo masses. The halo formation redshift was fixed to $z_{\rm 0} = 0$. 

  For small mass halos,  accretion stops abruptly  as the Minimal Mass
  reaches   the  parent halo  mass.   This   means  that star  forming
  progenitors are  not present anymore  in the  halo.  For  halos more
  massive than $M_{\rm min}$, accretion remains    active up to the   halo
  formation redshift.  The   accretion  rate actually diverges as   $z
  \rightarrow z_{\rm 0}$, as all the remaining diffuse  gas is accreted into
  the final virialized halo.  This diffuse mass accretion rate remains
  however finite for $M_{\rm 0} > M_{\rm min}$, and  we can compute its value as
  $z   \rightarrow  z_{\rm 0}$   \begin{eqnarray}   \dot{f}_{\rm acc}  =  -  f_{\rm b}
  \sqrt{\frac{2}{\pi}}                      \frac{\dot{\delta}_{\rm c}(t_{\rm 0})}
  {\sqrt{\sigma(M_{\rm min})^2-\sigma(M_{\rm 0})^2}}.   \label{asymptoticaccrate}
  \end{eqnarray}

  \subsection{Cosmic Cooling Rate}
  
  Our third baryon component  (namely cold atomic gas in centrifugally
  supported discs) is progressively  built up by accreting cooling gas
  into the very  center of their parent dark matter  halo.  We need to
  estimate the global  rate at which hot gas  is transferred into this
  dense and  cold component.  Using  EPS theory, we compute  this rate
  for individual halo  mass ($M_{\rm 0}$, $z_{\rm 0}$). Note that we recover the results for the Universe as a whole by taking the limit $M_{\rm 0} \rightarrow +\infty$ and $z_{\rm 0} \rightarrow -1$.
  
  We follow our basic methodology, assuming this time that all baryons
  are in the hot halo phase.   Using our simple cooling model detailed
  in  Appendix~\ref{coolingmodel}, we compute  the total amount of gas
  cooling from the hot  halo component during  a unit time interval by
  integrating  the  instantaneous  cooling   rate  over  the PS   mass
  distribution from $M_{\rm min}$  to $M_{\rm max}$, the Maximal Cooling Mass.
  Indeed, above this  mass (see Appendix~\ref{coolingmodel}), the halo
  enter in  the   slow regime  cooling.   We therefore  neglect   this
  contribution. It gives

  \begin{equation}
    \dot f_{\rm cool} = f_{\rm b}\frac{1}{t_{\rm orb}}
    \int^{\nu_{\rm max}}_{\nu_{\rm min}} \sqrt{\frac{2}{\pi}}\exp(-\nu^2/2)d\nu,
  \end{equation}
  where   $\nu_{\rm min}$   is   defined  by   Equation~\ref{extnuf},
  $\nu_{\rm max}$ corresponds to the Maximum Cooling Mass $M_{\rm max}$
  \begin{eqnarray}
    \nu_{\rm max}(M_{\rm 0},t_{\rm 0}) = \frac{\delta_{\rm c}(t) -\delta_{\rm c}(t_{\rm 0})}
       {\sqrt{\sigma(M_{\rm max})^2-\sigma(M_{\rm 0})^2}},
       \label{extnumax}
  \end{eqnarray}
  and $t_{\rm orb}$ is the orbital decay timescale (see Appendix~\ref{coolingmodel}).
  We then define the Cosmic Cooling Rate of hot halo gas into
  cold gaseous discs by 
  \begin{eqnarray}
    \dot{f}_{\rm cool} = \omega_{\rm cool}f_{\rm hot},
    \label{cool}
  \end{eqnarray}
  where the cooling rate, in units of Gyr$^{-1}$, is given by
  \begin{eqnarray}
    \omega_{\rm cool} = \frac{1}{t_{\rm orb}}
    \frac{\mbox{erfc}(\nu_{\rm min}/\sqrt 2) - \mbox{erfc}(\nu_{\rm max}/\sqrt 2)}
	 {\mbox{erfc}(\nu_{\rm min}/\sqrt 2)}.
  \end{eqnarray}
  The Cosmic  Cooling Rate depends  on the cosmological model,  on the
  thermal history of the background  and on the details of the cooling
  model. A  similar model  has been  proposed by \cite{VanDenBosch02},  in a
  different context.

  \subsection{Star formation models\label{SF}}
  \label{starformmodel}

  In  this   simple  analytical  model,  we   completely  discard  the
  description  of the  gaseous discs.  Predicting the  disc  sizes and
  surface density  profiles obtained  in the hierarchical  scenario of
  structure formation is beyond the scope of this paper. We are only
  interested in the global baryon budget, and more precisely in the 
  global star formation history.

  We  therefore consider star  formation in  a dark  matter halo  as a
  function of  the total amount  of cold gas  in that halo.   The star
  formation  rate  in  each  halo  is simply  given  by  $\dot{M}_*  =
  \omega_* M_{\rm cold} $ with  $\omega_*$ is the average star  formation rate
  in that halo. In order to compute this average time scale from first
  principles,  one needs  to integrate  the local,  density dependent,
  star formation rate over the cold gas density PDF.

  In  this  analytical  model,  however, we  consider  star  formation
  models  inspired  by  star  formation recipes  used  in  numerical
  simulations and by  semi-analytical models \citep{Somerville01}. The
  halo star formation rate is parameterized by
  \begin{eqnarray}
    \omega_*  = \frac{1}{t_*}  (1+z)^{\alpha_*/2},
  \end{eqnarray}
  where  $t_*$  is the  present  day  star  formation time  scale  and
  $\alpha_*$ is the acceleration parameter. In the literature, two basic
  quiescent models are usually discussed in galaxy formation studies.

  The  first model, usually  referred to  as a  `constant efficiency'
  model,  assumes that  the halo  star  formation time  is a  constant
  $\alpha_*  = 0$.   This model  corresponds to  numerical simulations
  with a constant star formation density threshold.

  The second model assumes that  $\alpha_* = 3$.  It is usually called
  an `accelerated  efficiency' model.   The star formation
  time  scale decreases  with redshift  (as  the mean  density of  the
  Universe   increases).    This   model  corresponds   to   numerical
  simulations  with  a   constant  star  formation  {\it  overdensity}
  threshold.   It is  also used  in semi-  analytical models  to mimic
  starbursts triggered by mergers \citep{Somerville01}.

  We compute  now  the global star  formation  rate,  using our  basic
  methodology. Since the    halo star formation  rate,  in  our simple
  scenario, does not  depend on halo  mass, we can integrate  over the
  EPS mass  function,  and obtain the Cosmic   Star Formation Rate  as
  \begin{eqnarray}     \dot{f}_{*} =  \omega_*   f_{\rm cold}. \label{star}
  \end{eqnarray}

  \begin{figure}
    \centering \includegraphics[width=\hsize]{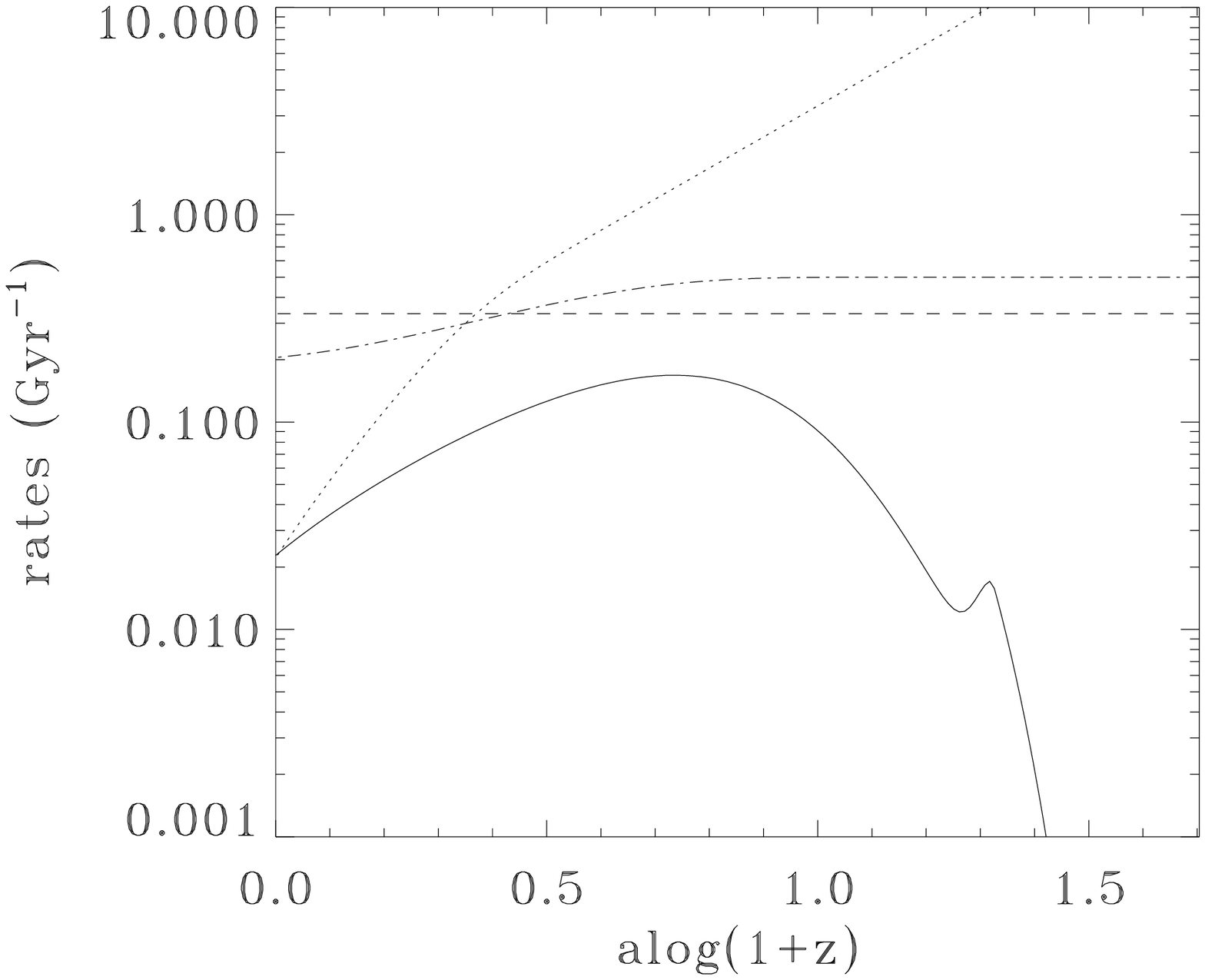}
    \cpt{Cosmic Accretion  Rate (solid line),  Cosmic Cooling Rate
      (dotted  line), Cosmic  Star  Formation Rate  (dashed line)  and
      Cosmic Outflow Rate (unbound  fraction, dot-dashed line) for the
      $\Lambda$CDM  cosmology  with  the following  model  parameters:
      $z_{\rm r}=20$, $t_* = 3$ Gyr, $\alpha_*  = 0$, $T_{\rm w} = 2\times 10^6$ K
      and   $\eta_{\rm w}  =   1.5$.   These  rates   were  computed   using
      $M_{\rm 0}=+\infty$  and $z_{\rm 0}=-1$,  and therefore  corresponds  to the
      universe as a whole.}
    \label{figrates}
  \end{figure}

  \subsection{Cosmic Winds}

  The  last  ingredient  in our  model,  but  not  the least,  is  the
  contribution of galactic winds to the overall baryon budget. It is a
  well known issue in current  models of galaxy formation that without
  feedback  processes, most  baryons would  end  up into  cold gas  or
  stars,  in  contradiction  with several  observational  constraints.
  This problem is known as the `overcooling problem' \citep{Blanchard92}.

  As discussed in  the introduction, the exact  nature of the dominant
  feedback process is still  unknown.  It is  most likely that various
  processes are in  competition, and their impact  on baryons may vary
  as a function of halo  mass. Following \cite{Springel03b}, we assume
  in our model that winds occur during star formation events, probably
  related to supernovae. We therefore assume  that cold gas is ejected
  from the disc with a typical wind velocity $u_{\rm w}$  and with a typical
  outflow  rate  \begin{eqnarray}   \dot{M}_{\rm wind}=\eta_{\rm w}  \dot{M}_*.
  \end{eqnarray}  The two   additional parameters  are  $\eta_{\rm w} \simeq
  1-5$, the wind efficiency, and  $u_{\rm w} \simeq 200-500$ km s$^{-1}$, the
  wind  velocity  \citep{Springel03b}.    These  wind  parameters  are
  typical   of   observed    outflows   in   star   forming   galaxies
  \citep{Martin99}.

  The fate of  this ejected gas depends on the halo  mass. If the wind
  velocity exceeds  the escape velocity  of the halo, the  ejected gas
  leaves the halo into the diffuse background, from where, eventually,
  it  will  be   accreted  again.   Such  winds  are   referred  to  as
  `unbound'.  If, on the  other hand, the  halo is too  massive, the
  ejected gas  remains in the hot  halo component, from  where it will
  eventually cool again. Such winds are referred to as `bound'.

  Assuming  again  that all   baryons are   locked  up  into  the cold
  component, we can   compute   the  global   wind outflow  rate    by
  integrating  over  the EPS  mass  function.   We finally  obtain the
  following  equations,  valid  in the  general  case \begin{eqnarray}
  \dot{f}_{\rm wind} = \omega_{\rm w} f_{\rm cold}. \label{wind} \end{eqnarray} where
  the  outflow  rate,   in    units  of   Gyr$^{-1}$,  is   given   by
  \begin{eqnarray} \omega_{\rm w} =  \eta_{\rm w} \omega_*. \end{eqnarray} We first
  compute  the unbound fraction.   It  corresponds to winds emitted by
  halos whose escape velocity is smaller than the wind velocity. Using
  the  EPS  distribution,    we    get \begin{eqnarray}   \zeta_{\rm w}    =
  \frac{\mbox{erfc}(\nu_{\rm min}/\sqrt 2) - \mbox{erfc}(\nu_{\rm w}/\sqrt 2)}
  {\mbox{erfc}(\nu_{\rm min}/\sqrt  2)}, \end{eqnarray}  where   $\nu_{\rm w}$ is
  defined  by  the  `Wind  Mass'  \begin{eqnarray}  \nu_{\rm w}(M_{\rm 0},t_{\rm 0}) =
  \frac{\delta_{\rm c}(t)                                    -\delta_{\rm c}(t_{\rm 0})}
  {\sqrt{\sigma(M_{\rm w})^2-\sigma(M_{\rm 0})^2}}. \label{nuwind} \end{eqnarray}
  This `Wind Mass'  (the halo  mass  above which winds  are bound)  is
  related  to the  wind velocity  by  noticing  that for typical  dark
  matter halos, $v_{\rm esc} \simeq 3 V_{\rm 200}$.  Using the standard Virial
  relation (Eq.~\ref{virial}), we  obtain the wind  Virial temperature
  \begin{eqnarray} k_{\rm B} T_{\rm w}  = \frac{1}{18} \mu m_{\rm H} u_{\rm w}^2. \label{twind}
  \end{eqnarray} The bound fraction is just  $1-\zeta_{\rm w}$.  The fate of
  the unbound gas depends now on the  parent halo mass $M_{\rm 0}$.  If $M_{\rm 0}
  > M_{\rm w}$,  the gas is recycled   into the halo  diffuse component, and
  ultimately into  the halo hot component as  $z \rightarrow z_{\rm 0}$.  If
  $M_{\rm 0} < M_{\rm w}$, the gas is  lost into the intergalactic medium, outside
  the  boundaries of the parent halo,  and never come back.  Note that
  in the  latter  case, $\zeta_{\rm w}$ is  always equal  to  one. This very
  crude  model turns  out to  be  surprisingly accurate  in predicting
  results obtained   by  numerical  simulations   (see  the  following
  sections).

  \begin{figure}
    \centering \includegraphics[width=\hsize]{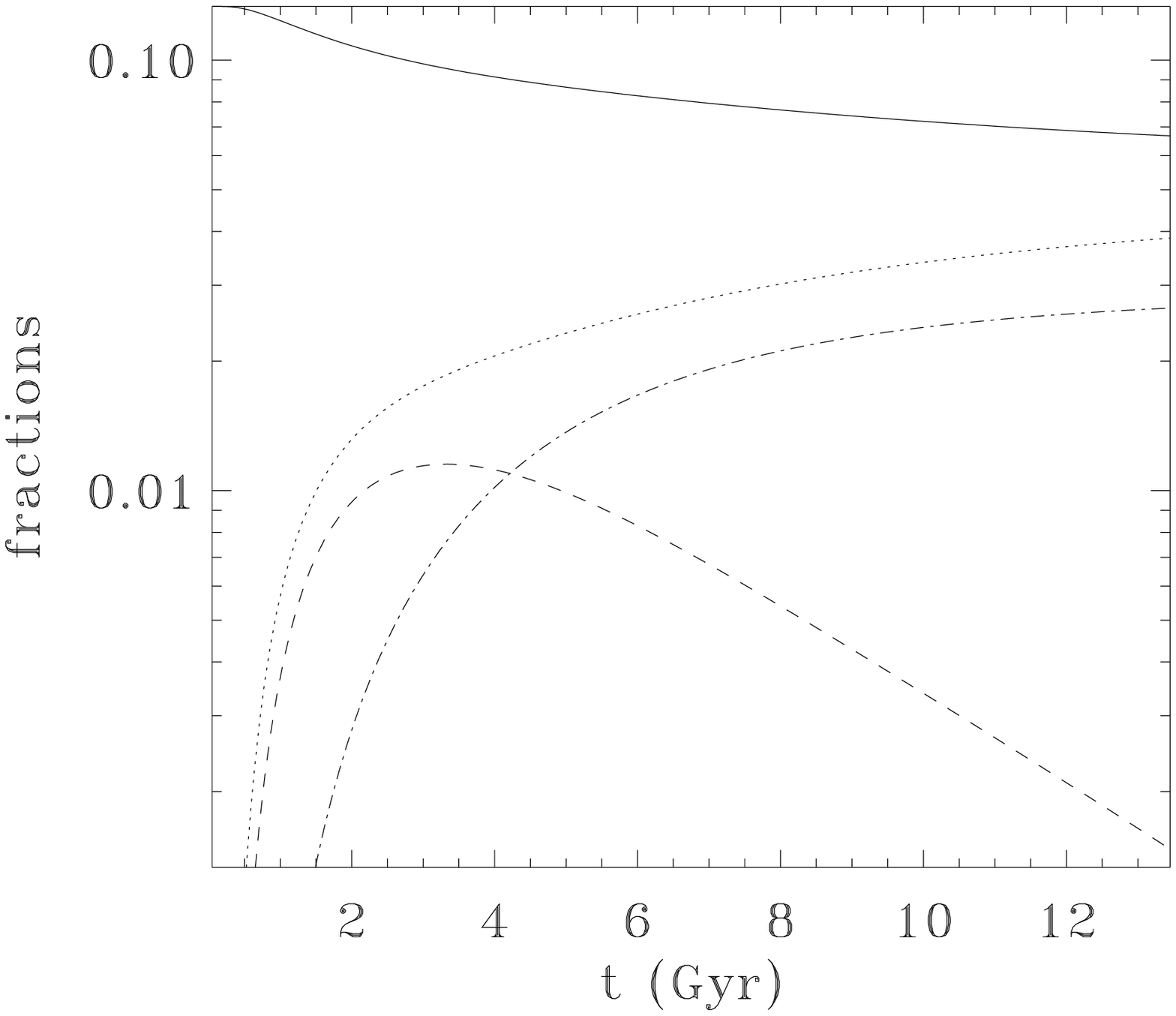}
    \cpt{History of  the mass  fraction in the  diffuse background
      (solid line),  in the  hot halo gas  (dotted line), in  the cold
      discs  (dashed line)  and  in stars  (dot-dashed  line) for  the
      $\Lambda$CDM  cosmology  with  the following  model  parameters:
      $z_{\rm r}=20$, $t_* = 3$ Gyr, $\alpha_*  = 0$, $T_{\rm w} = 2\times 10^6$ K
      and  $\eta_{\rm w}  =  1.5$.   These  fractions  were  computed  using
      $M_{\rm 0}=+\infty$  and $z_{\rm 0}=-1$,  and therefore  corresponds  to the
      universe as a whole.}
    \label{figfrac}
  \end{figure}

  \subsection{The Baryon Supply Chain}
  \label{chain}
  
  We are now in a position to compute the  baryon budget history.  The
  last sections  were devoted to computing mass  transfer rates
  between our 4 baryon components. The mass fraction in each component
  are the independent variables in our problem: $f_{\rm back}$, $f_{\rm hot}$,
  $f_{\rm cold}$ and  $f_*$, referring respectively to diffuse background,
  hot gas, cold discs and stars.

  The  methodology we follow  in this  paper allows  us to  compute in
  advance   3  important   mass  transfer   rates.   This   rates  are
  $\omega_{\rm acc}$,    $\omega_{\rm cool}$   and    $\omega_{*}$,   referring
  respectively to  the Cosmic Accretion Rate, the  Cosmic Cooling Rate
  and  the Global  Star Formation  Rate.   Our very  crude wind  model
  provides  us  with  an  additional  parameter,  namely  the  unbound
  fraction   $\zeta_{\rm w}$.   These   various   rates   are   plotted   in
  Figure~\ref{figrates} for our fiducial model and our notations are summarized in table \ref{table4}. 

\begin{table}
\begin{tabular}{|c|c|}
\hline
$z_{\rm r}$&reionization redshift\\
\hline
$t_*$&star formation time scale\\
\hline
$\eta_{\rm w}$&wind efficiency\\
\hline
$\zeta_{\rm w}$&unbound wind fraction\\
\hline
$f_*$&stellar fraction\\
\hline
$f_{\rm cold}$&cold gas fraction\\
\hline
$f_{\rm hot}$&hot gas fraction\\
\hline
$f_{\rm back}$&background gas fraction\\
\hline
$\omega_*$&star formation rate\\
\hline
$\omega_{\rm cool}$&cooling rate\\
\hline
$\omega_{\rm acc}$&accretion rate\\
\hline
\end{tabular}
\label{table4}
\cpt{Main notations}
\end{table}

  We  have to solve  a set  of  ordinary differential  equations, with
  pre-computed transition rates  between each component  of our baryon
  supply chain.    \begin{eqnarray}  \frac{df_{\rm   back}}{dt}    &=&
  \zeta_{\rm w}   \eta_{\rm w}  \omega_*  f_{\rm cold}   - \omega_{\rm
  acc}f_{\rm    back},   \label{fback}\end{eqnarray}  \begin{eqnarray}
  \frac{df_{\rm  hot}}{dt}    &=&    \omega_{\rm acc}f_{\rm   back}  -
  \omega_{\rm cool}f_{\rm  hot}   +  (1-\zeta_{\rm w})   \eta_{\rm  w}
  \omega_* f_{\rm cold}, \end{eqnarray} \begin{eqnarray} \frac{df_{\rm
  cold}}{dt} &=& \omega_{\rm cool}f_{\rm hot} - \omega_* f_{\rm cold}-
  \eta_{\rm  w} \omega_* f_{\rm cold}, \end{eqnarray} \begin{eqnarray}
  \frac{df_*}{dt} &=&    \omega_* f_{\rm cold}.    \end{eqnarray}   If
  $M_{\rm 0}   > M_{\rm w}$, one  sees  that the total  baryon mass is
  conserved. Note that Eq. \ref{fback} should be modified if $M_{\rm 0}<M_{\rm w}$: the wind contribution is set to $0$ and  therefore the total   baryon mass  in  the  parent halo is  not conserved  anymore.   This is  as   expected, since   winds  are now escaping outside the parent halo boundaries.

  Using  any time integration method   of sufficient accuracy, one can
  finally solve for  the previous  set  of differential equations.  We
  used in this  work a Backward  Euler scheme. Interestingly,  one can
  solve  formally the latter system   using matrix exponentials.  This
  type of equations are typical of galaxy  formation studies, like the
  early  work  of  \cite{Tinsley80}.  More recently, \cite{Pei99} have
  designed a similar approach, based on the observed galaxy luminosity
  functions,  while  here,  our equations  are  based  on  EPS theory.
  Figure~\ref{figfrac} shows the   baryon  budget  evolution for   our
  fiducial  model.    Before   applying   this analytical  model    to
  cosmological observations, we  need to determine its  validity range
  using high resolution numerical simulations.
  
  \section{Simulations versus Model}

  We  now  compare our   analytical  predictions to  the baryon budget
  history  obtained in our high-resolution hydrodynamical simulations.
  Recall that  we can compute analytically  the baryon history for the
  universe as a whole, but  also on a halo  by halo basis. In order to
  make  a  careful comparison, we   need  to extract   halos from  the
  simulated density field. We  use for that  purpose a halo  detection
  code based on  the Spherical Overdensity  algorithm \citep{Lacey93}.
  We also need to  carefully define the  effective mass resolution  of
  our  simulations, with respect   to star formation. This  additional
  mass  scale is a pure numerical  artifact, that can be accounted for
  explicitly in our  analytical model.  Using  this modified model, we
  will estimate how our results converge (or  not) to the correct halo
  model predictions.
  
  We need also  to estimate the halo star  formation time, (as defined
  in  the previous section)  in our  numerical simulations.   The star
  formation  algorithm  is based  on  a  simple  Schmidt law,  with  a
  specified (over-) density threshold. It is however more complex than
  the  approach  used  in the  halo  model.  We  will show  that  both
  approaches can  be related to  each other, with a  `shape factor'
  reflecting the probability distribution function of the cold phase
  density.

  We  finally  need to  estimate  the  unique  free parameter  in  our
  analytical  model: the orbital decay timescale. This  sort  of `model
  calibration'  will  be   performed  directly  using  our  simulation
  results.   We  will also  extend  this  `model versus  simulation'
  comparison  to  other  numerical  data  kindly  provided  to  us  by
  \cite{Springel03b},  for  which  galactic  winds were included.

  \subsection{Halo Detection}
  \label{halodetect}

  It  is  an  absolute necessity  to  define  a  halo in  a  numerical
  simulation as it  is defined in the theoretical  model. As explained
  before, in  order for the Press  \& Schechter approach  to be valid,
  the halo mass is the mass enclosed in radius $R_{\rm 200}$, enclosing an
  overdensity 200 times larger than the average background density. As
  noted  by  several  authors \citep{Jenkins01,White02},  this  rather
  large region encloses dark matter particles which are not completely
  relaxed  yet, and  also several  large  satellites flying  by. As  a
  consequence, the halo  mass turns out to be  highly dependent on the
  exact algorithm  used to detect automatically dark  matter halos in
  the simulation.

  These same authors suggest the  following strategy: in a first step,
  halos are detected with a  high density contrast ($\Delta = 600$ is
  our  choice  here) by  any  classical  algorithm  (we use  Spherical
  Overdensity in this paper).  Since the density contrast is high, the
  detected region is in a  well relaxed state and all algorithms agree
  more or less on the mass  and on the number of detected halos.  The
  halo center  is defined as the  center of mass of  this high density
  region only.  In  a second step, the halo radius  is increased up to
  $R_{\rm 200}$, in  order to obtain the  large halo mass  required by the
  Press \& Schechter prediction.

  We then  compute for  each  individual halo the total  stellar
  mass within $R_{\rm 200}$.  Since we have stored the formation epoch of
  each individual star particle, we can also compute the complete star
  formation history   of the  parent  halo.  This last  point  is very
  important: we do not compute star formation rates and gas content of
  individual galaxies. Since each  halo can host several galaxies (one
  central and several satellites),   the galaxy baryon budget will  be
  somewhat different  than the halo  baryon  budget.  Within  the halo
  radius, we also  compute the fraction  of cold gas,  defined as $T <
  10^5$ K,  and  $\rho >   10^5 \bar \rho$.     The remaining  gas  is
  considered as `hot  gas', even though its  temperature can  be lower
  than the Virial temperature of the halo.
  
  The effective star formation rate of each halo is computed by simply
  dividing the  total amount of star  created during the  last 10\% of
  the  halo age by  the elapsed  time. The  results presented  in this
  section will be based on  this analysis. Each individual halo baryon
  budget are  averaged into  mass bins, in  order to compare  with the
  halo model predictions.

  \subsection{Mass Resolution}

  Results  of cosmological  simulations  depend strongly  on the  mass
  resolution of the  code. The two main numerical  limitations are the
  box   length   and  the   number   of   particles.    As  shown   in
  Figure~\ref{resolmap}, the  more particles  we start with,  the more
  small  mass   halos  and  galaxy   satellites  we  obtain   in  the
  simulations.   Since we  are interested  here in  the  global baryon
  budget, each individual halo must be  able to allow gas to cool down
  and condense,  and ultimately form  stars. This is a  more stringent
  requirement  than just  reach the  Virial overdensity  of  $\Delta =
  200$.

  Since  star  formation  occurs  at  the  high  end  of  the  density
  distribution, we  take the star  formation density threshold  as the
  limiting factor that defines our  mass resolution. Let us assume for
  sake of simplicity  that each halo is a  pure isothermal sphere. The
  gas density profile is given by
  \begin{eqnarray}
    \rho = \frac{\Delta}{3}\bar \rho \left( \frac{r}{R_{\rm 200}} \right)^{-2}.
  \end{eqnarray}
  The radius above which star  formation occurs is given by $\rho(r) >
  \rho_{\rm 0} (1+z)^{\alpha_{\rm 0}}$.   If we require that within  this radius, we
  have at least 10 dark matter  particles of mass $m_{\rm p}$, we obtain the
  minimal halo mass as
  \begin{eqnarray}
    M_{\rm resol} = 10 \left[ 
      \frac{\rho_{\rm 0}(1+z)^{\alpha_{\rm 0}}}{\Delta/3\bar \rho(z)}
      \right]^{1/2} m_{\rm p}. 
    \label{massresol}
  \end{eqnarray}
  One  clearly sees  that the  higher the  density threshold  for star
  formation, the larger the minimal  mass will be.  For star formation
  density  thresholds constant  in comoving  units  ($\alpha_{\rm 0}=3$), this
  mass  scale is a  constant in  time. This  is the  case for  the AMR
  simulations      presented      here,      for     which,      using
  Equation~\ref{massresol}, we obtain  $M_{\rm resol} \simeq 400 m_{\rm p}$.  On
  the  other hand, for  star formation  density threshold  constant in
  physical units,  the mass  resolution scales as  $(1+z)^{-1.5}$. SPH
  simulations  presented  in  \cite{Springel03b}  were based  on  this
  second  approach. The  mass resolution  we  obtain in  this case  is
  $M_{\rm resol}  \simeq  1000(1+z)^{-1.5}  m_{\rm p}$.   At high  redshift,  $z
  \simeq  20$, the  corresponding mass  resolution  can be  as low  as
  $M_{\rm resol} \simeq 10 m_{\rm p}$.

  Simulated halos with  mass lower than $M_{\rm resol}$ will  not be able
  to form stars or,  equivalently, condensed cool gas. Therefore, they
  will  be part  of the  simulated diffuse  background. This  new mass
  scale  is  a pure  numerical  artifact,  that  strongly affects  our
  results.  We  take this  mass scale into  account in  our analytical
  model by setting the Minimal  Mass for star forming halos $M_{\rm min}$
  as  the   maximum  between  the  true  physical   Minimal  Mass  and
  $M_{\rm resol}$. As we  will see later in this section,  this trick is a
  very powerful tool  to account for finite resolution  effect in the
  simulation,  and  to  assess   the  convergence  properties  of  our
  numerical results.
  
  \subsection{Halo Star Formation Time}
  \label{halotstar}

  \begin{figure}
    \includegraphics[width=\hsize]{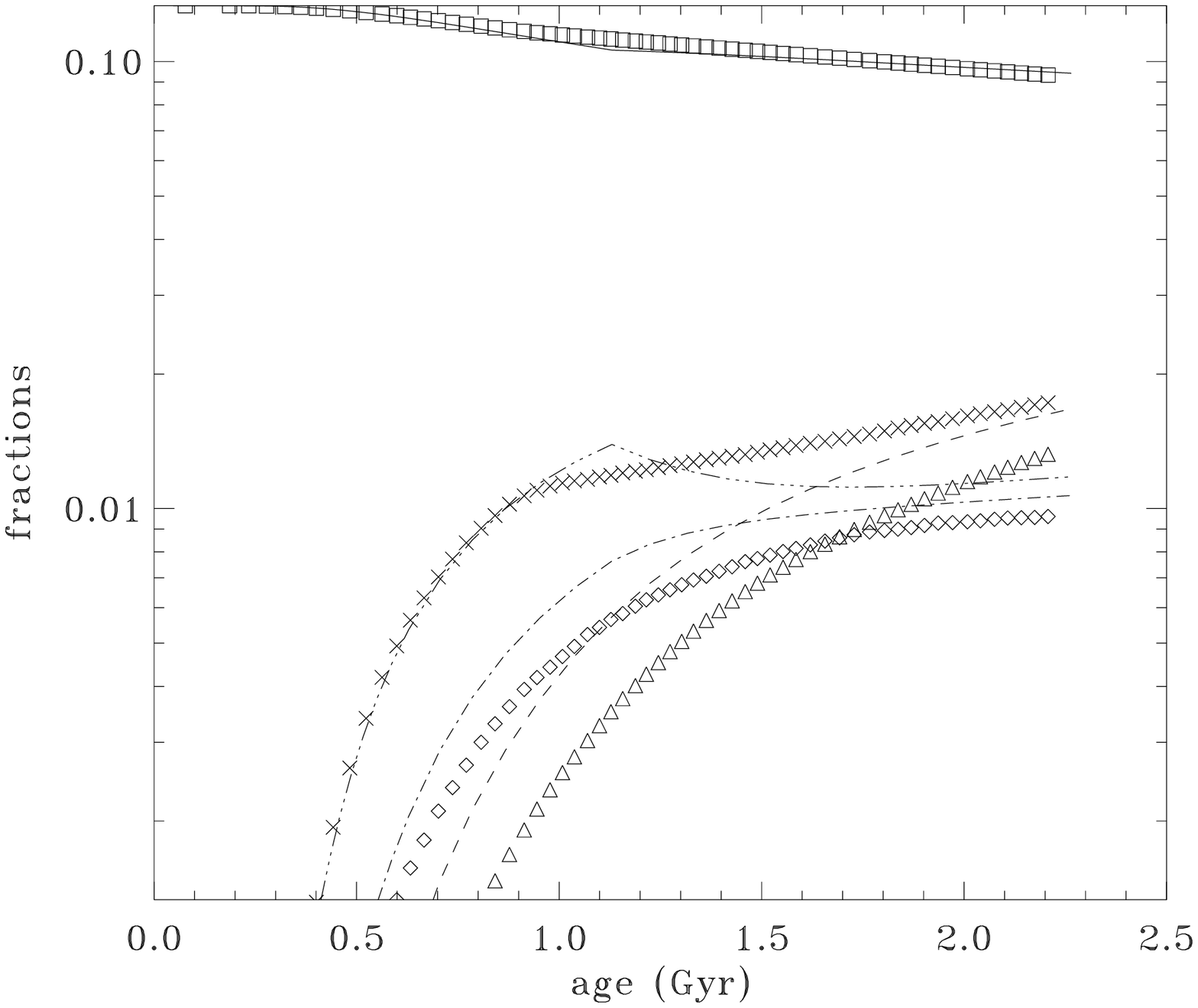}
    \cpt{Time  evolution  of  the  various baryon  phases  in  our
      highest  resolution run  L10N512S30. Symbols  are  mass fraction
      measured  in   the  simulation  (squares:   diffuse  background,
      crosses:   hot   gas,   diamonds:   cold  gas   and   triangles:
      stars). Lines  are the predictions of our  analytical model with
      $M_{\rm resol}   \simeq  2\times10^8$ h$^{-1}$M$_{\odot}$   (solid:  diffuse
      background,  dot-dot-dashed: hot gas,  dot-dashed: cold  gas and
      dashed: stars).}
    \label{frac_evol}
  \end{figure}

  \begin{figure}
    \includegraphics[width=\hsize]{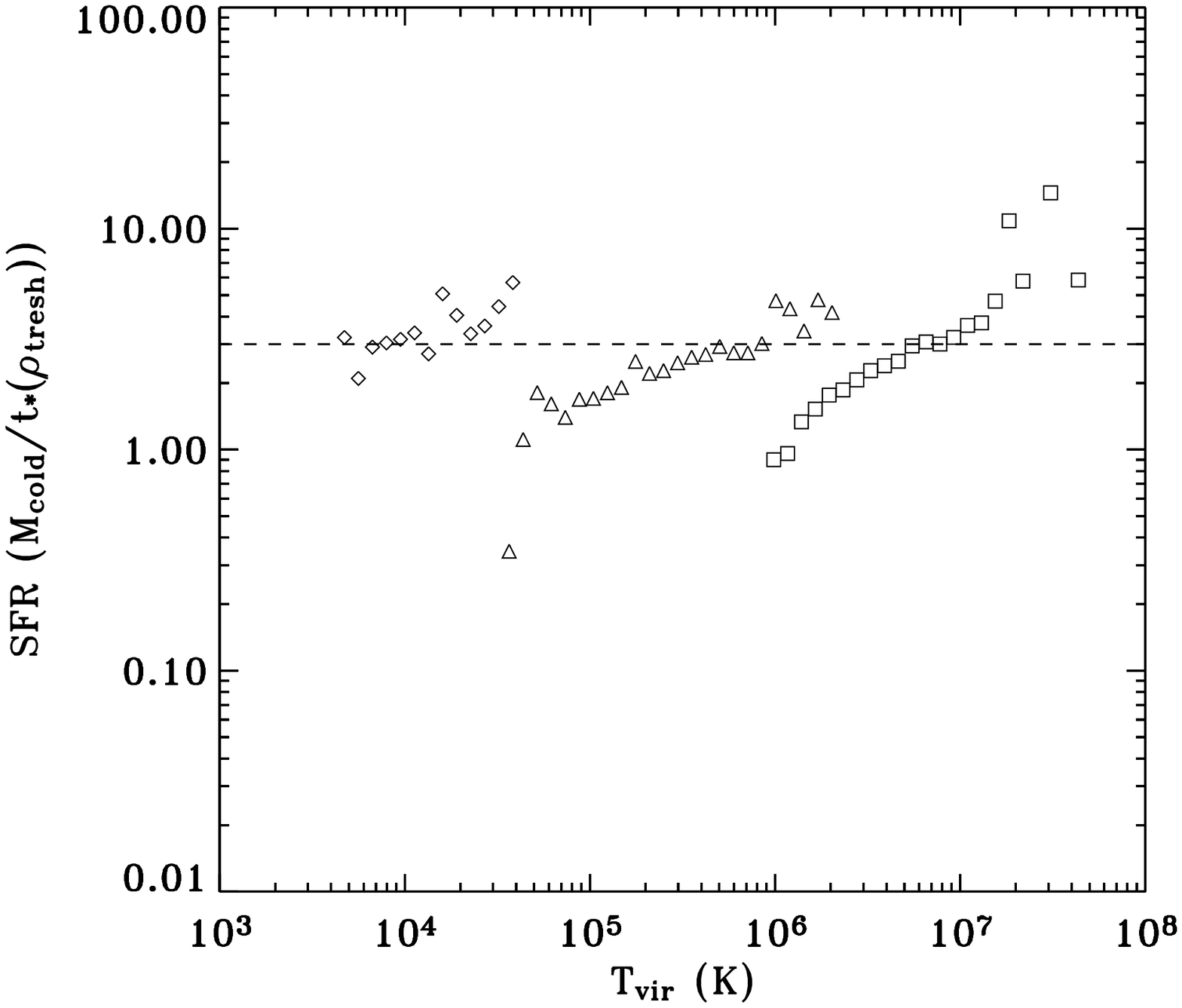}
    \cpt{  Average  star formation  rate  of  simulated halos  in
      various     Virial    temperature     bins,    in     unit    of
      $M_{\rm cold}/t_*(\rho_t)$. In our framework, this is a direct measure of
      the `shape factor' $F(\mu)$ of the underlying cold gas density
      distribution.    Numerical  data   suggest   a  constant   value
      represented here as the  dashed line $F(\mu) \simeq 3$. Diamonds
      are for  run L1N256S30 at $z=5.5$, triangles  for run L10N256S30
      at $z=2.5$ and squares for run L100N256S30 at $z=0$.}
    \label{beta}
  \end{figure}
  
  \begin{figure}          \includegraphics[width=\hsize]{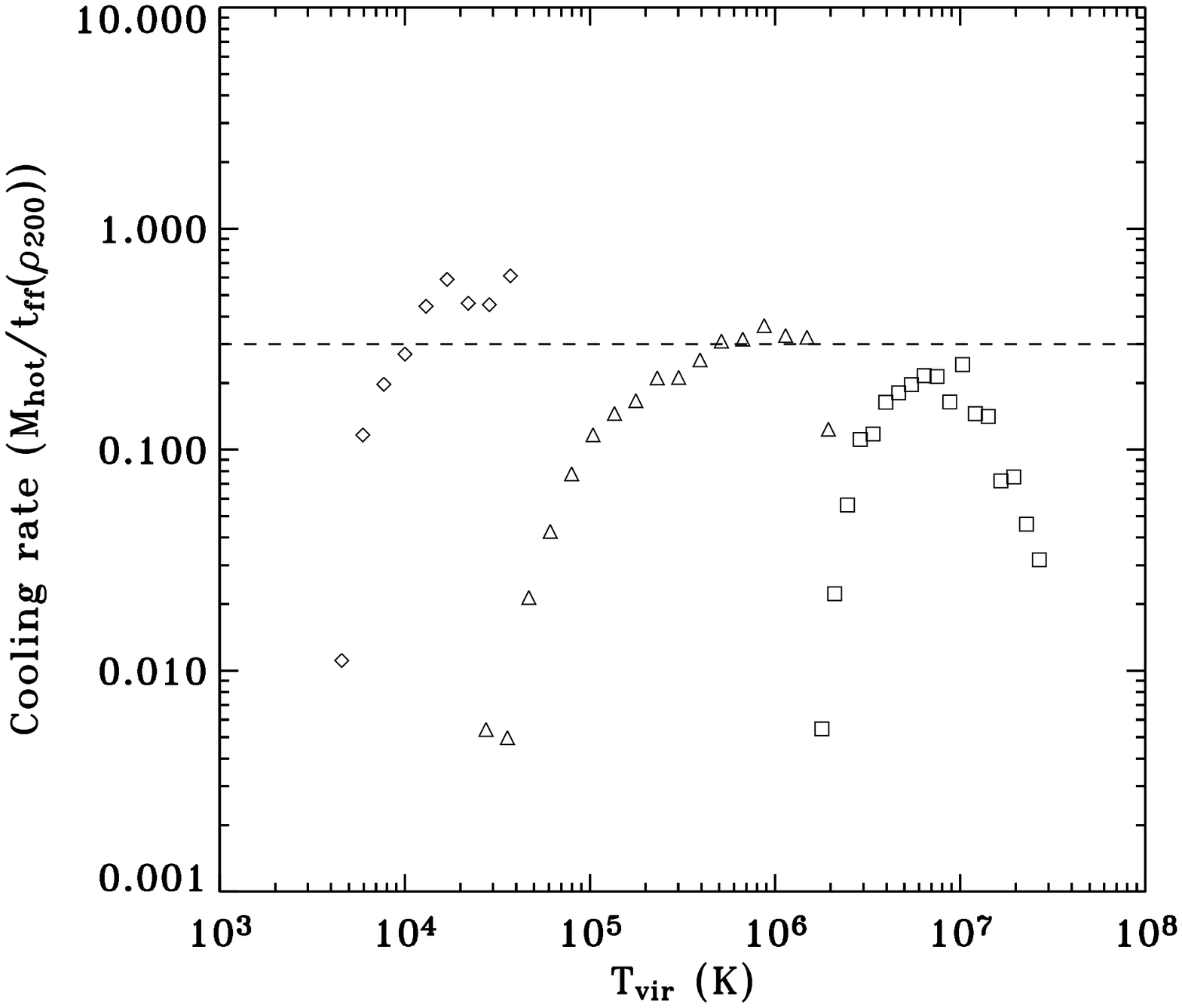}
    \cpt{  Average star  formation   rate  of simulated  halos  in
    various  Virial    temperature   bins,    in    unit of   $M_{\rm hot}
    R_{\rm 200}/V_{\rm 200}$ for  the high efficiency  runs.  This is a direct
    measure of the cooling rate  of hot gas  into dense cold discs. At
    high temperature ($T >  10^7$ K), we observe a  sudden drop due to
    inefficient  Bremstrhalung cooling.   At lower  temperatures,  the
    cooling rate has  a plateau around  $1/3$.  In our framework, this
    suggests    an  orbital    decay  timescale  $t_{\rm orb}     \simeq 3
    R_{\rm 200}/V_{\rm 200}$  for infalling gas clumps.   Diamonds are for run
    L1N256S3 at $z=5.5$, triangles  for run  L10N256S3 at $z=2.5$  and
    squares for   run L100N256S3  at   $z=2.5$ also.}    \label{alpha}
    \end{figure}
  
  The methodology we  use in this paper is  to describe star formation
  on  a  halo  by  halo  basis. We  completely  discard  the  detailed
  modeling  of exponential  gas  discs and  nuclear  bursts. This  is
  usually performed in semi-analytical  models of galaxy formation. In
  our AMR simulations, we do have however a higher level of complexity
  than  in the analytical  model.  Visual  inspection of  density maps
  shows the presence of gas  discs in centrifugal equilibrium, as well
  as  several   small  satellites  orbiting  a   central  galaxy  (see
  Fig.~\ref{zoom512}).  We  are aware of  the fact that  many physical
  ingredient are  probably missing in our  current numerical solution
  of galaxy  formation.  Nevertheless, we  need to establish  the link
  between our analytical model and our numerical implementation of star
  formation.

  Since star  formation in  the code  is based on  a Schmidt  law (see
  Eq.~\ref{starrateeq}),  we   can  compute  the   instantaneous  star
  formation rate in  any halo by integrating over  the entire cold gas
  present in that particular halo.
  \begin{eqnarray}
    \dot{M}_* = M_{\rm cold} \int_{\rho_{\rm t}}^\infty 
    \frac{\mu(\rho) d\rho}{t_*(\rho)},
  \end{eqnarray}
  where $\rho_{\rm t} = \rho_{\rm 0}(1+z)^{\alpha_{\rm 0}}$  is the star formation density
  threshold  and $\mu(\rho)$  is the  mass fraction  of cold  gas with
  density $\rho$. The exact form of the cold gas distribution function
  $\mu$  is beyond  the  scope of  this  paper.  It  is  likely to  be
  determined by the global surface  density as well as the small scale
  turbulence inside rotating  discs. We will make here  the very crude
  approximation  that $\mu$ is  self-similar in  the variable  $\rho /
  \rho_{\rm t}$, so we can simplify the last equation further more into
  \begin{eqnarray}
    \dot{M}_* = \frac{M_{\rm cold}}{t_*(\rho_{\rm t})} F(\mu),
    \label{shapefac}
  \end{eqnarray}
  where $F(\mu)$  is a dimensionless `shape factor'  that depends on
  the exact form of the cold gas density distribution
  \begin{eqnarray}
    F(\mu) = \int_{\rm 1}^\infty \mu(x) x^{1/2} dx.
  \end{eqnarray}
  We  are  now  in a  position  to  make  a  direct link  between  our
  analytical model and numerical simulations. We recognize in the last
  equation   the   halo   star    formation   rate   as   defined   in
  section~\ref{starformmodel}, with star formation parameters given by
  $\alpha_* = \alpha_{\rm 0}$ and $t_* = t_{\rm 0} / F(\mu)$. The shape factor $F$
  has the  effect of reducing  the effective halo star  formation time
  scale, relative  to the reference  time $t_{\rm 0}$. Indeed, if  very high
  density  gas  is present,  the  star  formation  rate is  likely  to
  increase accordingly.  
  
  Since we  can't predict the  value of this shape  factor, we have to
  measure it directly in the simulations. We plot in Figure~\ref{beta}
  the      halo star     formation      rate, in   units  of   $M_{\rm
  cold}/t_*(\rho_{\rm t})$ (see Eq.~\ref{shapefac}), and averaged over
  halos  of similar  mass.  This should  be equal  to the shape factor
  $F(\mu)$. For  3 different box sizes  and  at 3  different redshift,
  this factor is not  exactly a constant,   although it varies  slowly
  with mass.  This illustrates that our approach is only a first order
  approximation    of   our  simulation results.    Nevertheless,   we
  approximate this by  taking $F(\mu) \simeq 3$,  as suggested  by the
  dashed line in Figure~\ref{beta}.  This specifies how star formation
  in the simulations and star formation in  the model are connected to
  each other.
 
  \subsection{Halo orbital decay timescale}
  
  \begin{figure*}                                  \begin{tabular}{cc}
    \includegraphics[width=0.5\hsize]{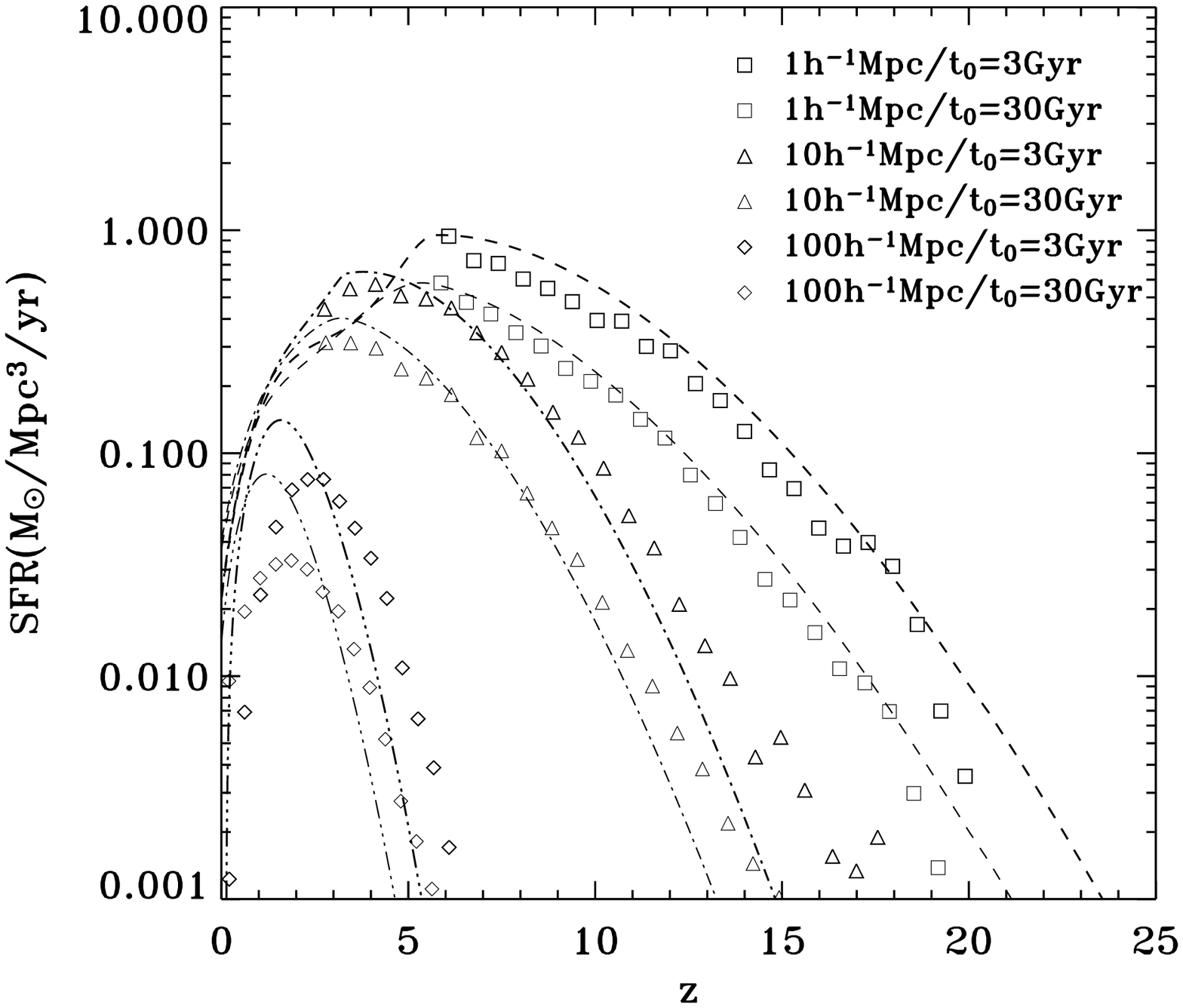}
    \includegraphics[width=0.5\hsize]{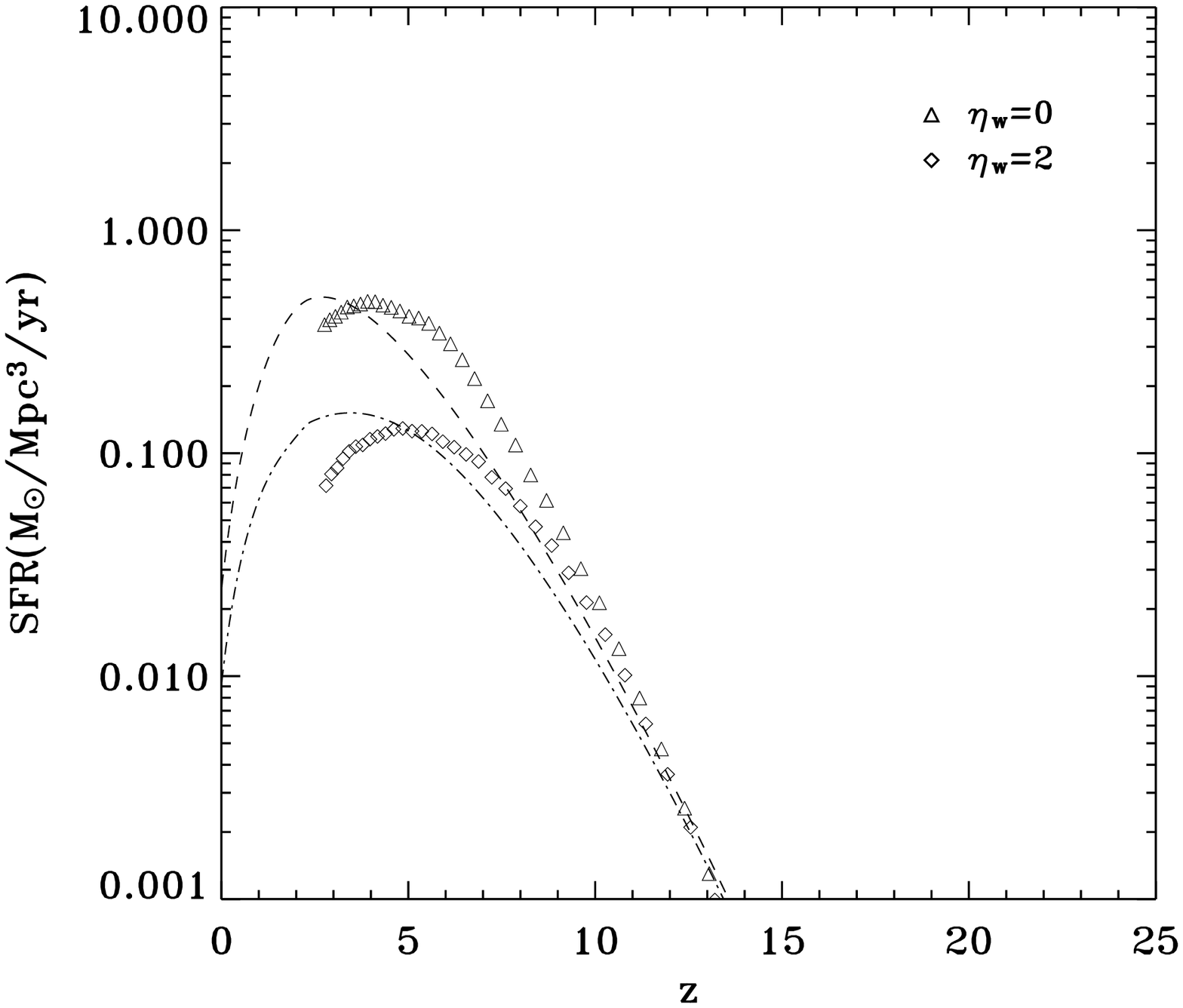}             \\
    \includegraphics[width=0.5\hsize]{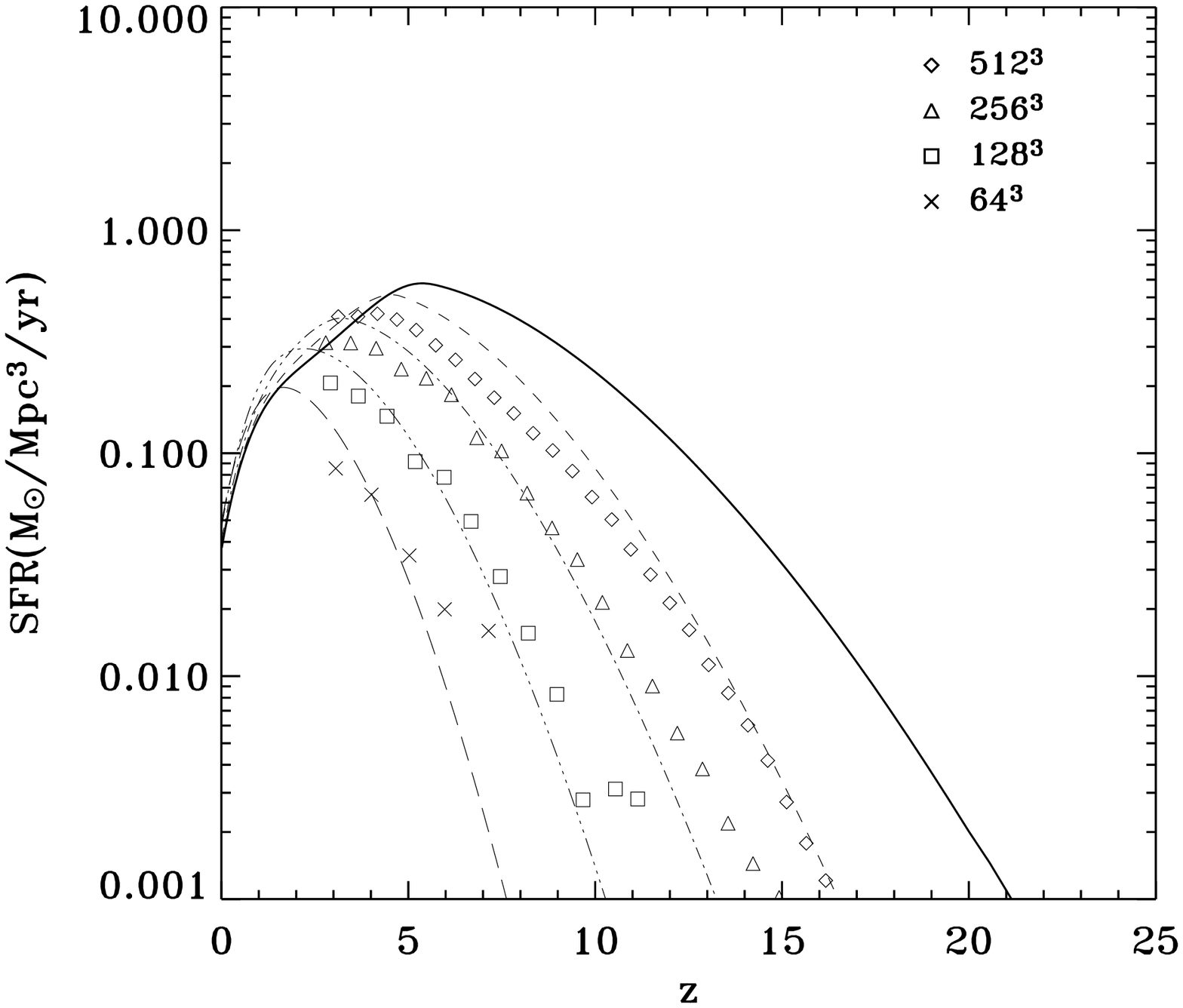}
    \includegraphics[width=0.5\hsize]{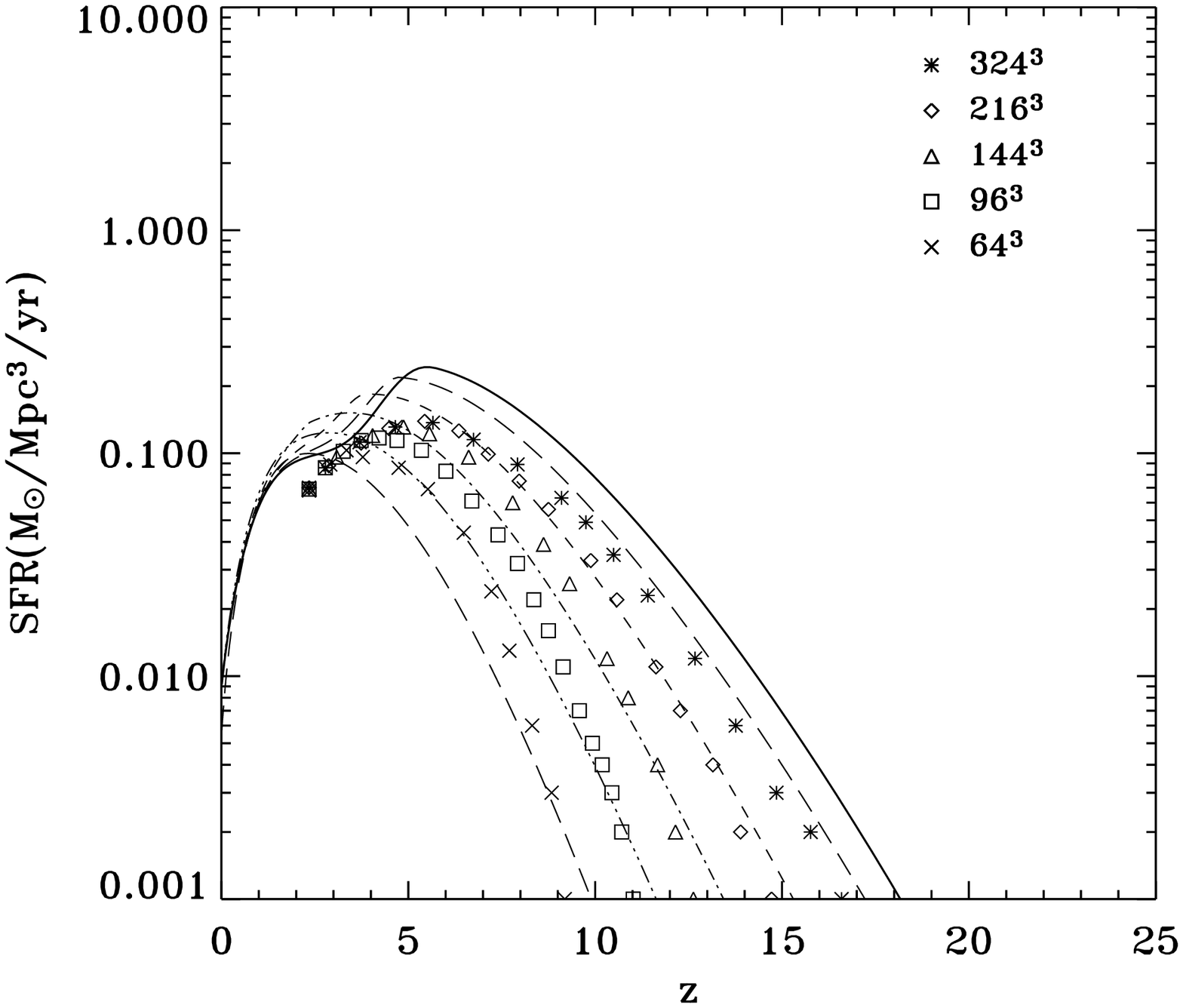}
    \end{tabular}  \cpt{Global comoving  star formation  rate as a
    function  of redshift.  In each   plot, symbols are for  numerical
    simulations, while  lines  are  for our corresponding   analytical
    prediction.  \emph{Upper  left  plot: high efficiency  series and
    low efficiency series}   with  run L1N256S3 (black  squares),  run
    L1N256S30 (grey  squares), run  L10N256S3  (black  triangles), run
    L10N256S30 (grey triangles),  run L100N256S3  (black diamonds) and
    run  L100N256S30  (grey  diamonds)   .   \emph{Lower   left  plot:
    convergence  study  series} with  run  L10N512S30  (diamonds), run
    L10N256S30 (triangles), run L10N128S30 (squares) and run L10N64S30
    (crosses).       \emph{Upper  right  plot:     \cite{Springel03b}
    simulations} with run 03 without winds (triangles) and Q3 including
    winds  (diamonds).   \emph{Lower right  plot:  \cite{Springel03b}
    simulations}  including winds    with  run  Q5   (stars),  run   Q4
    (diamonds),  run  Q3  (triangles), run Q2    (squares) and run  Q1
    (crosses).   The solid line    is  the  `fully  converged'   model
    prediction.  } \label{sfrdez} \end{figure*}

  The only unknown  parameter in our  analytical model is  the orbital
  decay   timescale    of   infalling   gas   clumps   (see   appendix
  \ref{coolingmodel}), before they  reach the high-density disc in the
  halo  center. When   cooling is very   fast,  for halos with  Virial
  temperature $T_{\rm min} < T_{\rm 200} <  T_{\rm max}$, we have assumed that the
  accretion rate into the disc is controlled by the orbital time scale
  of  infalling satellites.  Computing  this  time scale is beyond the
  scope  of this paper.  It is  probably  determined by details in the
  gravitational  dynamics and satellite   dynamical friction.    These
  aspects are all key  ingredients of semi-analytical models of galaxy
  formation.

  In order to determine this orbital decay timescale, we perform again
  a direct analysis of our numerical simulations.  Let us consider the
  case of very fast star formation $t_{\rm 0} = 3$ Gyr and $\alpha_{\rm 0}=3$.  In
  this  case,  the halo  star  formation rate is  almost equal (within
  10\%) to the halo cooling  rate. This can be  later confirmed by the
  analytical model.  We plot   in Figure~\ref{beta} the average   star
  formation rate of  halos within  different mass  range,  in units of
  $M_{\rm hot} R_{\rm 200}/V_{\rm 200}$.   In   our framework, this   quantity  is
  exactly equal to the ratio $(R_{\rm 200}/V_{\rm 200})/t_{\rm orb}$.  Here again,
  this ratio is  not perfectly a constant,  illustrating the fact that
  our  model is  only   a first order  approximation,  but  for the  3
  different box sizes and at 3 different redshifts, the curve exhibits
  a plateau around $t_{\rm orb} \simeq 3 R_{\rm 200}/V_{\rm 200}$.  We take this
  value as our canonical value in the analytical model.

  \subsection{Global Baryon Budget}

  We now present in  greater details our simulation results,  starting
  with   the    baryon   history for   the    universe  as   a  whole.
  Figure~\ref{frac_evol}  shows  the   baryon history  in  our highest
  resolution run L10N512S30.  The run  parameters correspond to a  low
  efficiency  star formation model.   Each  phase  is defined by   well
  defined limits   in the   $\rho$   -  $T$ diagram,  as   defined  in
  Section~\ref{results}. The various symbols in Figure~\ref{frac_evol}
  refer to baryon fractions  in different snapshots of the simulation,
  while  lines refer   to   the analytical   model  predictions,  with
  $M_{\rm resol}  \simeq  2 \times  10^8$  h$^{-1}$M$_{\odot}$,  as given by
  Equation~\ref{massresol}. The other  parameters of the model are set
  to their standard values ($F(\mu)=3$  and $R_{\rm orb}=3R_{\rm 200}$).   The
  agreement between the simulation and the  model is very good (within
  a factor of 2), given the simplicity of the latter.

  \begin{figure*}                                  \begin{tabular}{cc}
    \includegraphics[width=0.5\hsize]{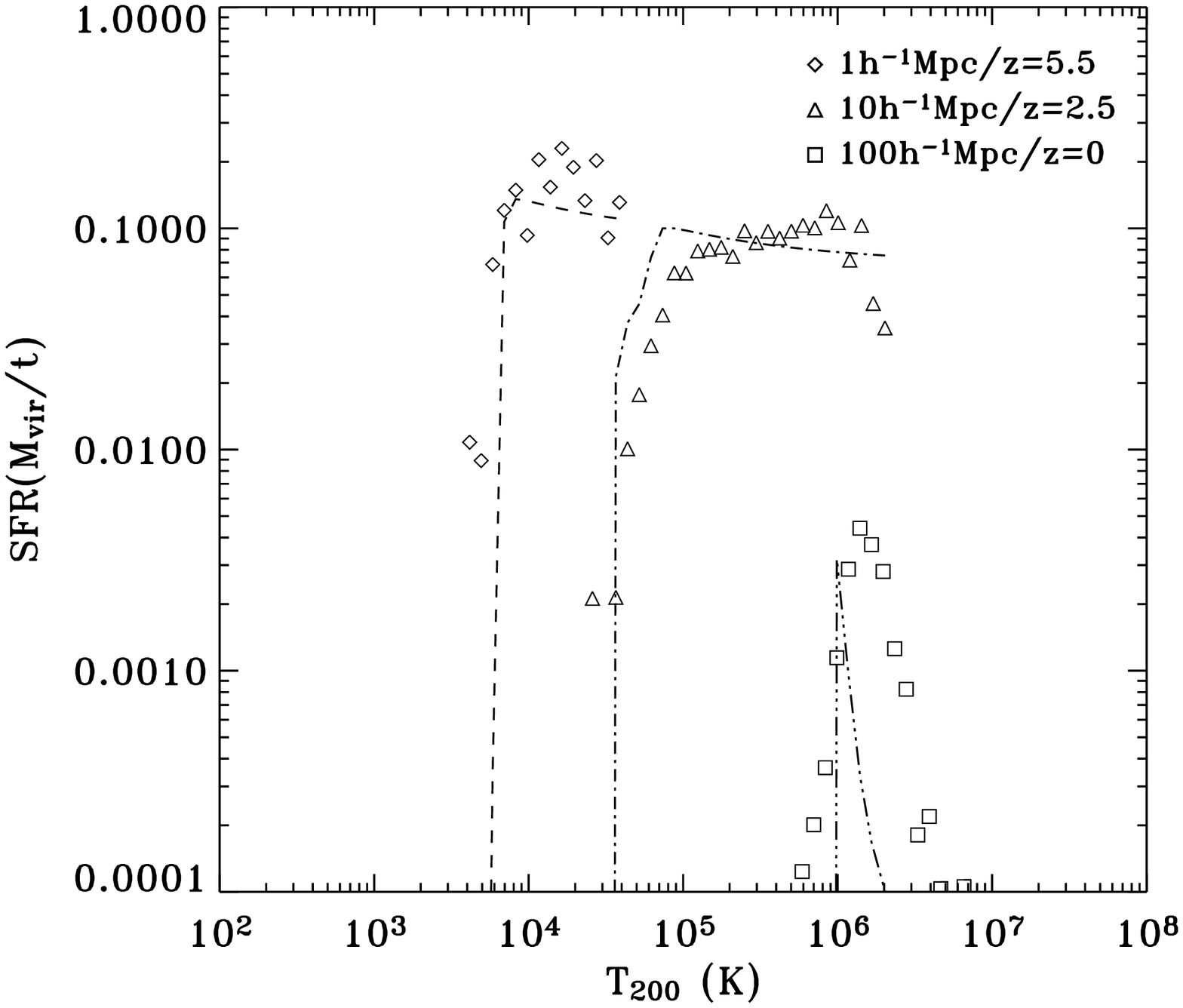}
    \includegraphics[width=0.5\hsize]{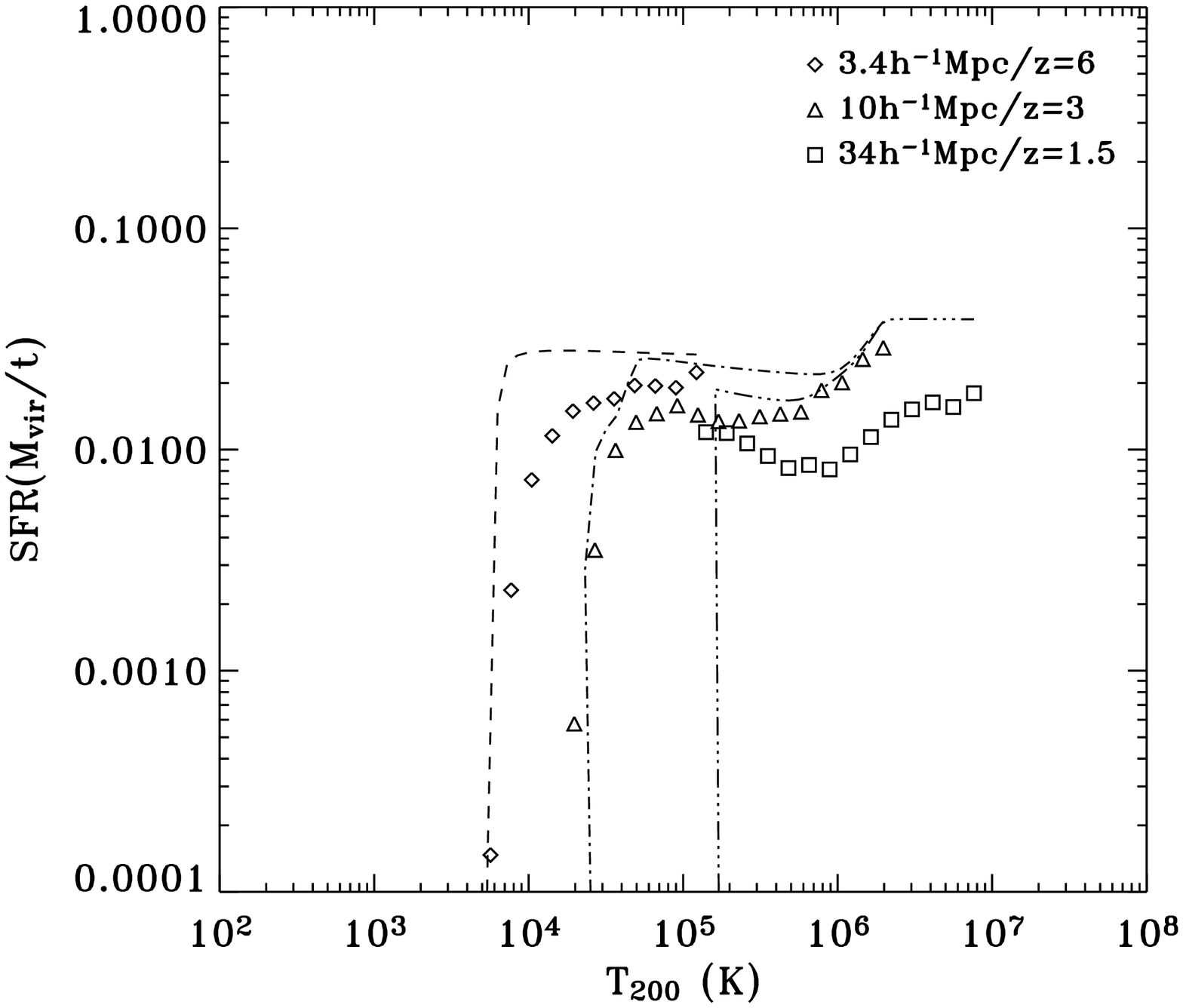}        \\
    \includegraphics[width=0.5\hsize]{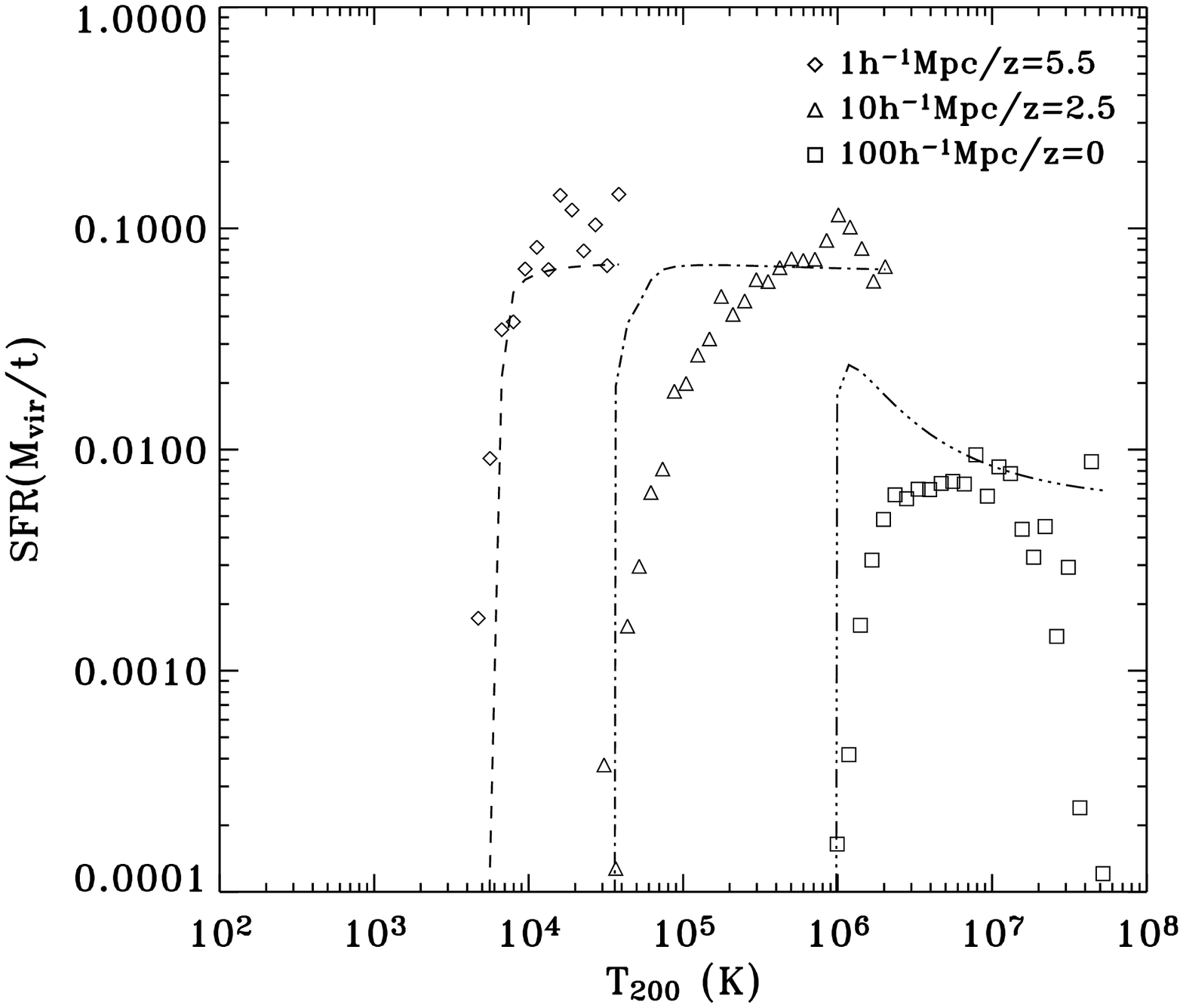}
    \includegraphics[width=0.5\hsize]{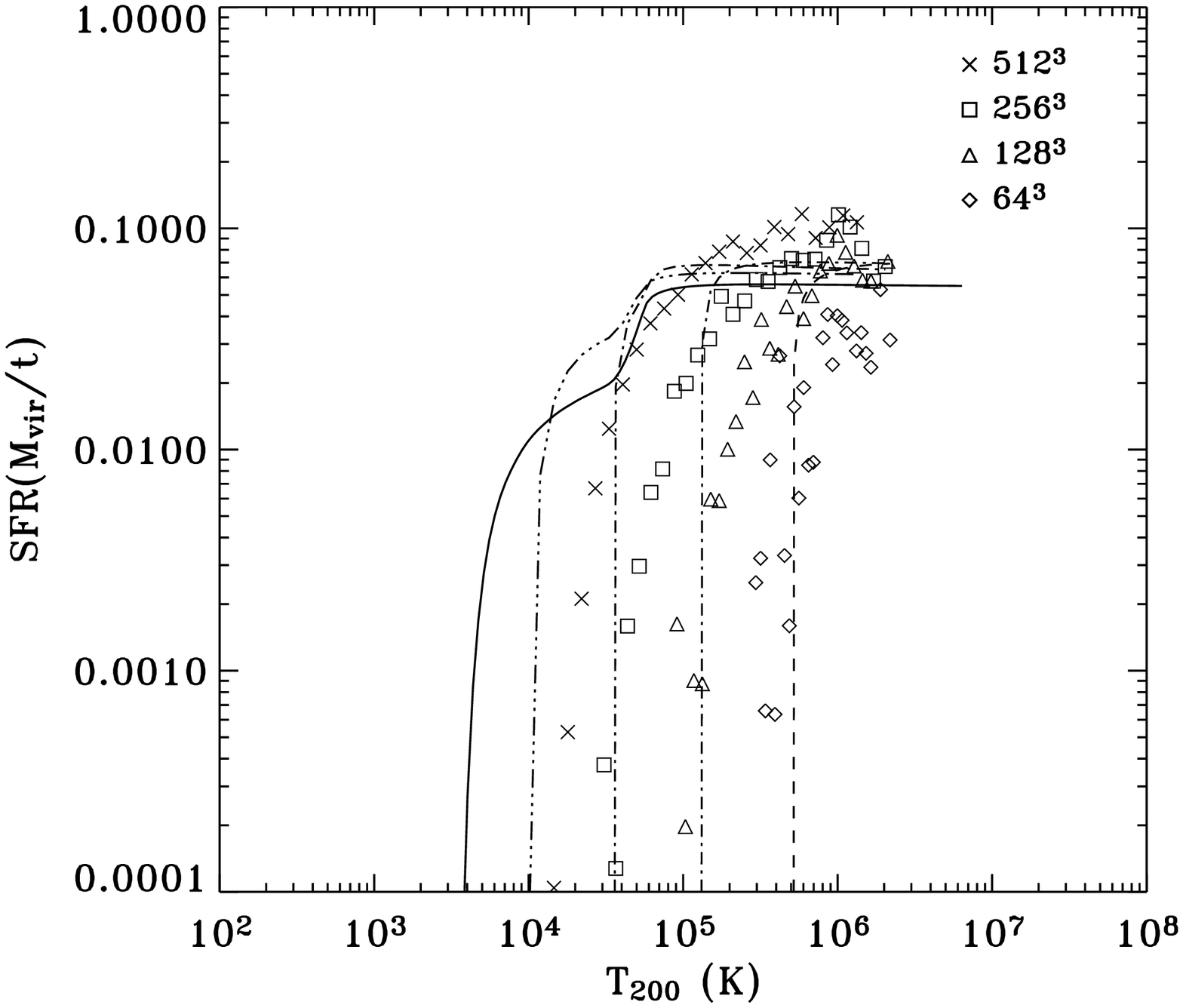}
    \end{tabular} \cpt{Halo star    formation rate, in units    of
    $M_{\rm 200}/t$, as a function of the  halo Virial temperature. In
    each plot, symbols are for numerical  simulations, while lines are
    for our  corresponding  analytical   prediction. \emph{Upper  left
    plot: high   efficiency series}   with  run L1N256S3   at  $z=5.5$
    (diamonds),    run   L10N256S3  at  $z=2.5$  (triangles)   and run
    L100N256S3 at $z=0$. \emph{Lower left plot: low efficiency series}
    with   run  L1N256S30 at  $z=5.5$    (diamonds), run L10N256S30 at
    $z=2.5$  (triangles)   and run L100N256S30   at $z=0$. \emph{Lower
    right plot:  convergence  study series} at $z=2.8$  with L10N64S30
    (diamonds),  run L10N128S30  (triangles), run L10N256S30 (squares)
    and L10N512S30 (crosses).  The solid line is the `fully converged'
    model    prediction.  \emph{Upper  right plot:  \cite{Springel03b}
    simulations} including winds, with run R4 (diamonds) at $z=6$, run
    Q4 at   $z=3$   (triangles) and run    D4  at $z=1.5$  (squares).}
    \label{sfrdem} \end{figure*}

  We  now examine more  closely the  global star  formation rate  as a
  function of  redshift, measured in all our  simulations, and compare
  our various results to the  analytical model. This quantity is a key
  prediction of hierarchical model of galaxy formation.  It translates
  more  or less  directly  into galaxy  colors  and luminosities,  and
  provides  a stringent test  of the  current cosmological  theory. As
  star particles  are created  during the course  of a  simulation, we
  keep track of  their birth epoch. It is  straightforward to compute,
  using  the last  output  only, the  star  formation epoch  histogram
  (history).

  Figure~\ref{sfrdez} shows this global star formation history for our
  `convergence  study'   simulation  suite:  L10N64S30,  L10N128S30,
  L10N256S30 and  L10N512S30. Numerical results are  shown as symbols,
  while  analytical predictions  are shown  as lines.   The analytical
  model predictions are computed  with a Minimal Mass corresponding to
  the    mass    resolution    of    each    run,    as    given    by
  Equation~\ref{massresol}.  Since our  star formation recipe is based
  on  a  constant  overdensity   threshold,  the  mass  resolution  is
  $M_{\rm resol} \simeq 400 m_{\rm p}$. The solid line stands for the analytical
  prediction, without  any finite resolution  effects ($M_{\rm resol}=0$).
  This gives  an indication on how  our results have  converged to the
  `true' star formation history in this particular model.

  Around  $z \simeq  3-5$, our  highest resolution  run  L10N512S30 is
  close to the correct value.  At very high redshift however, the star
  formation  rate is  lower than  the expected  value by  a  factor of
  ten. The mass resolution  $M_{\rm resol}$ is indeed significantly higher
  than $M_{\rm min}$  at redshift  $z > 10$:  this explains the  origin of
  this discrepancy.  As illustrated by the Figure~\ref{resolmap}, low
  resolution  runs  do  miss  the  formation of  dwarf  galaxies  that
  contribute significantly to the global star formation history.

  This  first series  of simulations  was  performed for  a box  length
  $L=10$ h$^{-1}$ Mpc.  By $z  \simeq 3$, the non-linear scale becomes
  comparable to the box size. We  have to stop the simulations, as our
  realizations  are not  representative  of the  universe  as a  whole
  anymore.

  \begin{figure*}                                  \begin{tabular}{cc}
    \includegraphics[width=0.5\hsize]{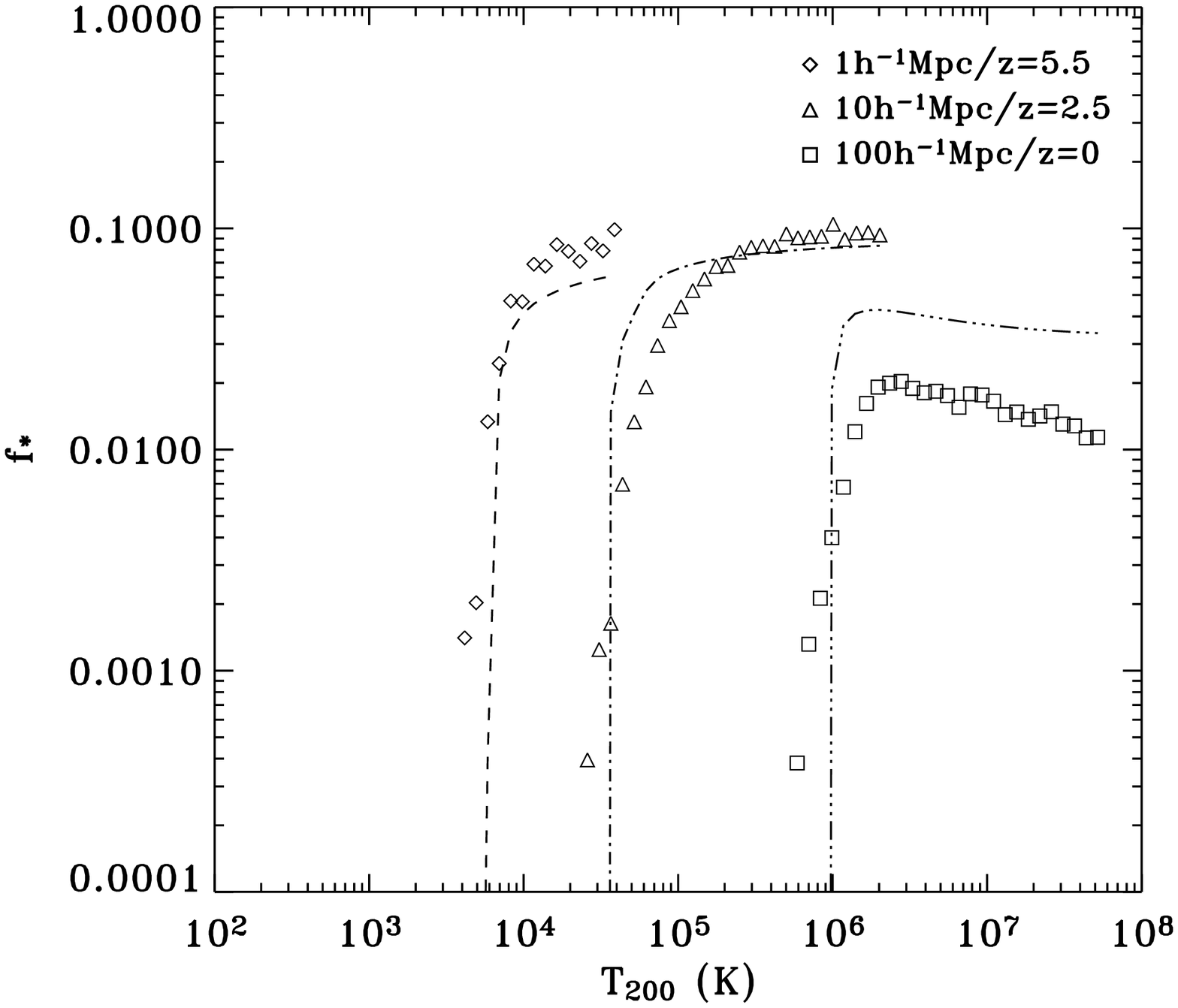}
    \includegraphics[width=0.5\hsize]{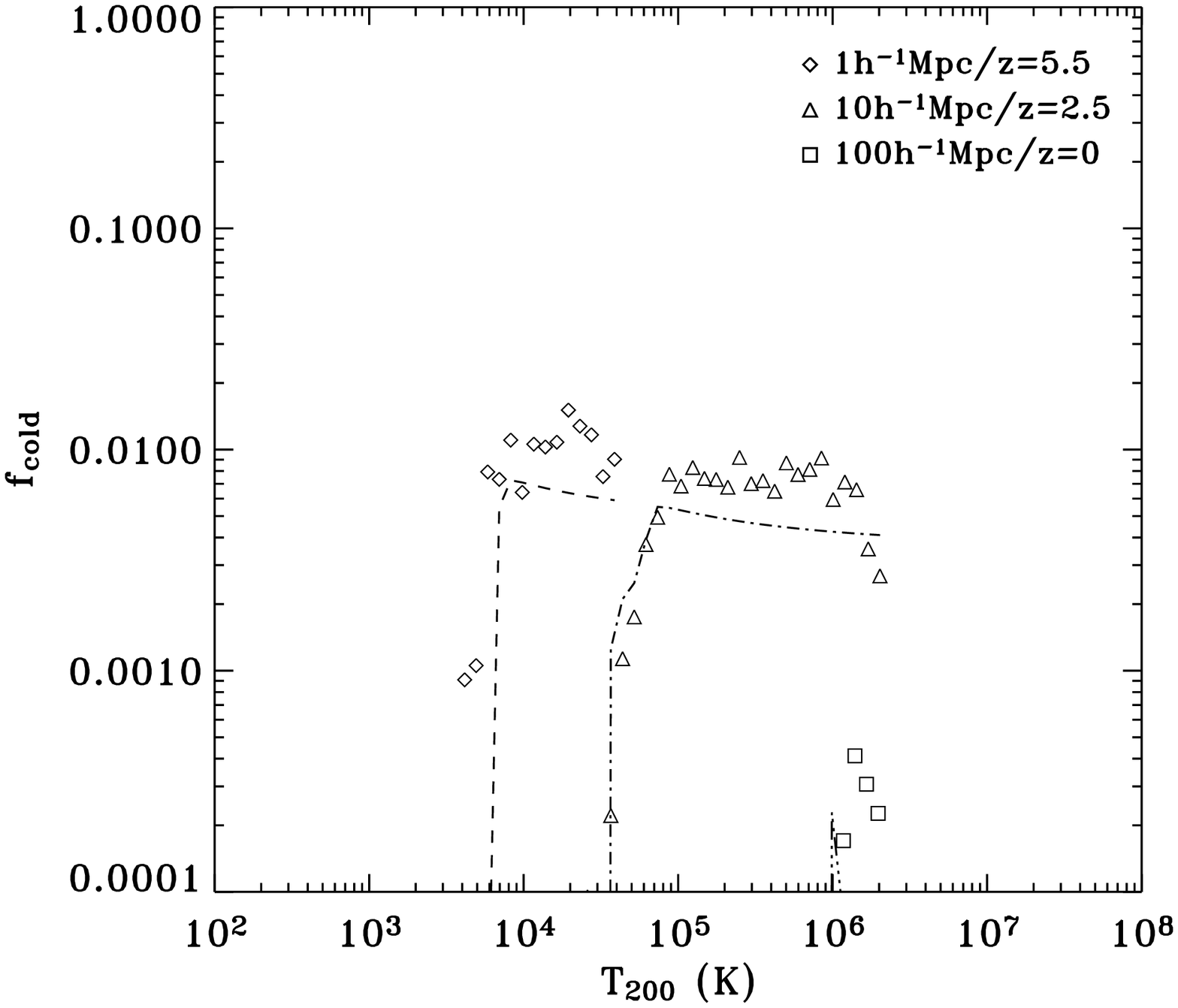}                    \\
    \includegraphics[width=0.5\hsize]{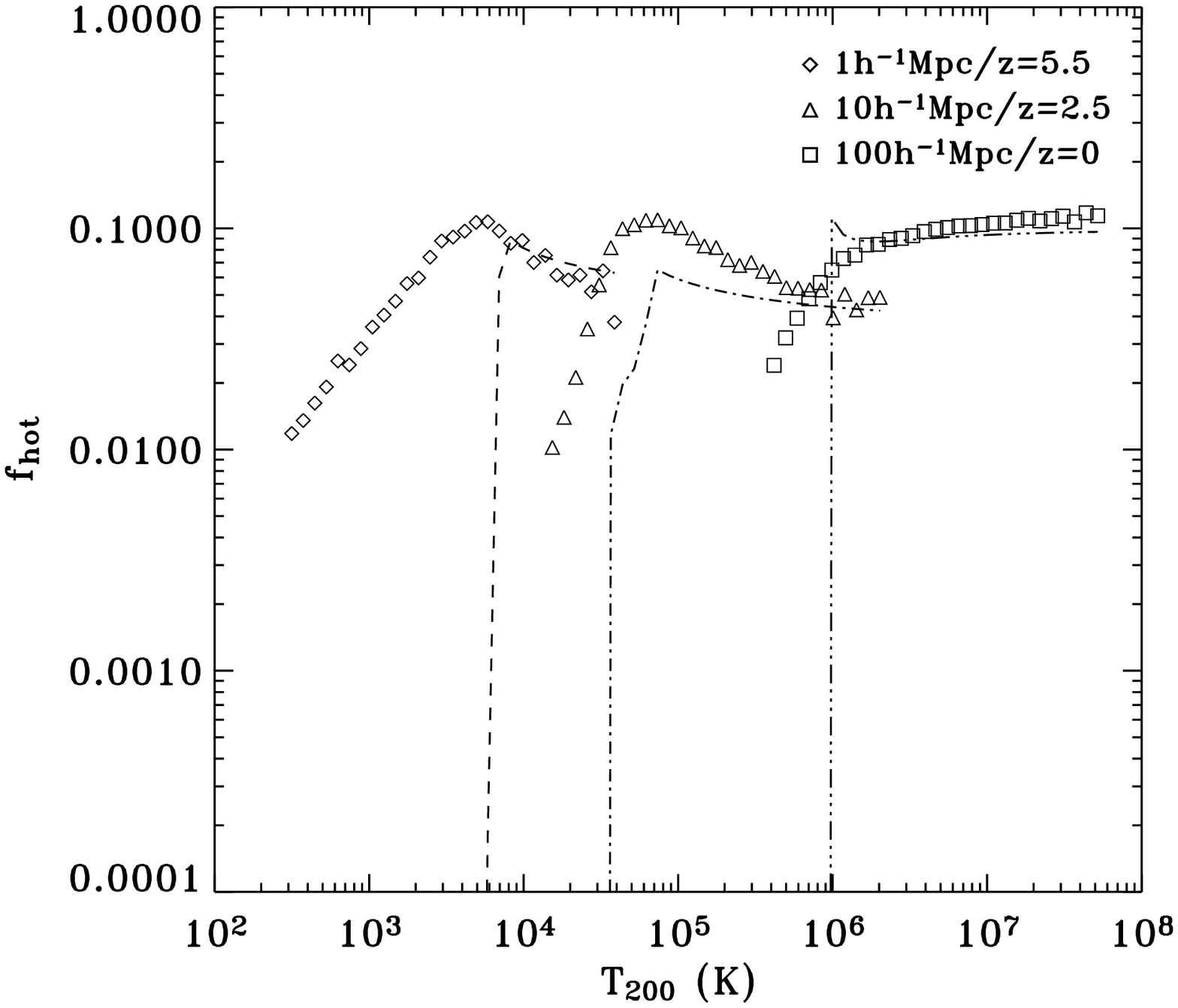}
    \includegraphics[width=0.5\hsize]{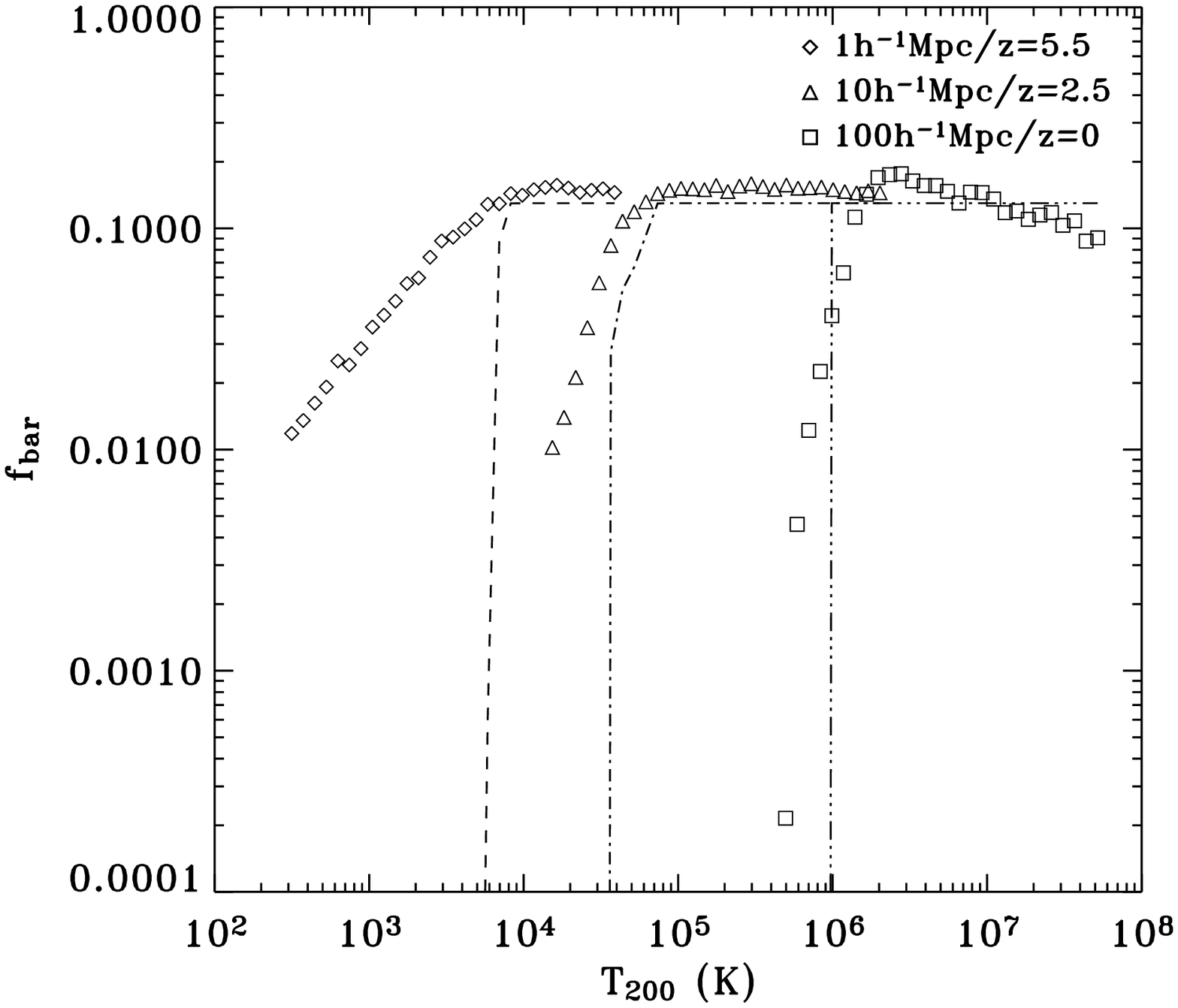}       \end{tabular}
    \cpt{Mass fraction in each baryon  phase as a function of halo
    Virial temperature    for the  \emph{high  efficiency   simulation
    series}. In each plot, diamonds refer  to run L1N256S3 at $z=5.5$,
    triangles  to  run   L10N256S3  at $z=2.5$    and squares  to  run
    L100N256S3 at $z=0$.  Lines are  for the corresponding  analytical
    model. The \emph{upper left  plot}  shows the star mass  fraction;
    the \emph{lower left  plot} shows the hot  gas mass  fraction; the
    \emph{lower right  plot} shows the total  baryon  fraction and the
    \emph{upper   right  plot}     shows  the   cold  gas   fraction.}
    \label{fracdem} \end{figure*}

  \begin{figure*}                                  \begin{tabular}{cc}
    \includegraphics[width=0.5\hsize]{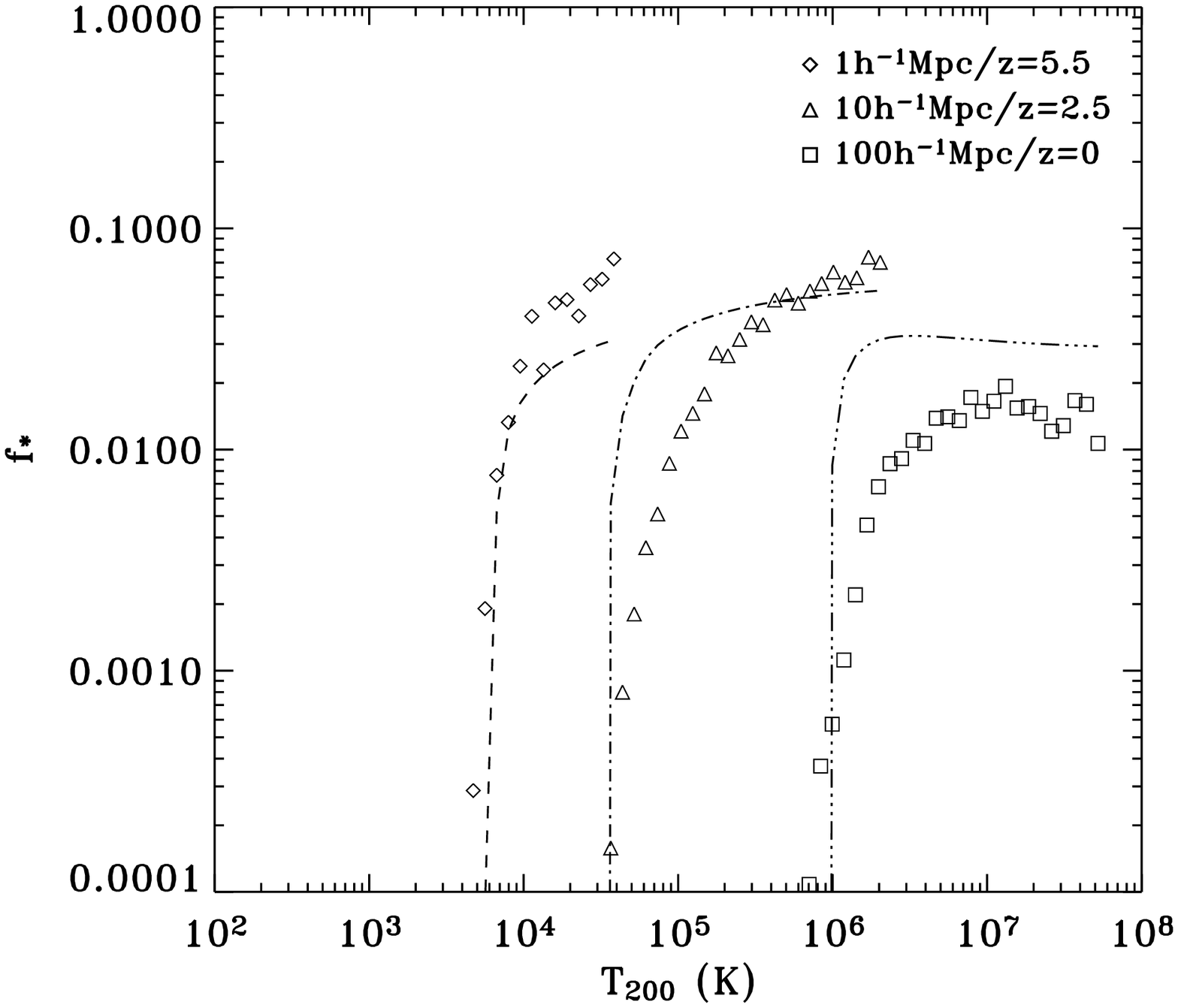}
    \includegraphics[width=0.5\hsize]{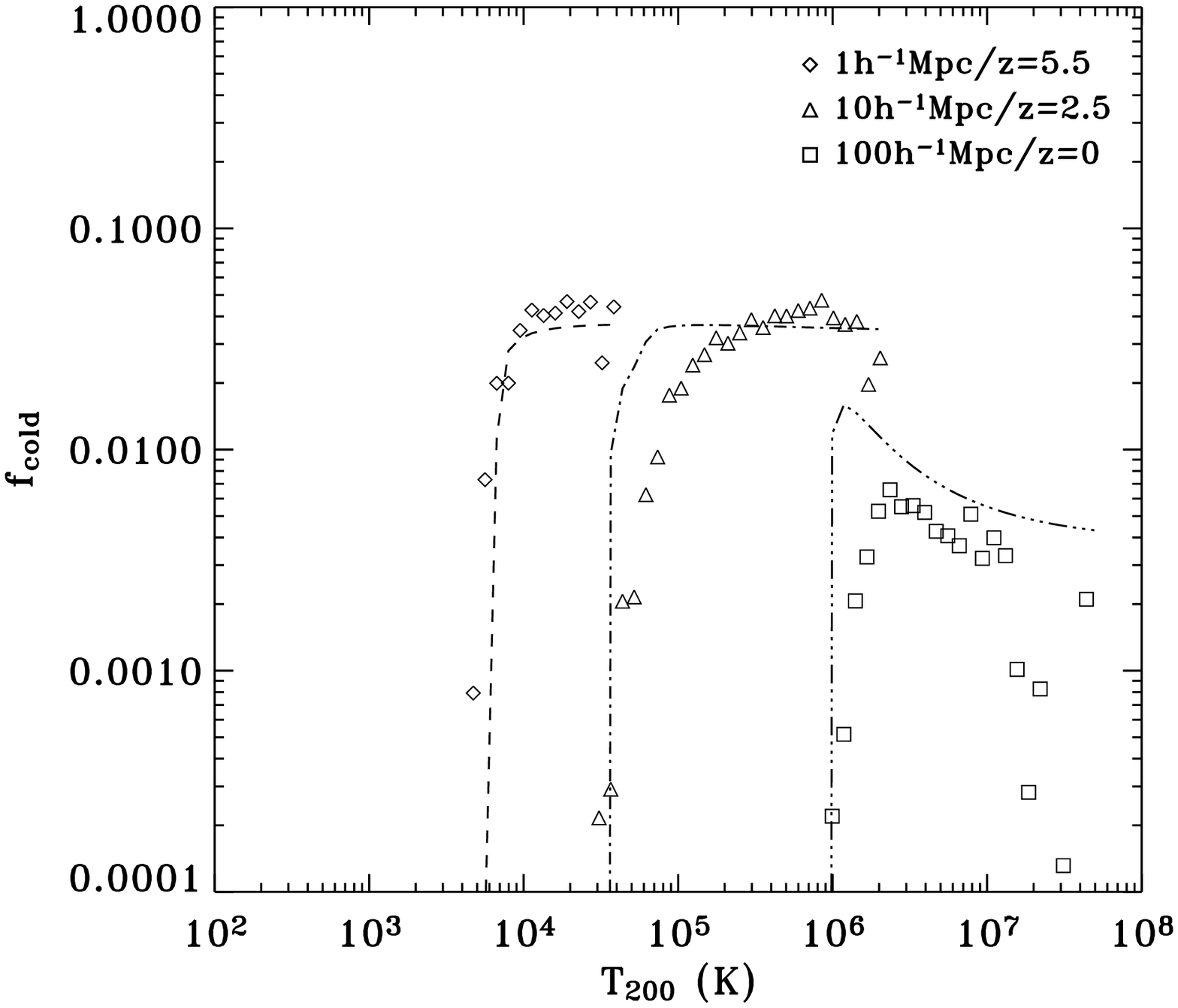}                \\
    \includegraphics[width=0.5\hsize]{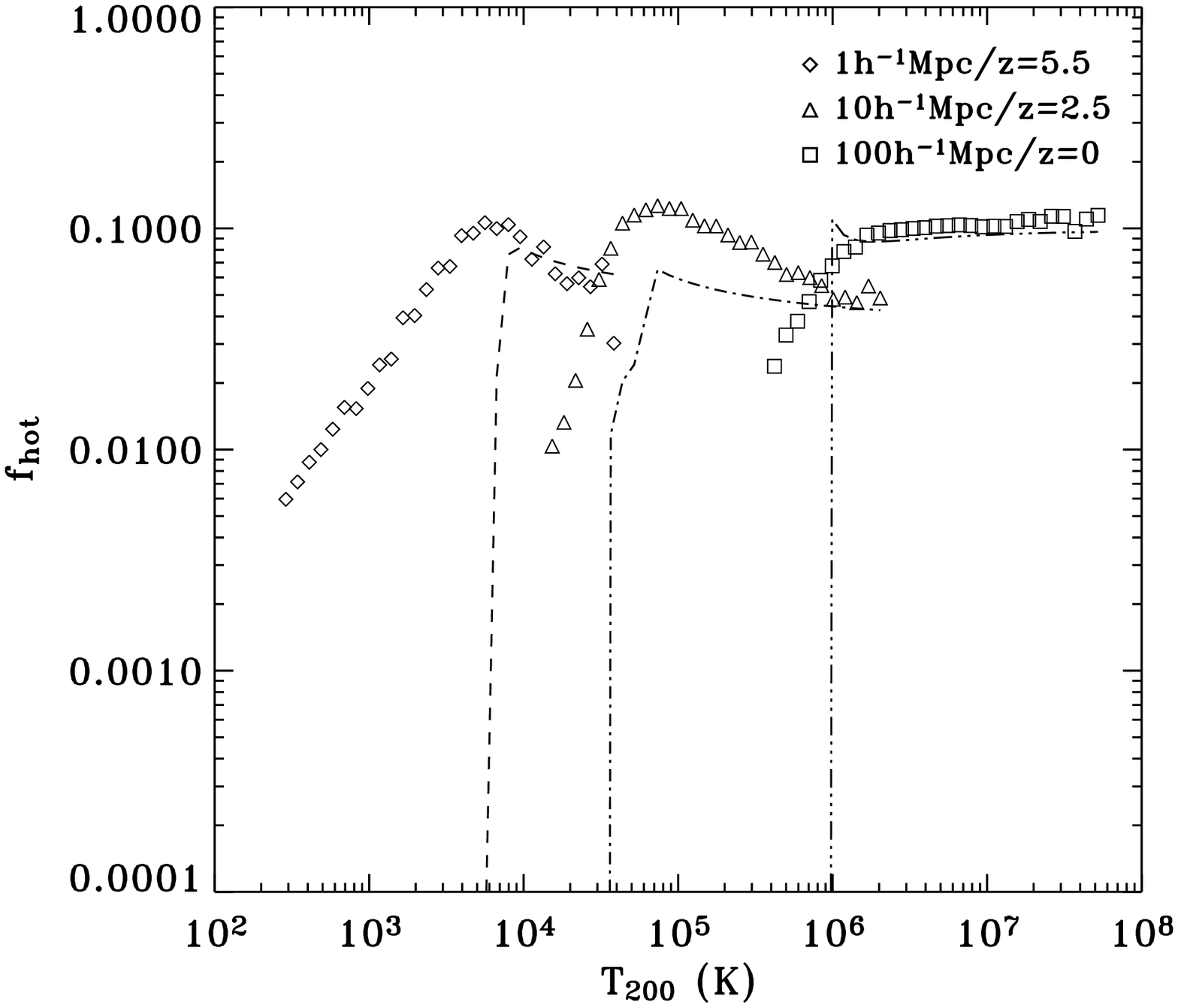}
    \includegraphics[width=0.5\hsize]{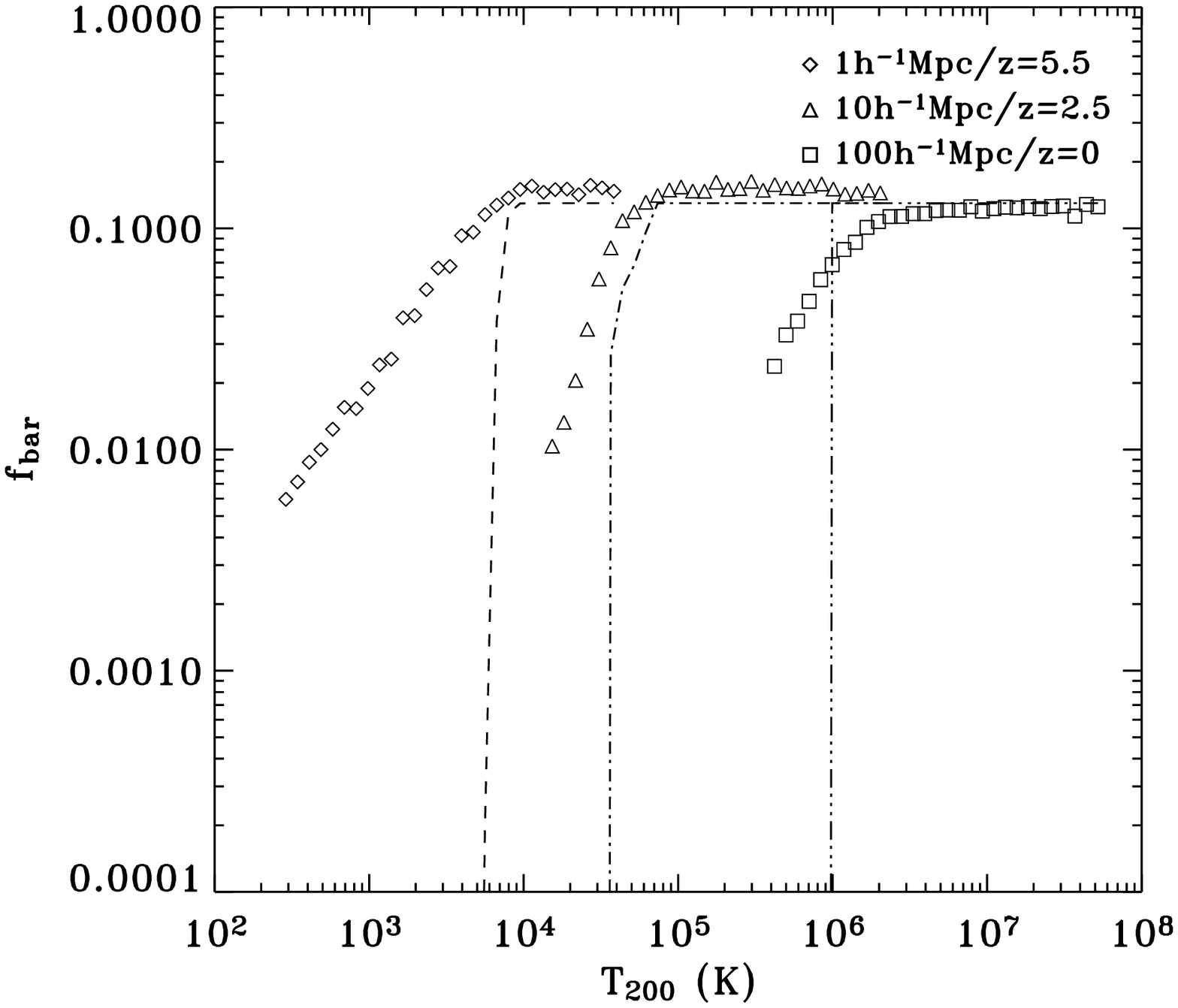}   \end{tabular}
    \cpt{Mass fraction in each baryon  phase as a function of halo
    Virial   temperature   for the   \emph{low  efficiency  simulation
    series}. In each plot, diamonds refer to run L1N256S30 at $z=5.5$,
    triangles  to  run  L10N256S30  at $z=2.5$   and   squares  to run
    L100N256S30 at $z=0$. Lines  are for the  corresponding analytical
    model.  The \emph{upper left plot} shows   the star mass fraction;
    the  \emph{lower left plot} shows the  hot  gas mass fraction; the
    \emph{lower right plot} shows  the  total baryon fraction and  the
    \emph{upper  right   plot}   shows  the     cold  gas   fraction.}
    \label{fracdemc001} \end{figure*}
  
  \begin{figure*}                                  \begin{tabular}{cc}
    \includegraphics[width=0.5\hsize]{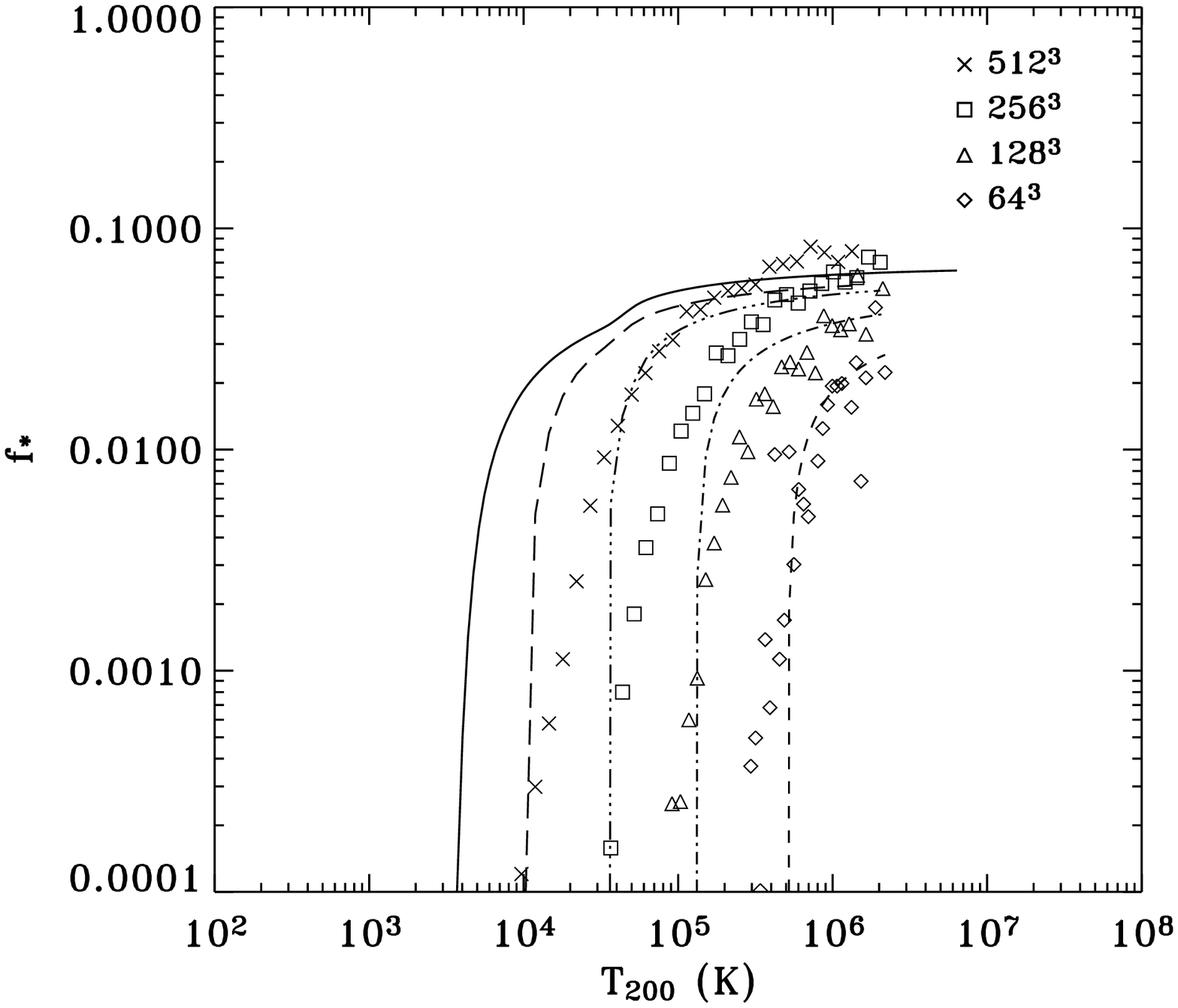}
    \includegraphics[width=0.5\hsize]{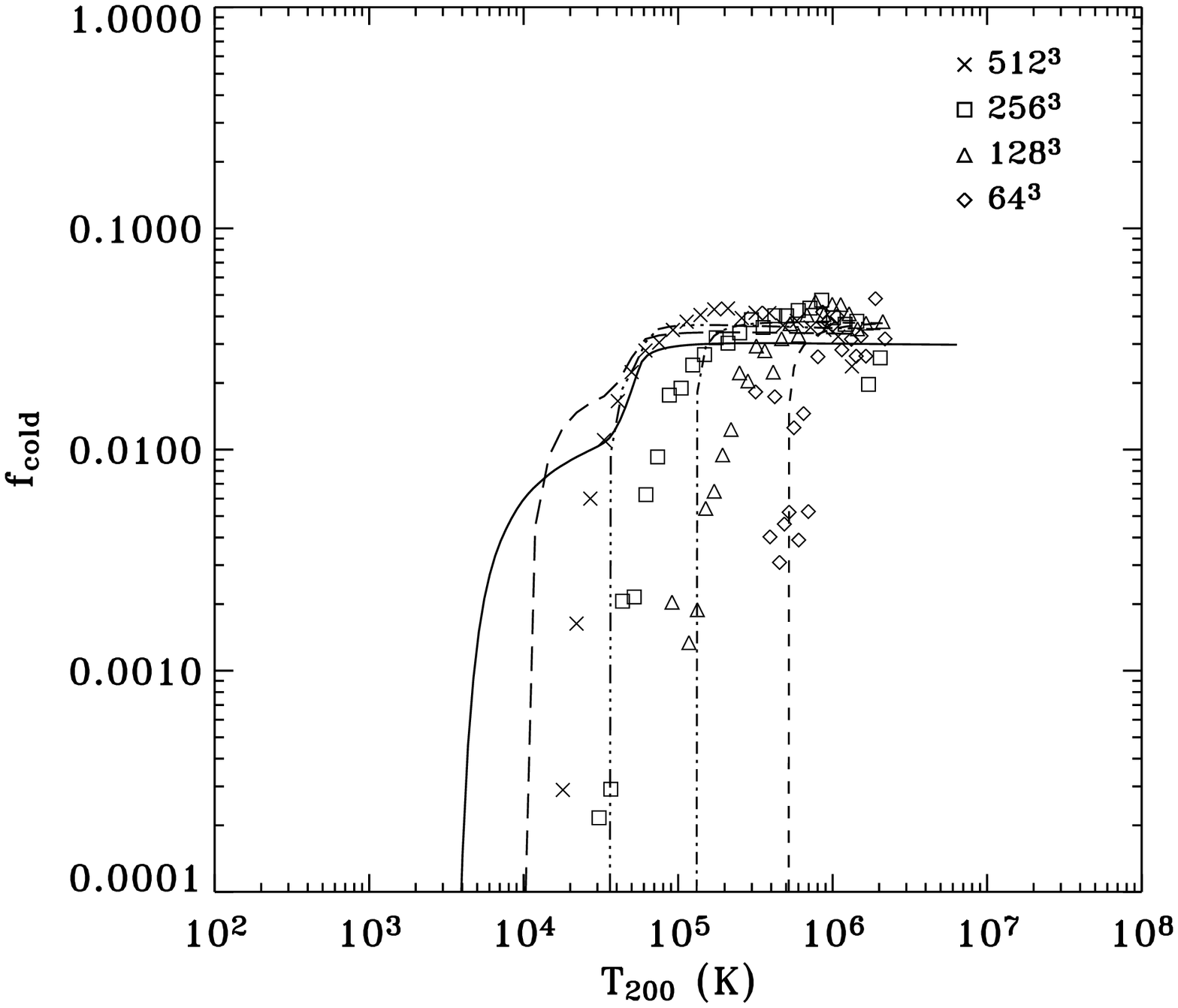}               \\
    \includegraphics[width=0.5\hsize]{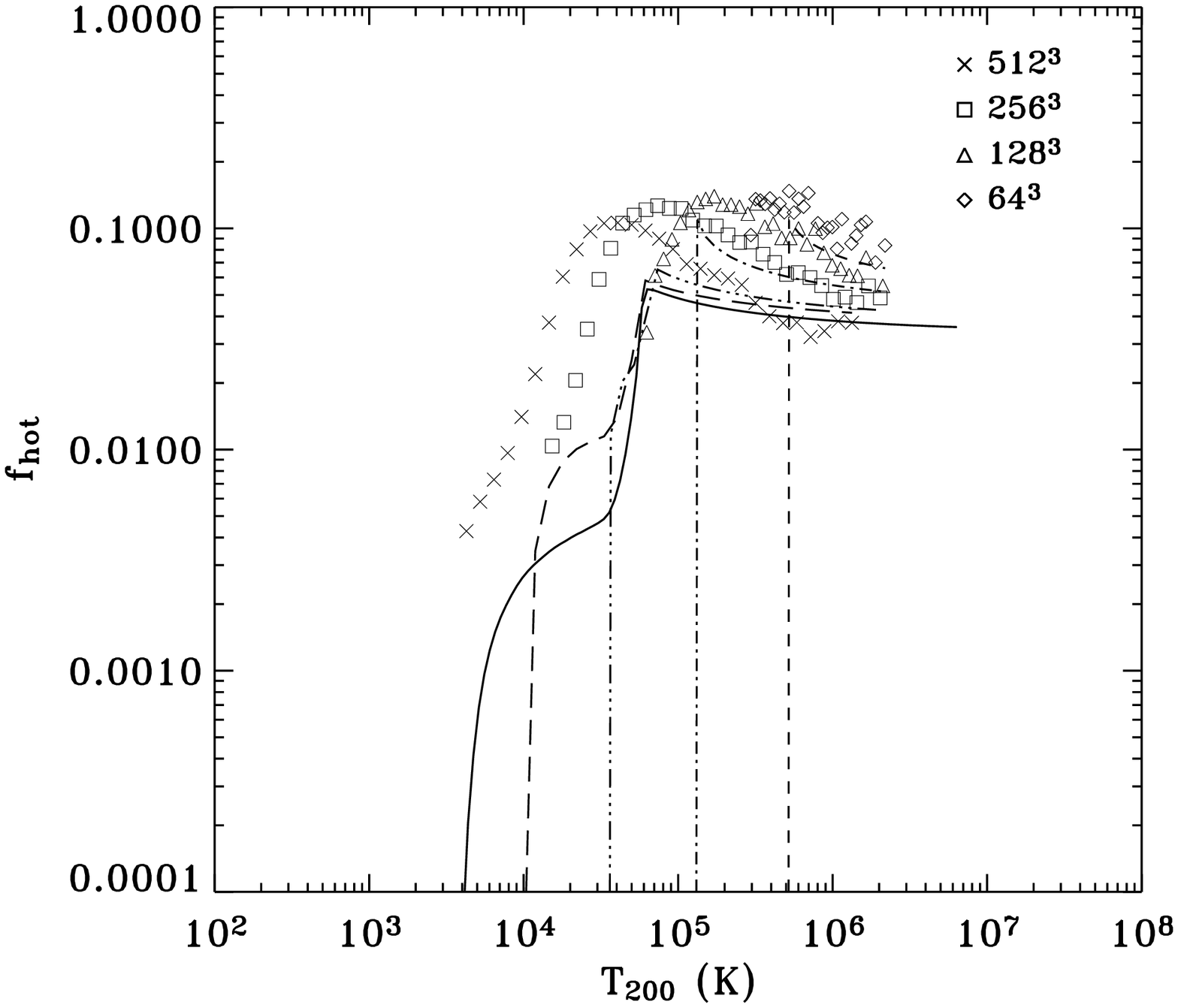}
    \includegraphics[width=0.5\hsize]{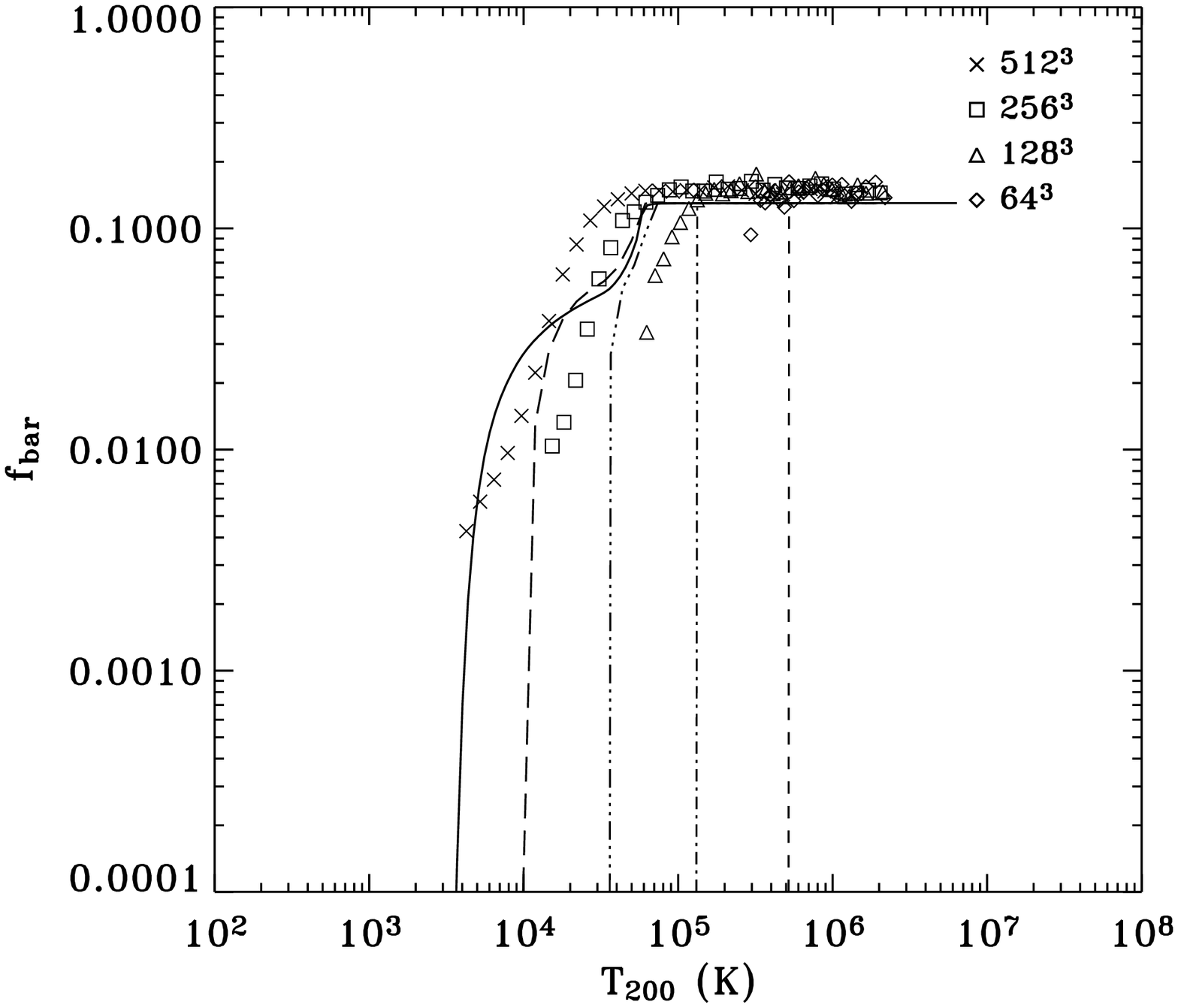}  \end{tabular}
    \cpt{Mass fraction in each baryon  phase as a function of halo
    Virial  temperature for  the  \emph{convergence  study  simulation
    series}  at     $z=2.8$.  In each   plot,   diamonds  refer to run
    L10N64S30,  triangles  to  to   run  L10N128S30, squares   to  run
    L10N256S30   and crosses to  run L10N512S30.    Lines  are for the
    corresponding analytical  model.  The \emph{upper left plot} shows
    the  star mass fraction; the \emph{lower  left plot} shows the hot
    gas mass  fraction;  the \emph{lower  right  plot} shows the total
    baryon fraction and the \emph{upper right plot} shows the cold gas
    fraction.  The solid   lines   are  the  `fully  converged'  model
    predictions.}  \label{fracdemresol} \end{figure*}
  
  Our  second series  of  simulations was  designed  to explore  larger
  ($L=100$ h$^{-1}$  Mpc) and smaller ($L=1$ h$^{-1}$  Mpc) scales, as
  well as another, more efficient, star formation scenario with $t_{\rm 0} =
  3$ Gyr.  The  corresponding Schmidt law is therefore  ten times more
  effective  than in  the  first  case. The  resolution  was fixed  to
  $256^3$ particles. All six  results are shown in Figure~\ref{sfrdez}
  with  symbols, while  the corresponding  analytical  predictions are
  overploted  with lines. Note  that the  smaller box  ($L=1$ h$^{-1}$
  Mpc) has a mass resolution  smaller than $M_{\rm min}$. It has therefore
  converged to  the `true'  star formation history.   The analytical
  model is indeed  in good agreement with our  numerical results. This
  very small  scale simulations have  to be stopped even  earlier than
  the  previous ones  ($z \simeq  5$). The  largest box  size ($L=100$
  h$^{-1}$ Mpc),  on the  other hand, is  strongly affected  by finite
  mass resolution  effect. The  star formation history  is drastically
  different  than the  other  two. Star  formation  begins very  late,
  around $z  \simeq 5$, and the  peak value is one  order of magnitude
  lower than  the `true' star  formation rate. Our  analytical model
  provides  again a good  fit to  numerical data,  when the  poor mass
  resolution is taken into account to define the diffuse background.

  Let  us  now  compare  the  two different  star  formation  scenario
  (inefficient with $t_{\rm 0}=30$ Gyr and efficient with $t_{\rm 0}=3$ Gyr). Both
  star formation history are parallel to each other, but with a factor
  of  two difference  only.   For the  very  efficient scenario,  star
  formation is  limited by accretion  and cooling, rather than  by the
  Schmidt law. As discussed in  the next section, very efficient star
  formation  models  give  SFR   quasi  independent  of  the  physical
  parameters.  This  is confirmed  by our numerical  simulations.  The
  agreement between  the simulation and  the model is  extremely good,
  except at low  redshift ($z <3$), for our large  box runs, where the
  model  significantly  overestimates the  star  formation rate.  This
  disagreement  might be  reduced by  improving the  analytical model,
  along  several lines  we have  outlined  in this  paper.  Note  that
  finite volume  effect might  also have an  additional effect  on the
  numerical simulation predictions, but we do not try to include those
  subtleties in the analytical model.
  
  From  now  on, we  have  analyzed  AMR  simulations for  which  star
  formation is  triggered in region  where the gas density  exceeds an
  {\it overdensity threshold}.  In this case, we naturally compare our
  numerical  results  to  the  `accelerated  efficiency'  analytical
  model.   We  now  test  the  model predictions  for  the  `constant
  efficiency' analytical model, using  SPH results kindly provided to
  us  by \cite{Springel03b}.   Our interest  to these  SPH  results is
  twofold: first, star formation is triggered in regions where the gas
  density exceeds  a {\it physical density  threshold}. This scenario,
  as  already discussed in  Section~\ref{halotstar}, corresponds  to a
  `constant    efficiency'    star    formation   model.     Second,
  \cite{Springel03b} have  included galactic winds  in their numerical
  modeling,  giving us the  opportunity to  test our  simple feedback
  model.

  In  Figure~\ref{sfrdez}   are shown \cite{Springel03b}  results when
  galactic winds    are turned  off.   The   only difference  with our
  simulations,  apart from   the overall numerical techniques,
  comes from the  star  formation details.   They  have considered the
  following  parameters $t_{\rm 0}  =  2.1$   Gyr, $n_{\rm 0}=0.1$  cm$^{-3}$  and
  $\alpha_{\rm 0} =  0$.  For the  analytical model, we  use the same `shape
  factor' $F(\mu)=3$, as   determined earlier using  our AMR  results.
  This translates  into the following   parameters for the  analytical
  model: $t_* = 0.7$ Gyr and $\alpha_* = 0$.   Taking into account the
  finite  mass resolution  (using Equation~\ref{sfrdez}),   we can now
  compare our analytical prediction to  SPH results.  The agreement is
  clearly  very good: this is rather  encouraging for our model, since
  our main unknown parameter, the  mean orbital length, was kept fixed
  to  its canonical value   $R_{\rm orb}=3R_{\rm 200}$, calibrated on  our AMR
  results.

  We now analyze SPH results when galactic winds are turned on. Recall
  that winds are assumed to eject  cold gas from the rotating discs at
  a  typical wind  velocity  $u_{\rm w}  \simeq 250-500$  km/s  and with  an
  efficiency  parameterized   by  $\eta_{\rm w}$.   This   approach,  directly
  inspired by \cite{Springel03b},  allows a straightforward comparison
  between  our analytical  model  and SPH  results  with winds.   This
  comparison  is shown in  Figure~\ref{sfrdez}.  The  analytical model
  parameters were set to $T_{\rm w} =  2\times 10^6$ K and $\eta_{\rm w} = 3$. The
  Virial temperature  $T_{\rm w}$ (below which winds are  unbound from their
  parent halo)  corresponds closely to  the wind velocity  $u_{\rm w} \simeq
  500$ km/s  used by \cite{Springel03b}  in their SPH  simulation (see
  Eq.~\ref{twind}). The wind  efficiency parameter, however, was chosen
  50\% higher than the value $\eta_{\rm w}  = 2$ used in the SPH simulation.
  As suggested  by \cite{Springel03b},  the global wind  efficiency (at
  the halo scale) is higher than the local wind efficiency (at the star
  forming regions scale) due to gas entrainement: additional cold gas,
  lying outside  star forming regions from which  winds originate, can
  be expelled by the  ram pressure of the outflow. Figure~\ref{sfrdez}
  shows again a good agreement  between SPH simulation results and our
  analytical model.

  We   consider   now  SPH   results   from   the   `Q  series'   in
  \cite{Springel03b} paper,  in order to study  finite mass resolution
  effects.   Wind parameters  are fixed  to $T_{\rm w}=2\times  10^6$  K and
  $\eta_{\rm w}  = 3$.   Note that  in this  case, due  to a  star formation
  strategy based  on a constant  physical density, $M_{\rm resol}$  is now
  varying  with  time   $M_{\rm resol}  \simeq  1000(1+z)^{-1.5}m_{\rm p}$.   In
  Figure~\ref{sfrdez}   are   plotted  \cite{Springel03b}   simulation
  results together  with our  model predictions, when  mass resolution
  effects are  taken into  account. At high  redshift, we  recover the
  correct convergence towards the asymptotic ($M_{\rm resol}=0$) converged
  curve. At  intermediate redshift, however, the  agreement is getting
  worse,  although both  curves remain  close to  each other  within a
  factor of  2.  One possible reason  is that Equation~\ref{massresol}
  is  not  accurate  enough   to  estimate  the  effective  halo  mass
  resolution of SPH simulations, especially in presence of winds.
  
  We conclude  that the global  baryon history we obtain  in numerical
  simulations  is correctly  described by  our analytical  model, {\it
  when finite  mass resolution effects  are taken into  account}. This
  was tested  for 2  different numerical technics  (AMR and  SPH), two
  different star formation  scenarios (`constant' and `accelerated'
  efficiency),  with and without  galactic winds.   The main  effect of
  insufficient mass resolution is  to artificially decrease the average
  age of stars in the universe and to lower the star formation rate at
  peak value.

  \subsection{Halo Baryon Budget}

  We present  now our simulation results concerning  the baryon budget
  in  individual  halos.   Halos  are detected  in  the  simulations
  according to the  method explained in Section~\ref{halodetect}.  The
  Extended Press \& Schechter (EPS)  theory give us the opportunity to
  apply our analytical method to  an `average' halo of mass $M_{\rm 0}$ at
  redshift $z_{\rm 0}$.  The  model predictions can then be  compared to the
  {\it average}  baryon fractions (in  each of the 4  different baryon
  phases), the  average being  taken over all  halos in a  given mass
  range, centered around $M_{\rm 0}$.

  Using all star particles found within $R_{\rm 200}$, we compute the star
  formation  history  within each  halo.   We  then  compute the  star
  formation    rate   as   explained    in   Section~\ref{halodetect}.
  Figure~\ref{sfrdem} shows  our various numerical  results, including
  the SPH simulations with winds provided to us by \cite{Springel03b},
  together with the analytical  model predictions. We have plotted the
  halo  star formation  rate as  a  function of  the halo  temperature
  $T_{\rm 200}$,  in  units of  $M_{\rm 200}/t$.  This definition  corresponds
  closely  to   the  `specific  star  formation   rate'  defined  by
  \cite{Springel03b}. 

  Each curve shows a sharp  cut-off at the low mass end, corresponding
  to the Minimal Mass for  each run.  The `convergence study' series
  clearly illustrates that  this Minimal Mass is in  fact equal to the
  mass   resolution   $M_{\rm resol}$,  except   for   our  smallest   box
  size. Another  non trivial  effect of finite  mass resolution  is to
  increase  the halo  star formation  rate. This  is due  to  a higher
  Cosmic Accretion Rate ($\dot{f}_{\rm acc} \propto \sigma(M_{\rm min})^{-1}$,
  see  Eq.~\ref{asymptoticaccrate}), as  $M_{\rm min}$ is  increased.  Our
  analytical  model predictions are  in good  agreement with  the halo
  star formation rates, even when winds are present, as soon as finite
  mass resolution is explicitly accounted for in the model.  The role
  of winds is  to remove cold gas from small mass  halos $T_{\rm 200} < 2
  \times  10^6$ K,  so  that  the halo  star  formation rate  decrease
  accordingly.

  When we compare the high star formation efficiency series with $t_{\rm 0} =
  3$ Gyr with the low star  formation efficiency series with $t_{\rm 0} = 30$
  Gyr, we see that the former has a higher halo star formation rate at
  high  redshift, but a  much lower  halo star  formation rate  at low
  redshift, as  cold gas is almost completely  consumed.  This complex
  behavior is  well reproduced by  the analytical model.   Both series
  show a  sharp decline of  the halo star  formation rate at  the high
  mass end: the cooling efficiency decreases for high mass halos and the
  mass accretion  rate on cold  discs vanishes.  Note that  we assumed
  here a  zero metallicity  plasma. Gas cooling  is likely to  be more
  efficient around $T_{\rm 200} \simeq 10^7$ K if metals are present.

  We now  present the  baryon budget inside  dark matter halos,  as a
  function of  the halo Virial  temperature. In each halo,  we compute
  the total baryon fraction, which  can be further decomposed into hot
  gas,   cold    gas   and    stars,   using   the    definitions   of
  Section~\ref{results}   and  \ref{halodetect}.  Figure~\ref{fracdem}
  shows this halo baryon budget for the `high efficiency' simulation
  series,   Figure~\ref{fracdemc001}   for   the  `low   efficiency'
  simulation    series   and    Figure~\ref{fracdemresol}    for   the
  `convergence study' simulation series.

  In   each   case, there is good      agreement between numerical and
  analytical results.  The total baryon fraction is close to the
  average value   $f_{\rm b}$ for halos   more  massive than  $M_{\rm min}$  and
  vanishes for smaller halos.  Note that the  total baryon fraction is
  actually slightly {\it above} the universal value for massive halos.
  Dissipative  collapse    of   baryons  condense   more     mass than
  collisionless collapse of dark matter. The analytical model predicts
  a  sharp   transition for $M  <   M_{\rm resol}$ where  the total baryon
  fraction  vanishes.  In  the   simulations, this transition  is much
  smoother,   but  its   location   is   correctly predicted    around
  $M_{\rm resol}$.  On  the other hand, when the  mass resolution is small
  enough, the low mass  end of the  analytical curve has also a smooth
  transition.  This  comes from halos   whose  mass was  greater  than
  $M_{\rm min}$ at early time, but as reionization heats up the background
  temperature,  these  same halos  end   up with  a  mass smaller than
  $M_{\rm min}$.  The   total baryon fraction  at   the final redshift  is
  therefore smaller than the universal value but greater than zero.

  The hot gas fraction is also strongly affected by resolution effect.
  As we  resolve smaller and  smaller mass progenitors, more  and more
  gas    is    condensed    into    cold    discs    and    eventually
  stars. Figure~\ref{fracdemresol} shows the  hot gas fraction for the
  `convergence study'  series and one clearly sees  that our poorest
  resolution run overestimates the hot  gas fraction by a factor of 3.
  As  a consequence,  the cold  gas and  star mass  fraction gradually
  increase as the  mass resolution is improved. The  curve showing the
  cold  gas fraction  as  a  function of  Virial  temperature is  very
  similar  to the  halo  star  formation rate,  as  expected from  our
  analytical  model,  although  they  have  been  computed  completely
  differently in the simulation  analysis.  In the `high efficiency'
  series, our largest box  run L100N256S3 has almost entirely consumed
  the cold gas of the most massive halos, where cooling has virtually
  stopped  (see  Fig.~\ref{fracdem}).  The  corresponding  run with  a
  `low efficiency' star formation scenario still shows some surviving
  cold gas at the high mass end of the halo distribution.

  We conclude at the end of this section that our analytical model has
  proven to successfully   reproduce the complex behavior  observed in
  our   simulations  and in  simulations   of \cite{Springel03b} (with
  winds) {\it if finite resolution effects are taken into account}. We
  note however a slight tendency of the model to overestimate the star
  formation rate at low redshift.   The baryon budget was analyzed  in
  great details on a halo by halo basis (and in  an average sense). We
  emphasized the important role   played by the Minimal  Mass  $M_{\rm
  min}$ in controlling the baryon history  in each mass range. We also
  described how cooling and winds introduce characteristic features in
  the various plots showing the baryon evolution as  a function of the
  halo Virial temperature.

  \section{Observations versus Model}

  \begin{figure}     \includegraphics[width=\hsize]{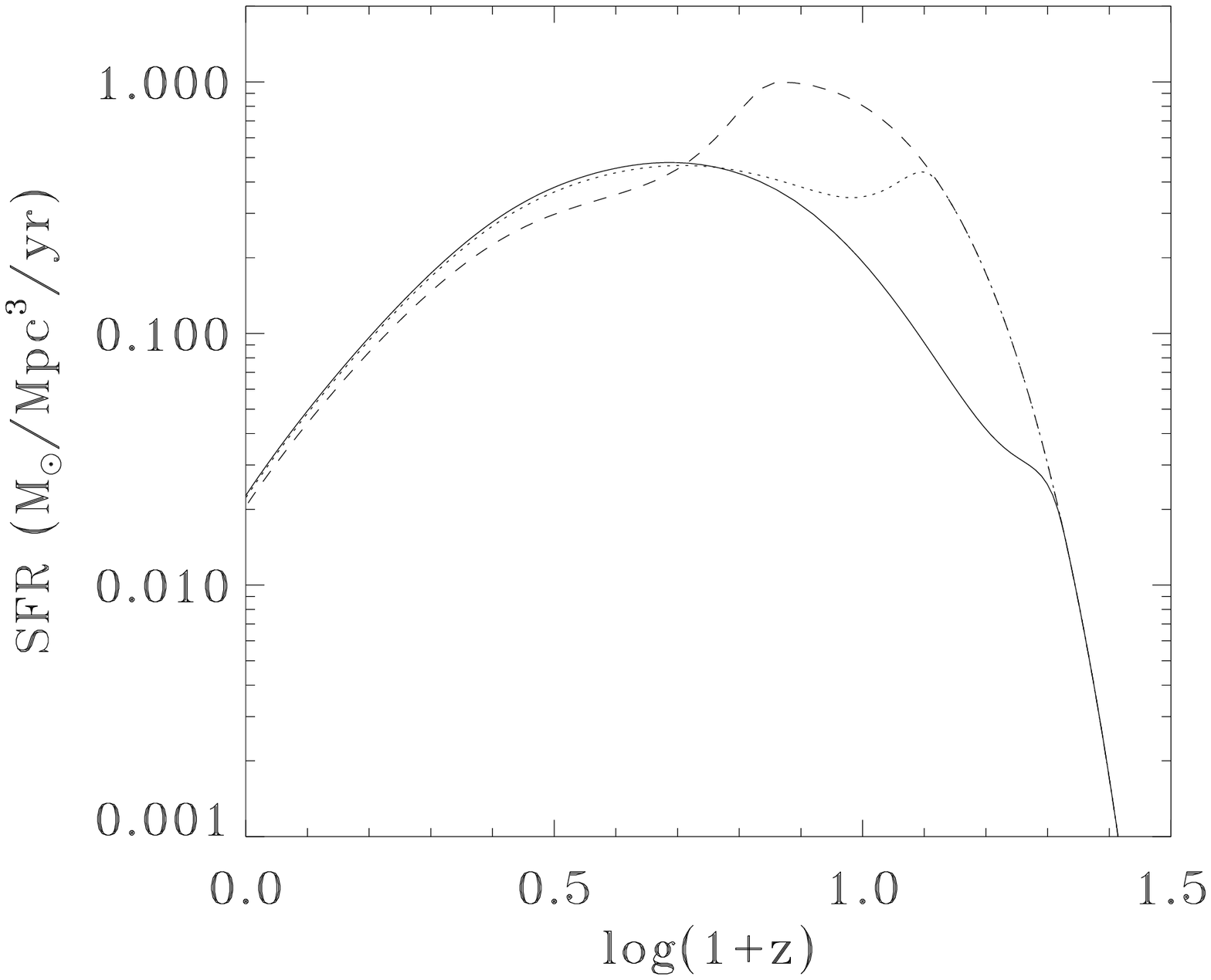}
      \cpt{Cosmic   Star  Formation  Rate   for   different  model
      parameters: $t_*=0$  and $\eta_{\rm w}=0$, $z_{\rm r}=20$ (solid
      line), $z_{\rm   r}=12$  (dotted line),  $z_{\rm  r}=6$  (dashed
      line).}  \label{sfrdez_zr} 
    \end{figure}

  \begin{figure*}                                 \begin{tabular}{ccc}
    \includegraphics[width=0.5\hsize]{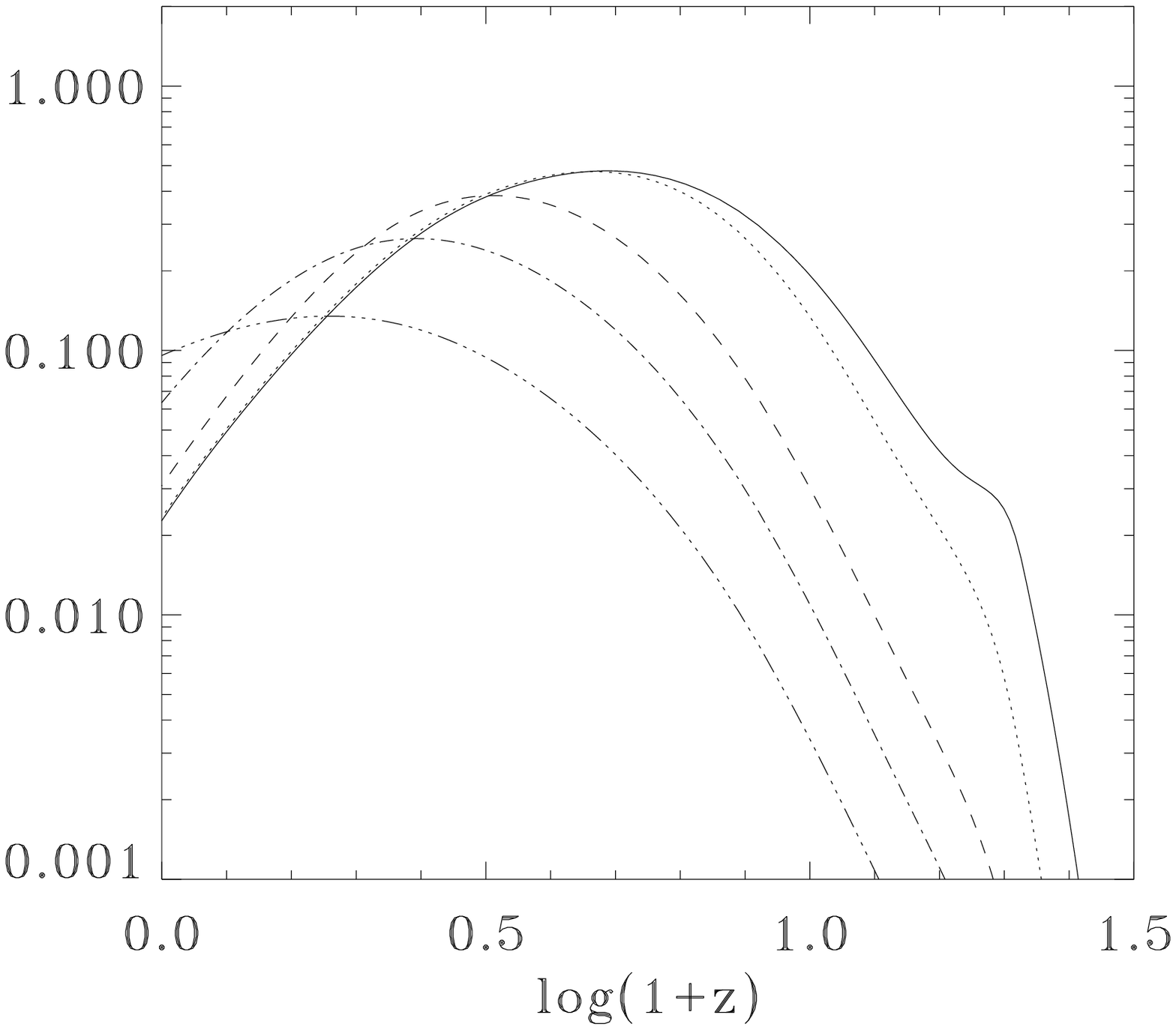}
    \includegraphics[width=0.5\hsize]{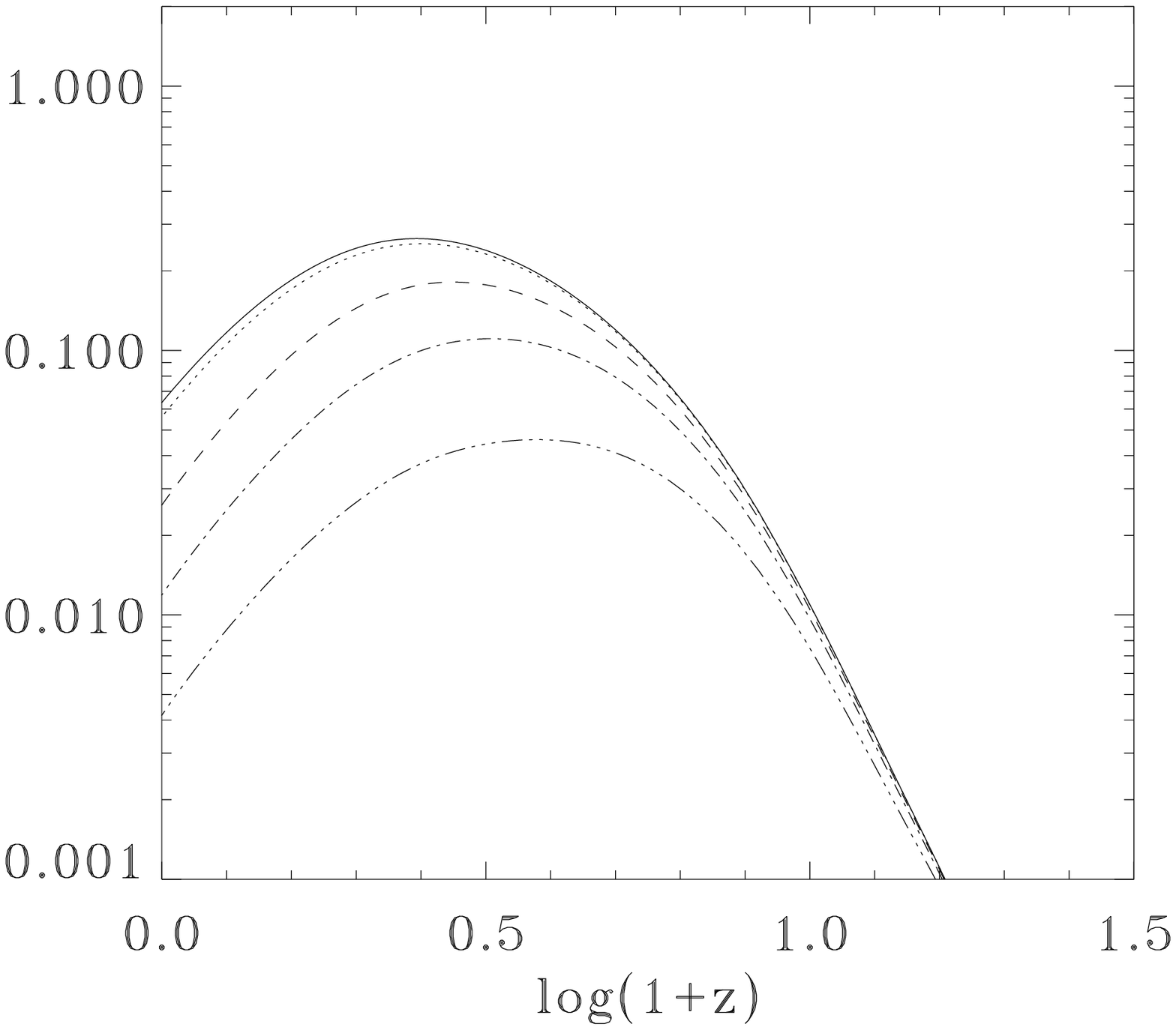}
    \end{tabular}  \cpt{Cosmic Star  Formation Rate for  different
    model parameters. Left plot: $z_{\rm r}=20$ and $\eta_{\rm w}=0$,
    $t_*=0$  (solid line),  $t_*=0.1$ Gyr (dotted   line), $t_*=1$ Gyr
    (dashed    line),  $t_*=3$ Gyr    (dot-dashed line),  $t_*=10$ Gyr
    (dot-dot-dashed line). Right plot: $z_{\rm r}=20$ and $t_*=3$ Gyr,
    $\eta_{\rm w}=0$ (solid  line), $\eta_{\rm w}=0.1$ (dotted  line),
    $\eta_{\rm   w}=1$  (dashed  line), $\eta_{\rm  w}=3$  (dot-dashed
    line),   $\eta_{\rm        w}=10$        (dot-dot-dashed   line).}
    \label{sfrdezparams} \end{figure*}

  From now on,  we assume that our analytical  model gives an accurate
  (within  a  factor  of 2)  modeling  of  the  baryon history  in  a
  hierarchical   universe.   The   careful  comparison   to  numerical
  simulations  we performed in  the previous  section was  a necessary
  step  to  calibrate   our  model  and  to  estimate   its  level  of
  accuracy. We  are now in a  position to use  this model as a  tool to
  analyze current observational constraints  put on the baryon history
  in the universe.

  \subsection{Model parameters study}

  We briefly recall how the analytical predictions depend on the model
  parameters. We  consider that all  cosmological parameters are fixed
  to  their `concordance' $\Lambda$CDM  value, as  we did throughout
  this  paper. We  also assume that  the  reionization temperature  is
  given by  $T_{\rm r} \simeq 6 \times 10^3$  K.  We finally use the cooling
  model  of Section~\ref{coolingmodel},  valid for  a primordial, zero
  metallicity, H and  He plasma and a wind  velocity $u_{\rm w} \simeq  500$
  km/s.  We end up with 3 main parameters that we allow to vary in our
  model:  the reionization redshift   $z_{\rm r}$,  the star  formation time
  scale  (or consumption time  scale) $t_*=M_{\rm cold}/\dot{M}_*$ and the
  wind    efficiency $\eta_{\rm w}=\dot{M}_{\rm wind}/\dot{M}_*$.  We   consider   in  this section  only
  `constant efficiency'  star  formation models  $\alpha_*=0$.   The
  alternate   scenario with `accelerated  efficiency' star formation
  and $\alpha_*=3$ gives  similar,  though  slightly poorer,  results.
  Using  current observational  constraints,  it    is not easy     to
  discriminate between these two models.
  
  We now present  the  model predictions  for  the
  Cosmic Star Formation Rate, for different values of $z_{\rm r}$, $t_*$ and
  $\eta_{\rm w}$.  Figure~\ref{sfrdez_zr}  corresponds  to an `infinite
  efficiency' star  formation    model  ($t_*=0$   Gyr),  for  various
  reionization   redshifts.  In this extreme    case,  the cosmic star
  formation rate (CSFR)  is equal to  the mass accretion  rate of cold
  discs, since star  formation is instantaneous.  At reionization, the
  background gas is promptly heated to $10^4$ K and the star formation
  rate drops. We  notice  that at redshifts lower  than  6, all models
  agree with one another, leading  us to the  conclusion that, as soon
  as  reionization proceeds  early enough,  this parameter  is in fact
  irrelevant to the low redshift universe.

  Figure~\ref{sfrdezparams}  shows the     influence of the  two
  remaining (and really interesting) parameters. Star formation inside
  cold, centrifugally supported, galactic discs acts merely as a delay
  with   respect to diffuse   gas accretion.  The  epoch  of peak star
  formation is delayed from $z \simeq 4$ for $t_*=0$ down to $z \simeq
  1$ for $t_*=10$ Gyr. It is worth noticing that  the amplitude of the
  global star formation rate  is mainly determined by the cosmological
  accretion rate,  and that the small scale,  poorly known, physics of
  star formation introduces a  rather small modification to the curve.
  When  winds are included in  the model, they lower significantly the
  amplitude of the CSFR.  They  also have  the  non trivial effect  of
  {\it advancing} the epoch of peak star  formation, from $z \simeq 1$
  for $\eta_{\rm w} = 0$ up to $z \simeq 3$ for $\eta_{\rm w}  = 10$. They play an
  important role  at low  redshift whereas at   high redshift the CSFR
  seems  independent    of  $\eta_{\rm w}$.   Indeed,  for  the   `constant
  efficiency'  recipe the star formation  rate at high redshift is so
  small   compared   to  the accretion  rate   that   winds (which are
  proportional to the star  formation rate) doesn't affect the amount
  of cold gas.
 
  Computing the  baryon history using  our analytical model is  a very
  fast  operation. This gives  us the  opportunity to  compute various
  observational  quantities for  a  grid of  model parameters.   Using
  various observations,  we will now  try to constrain  our parameter
  space  and  shed  light   to  these  2  important  galaxy  formation
  ingredients: $t_*$ and $\eta_{\rm w}$.
  
  \subsection{Cosmic Star Formation Rate}

\begin{figure*}                                  \begin{tabular}{cc}
    \includegraphics[width=0.46\hsize,height=0.37\hsize]{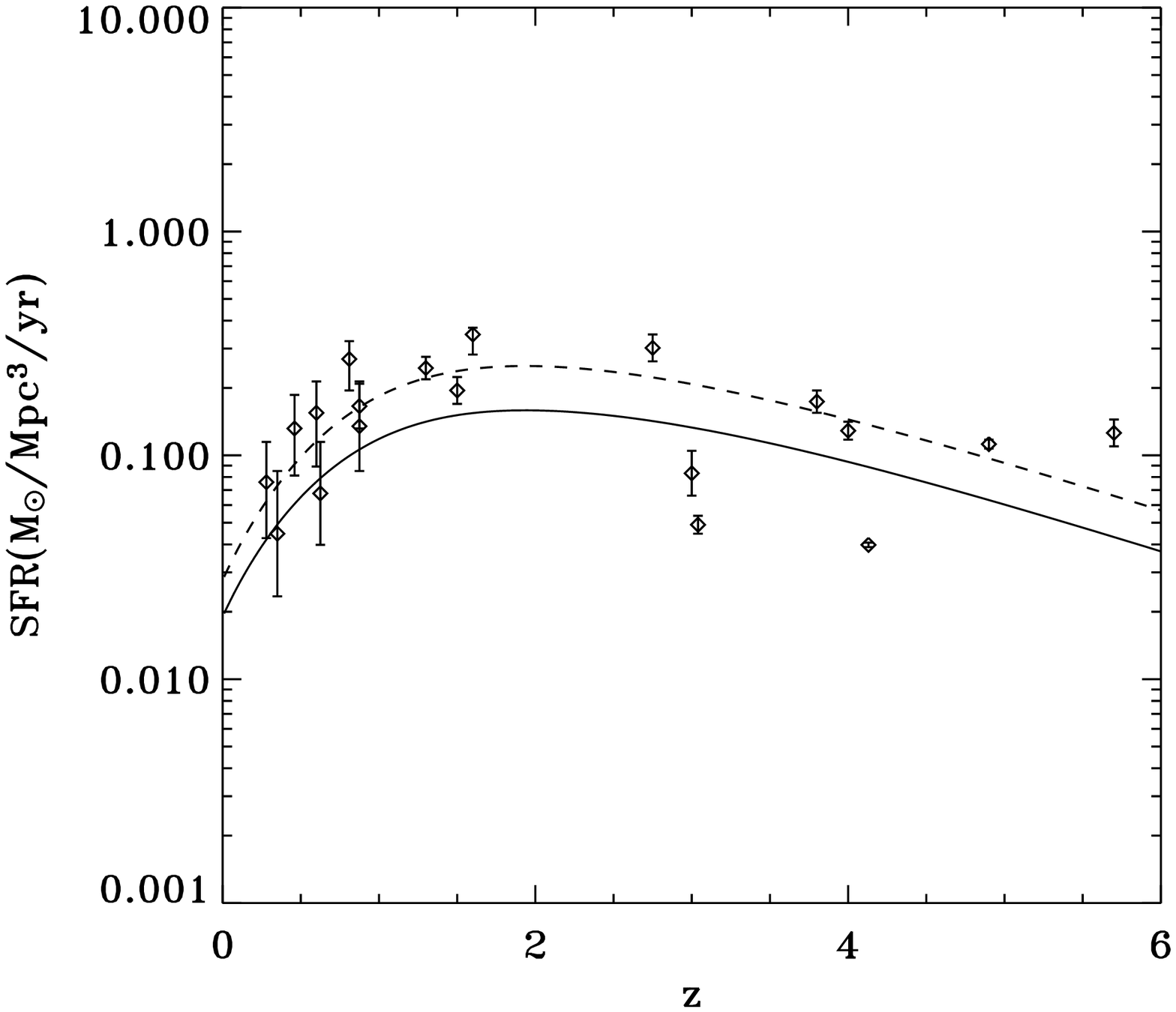}                &
    \includegraphics[width=0.46\hsize,height=0.37\hsize]{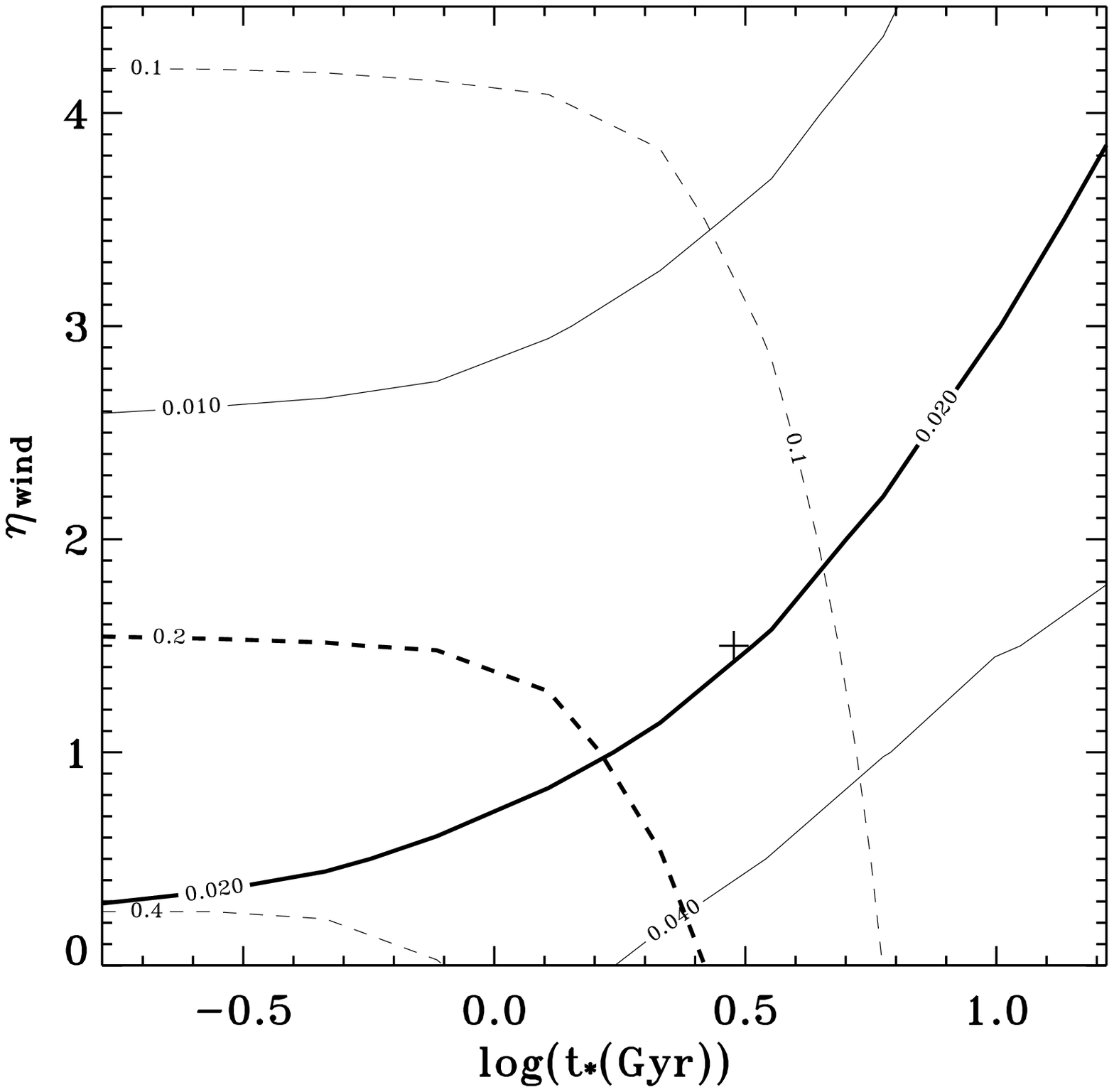}\\
    \includegraphics[width=0.46\hsize,height=0.37\hsize]{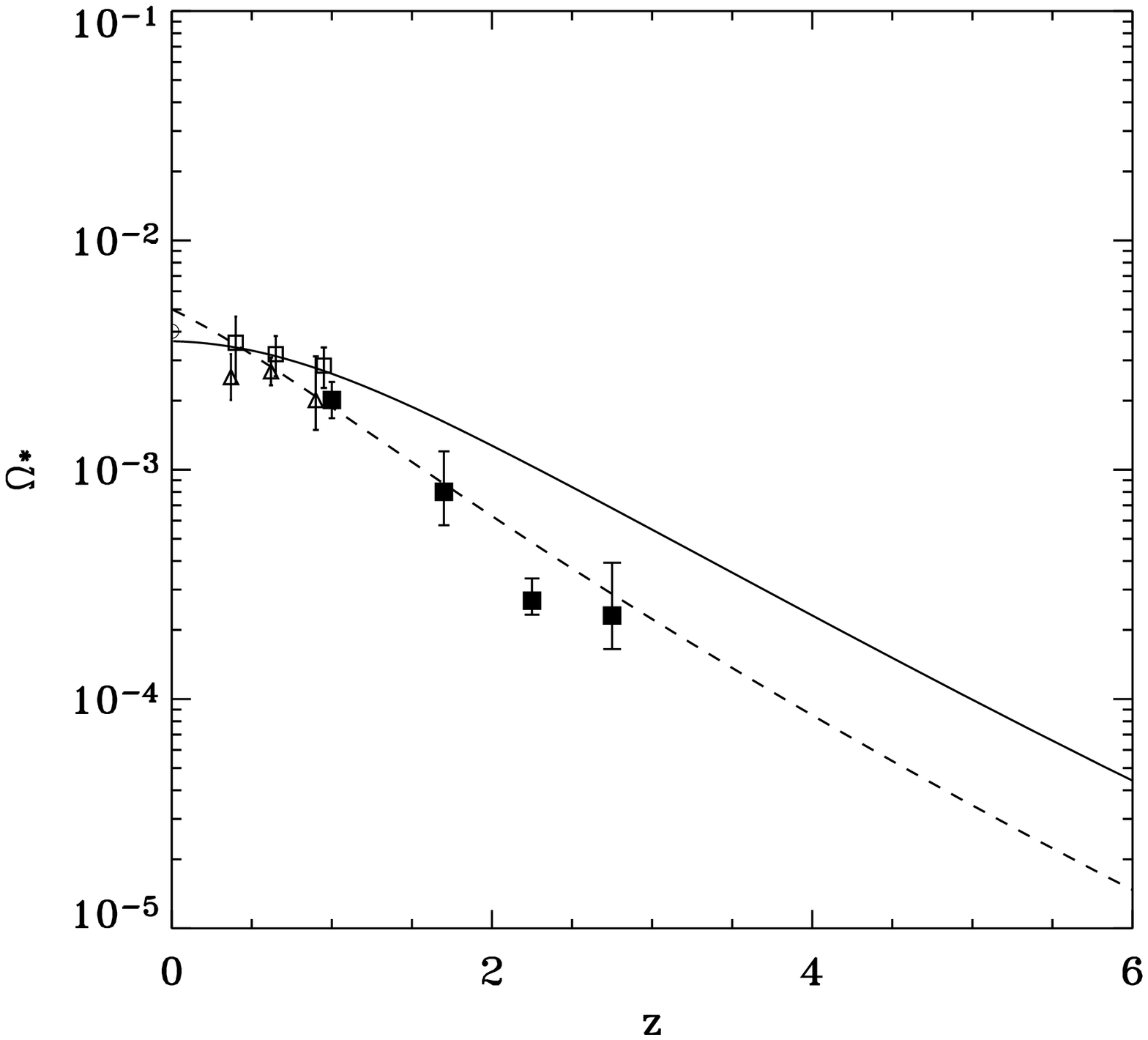}                 &
    \includegraphics[width=0.46\hsize,height=0.37\hsize]{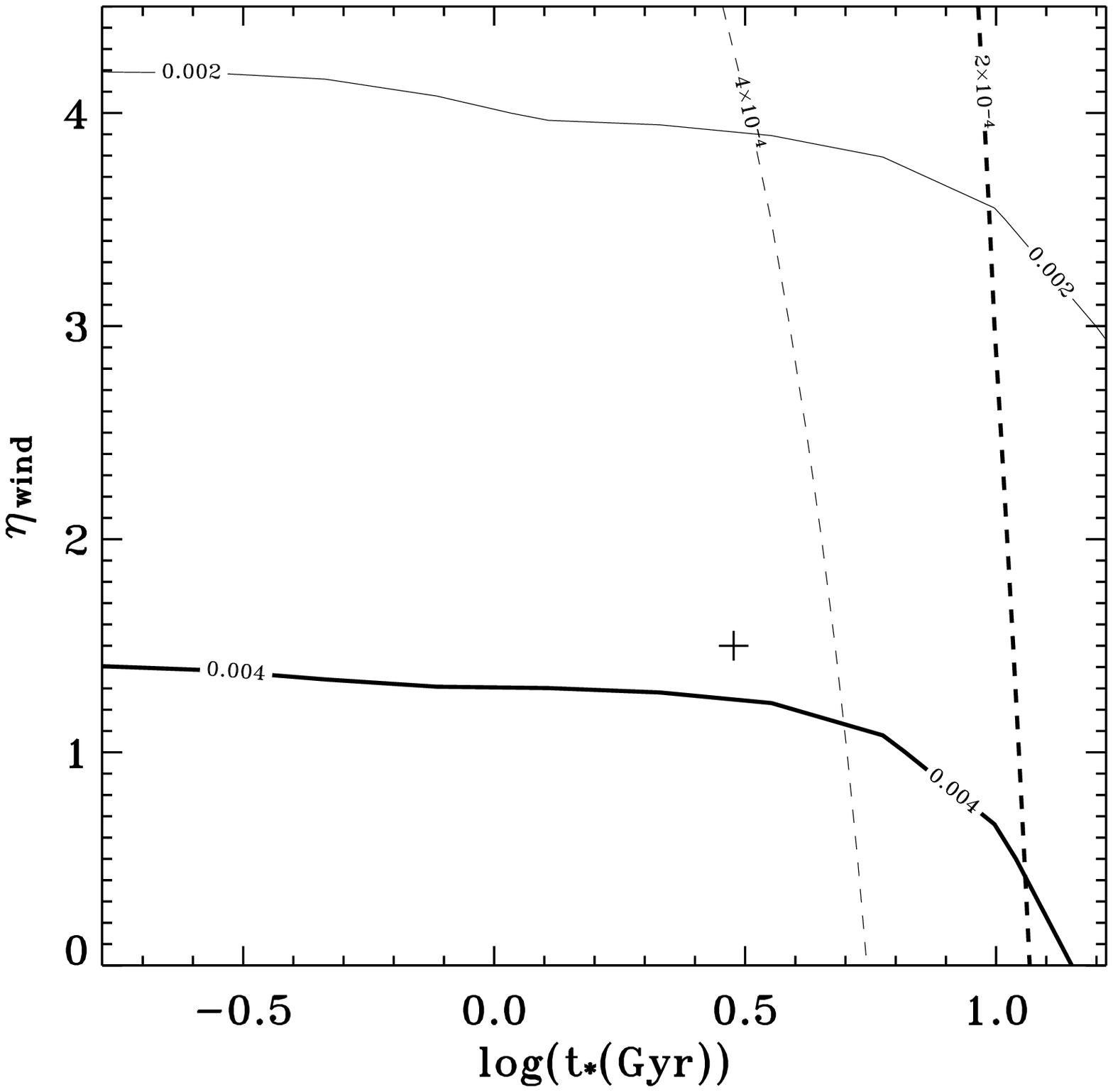}\\
    \includegraphics[width=0.46\hsize,height=0.37\hsize]{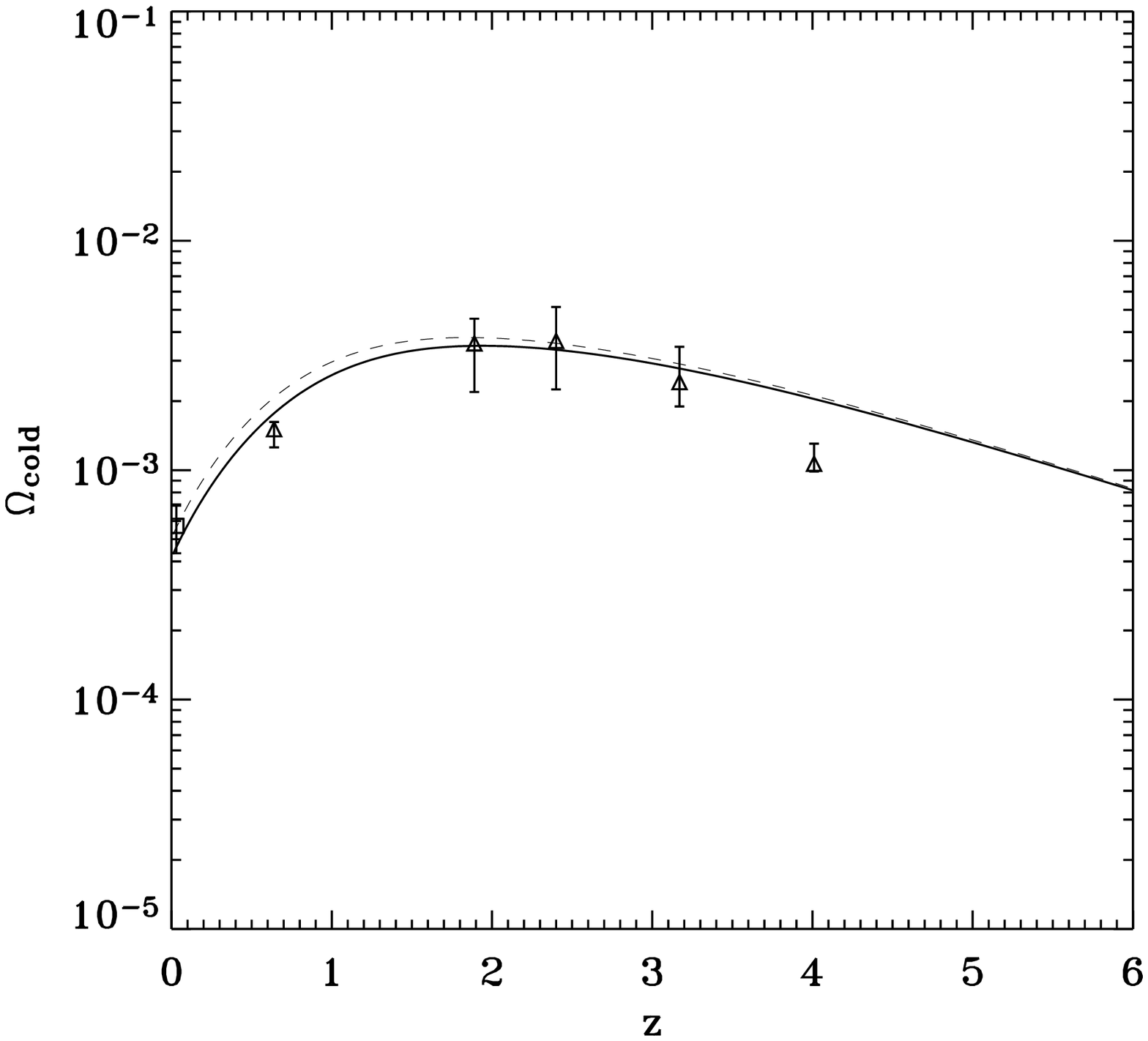}              &
    \includegraphics[width=0.46\hsize,height=0.37\hsize]{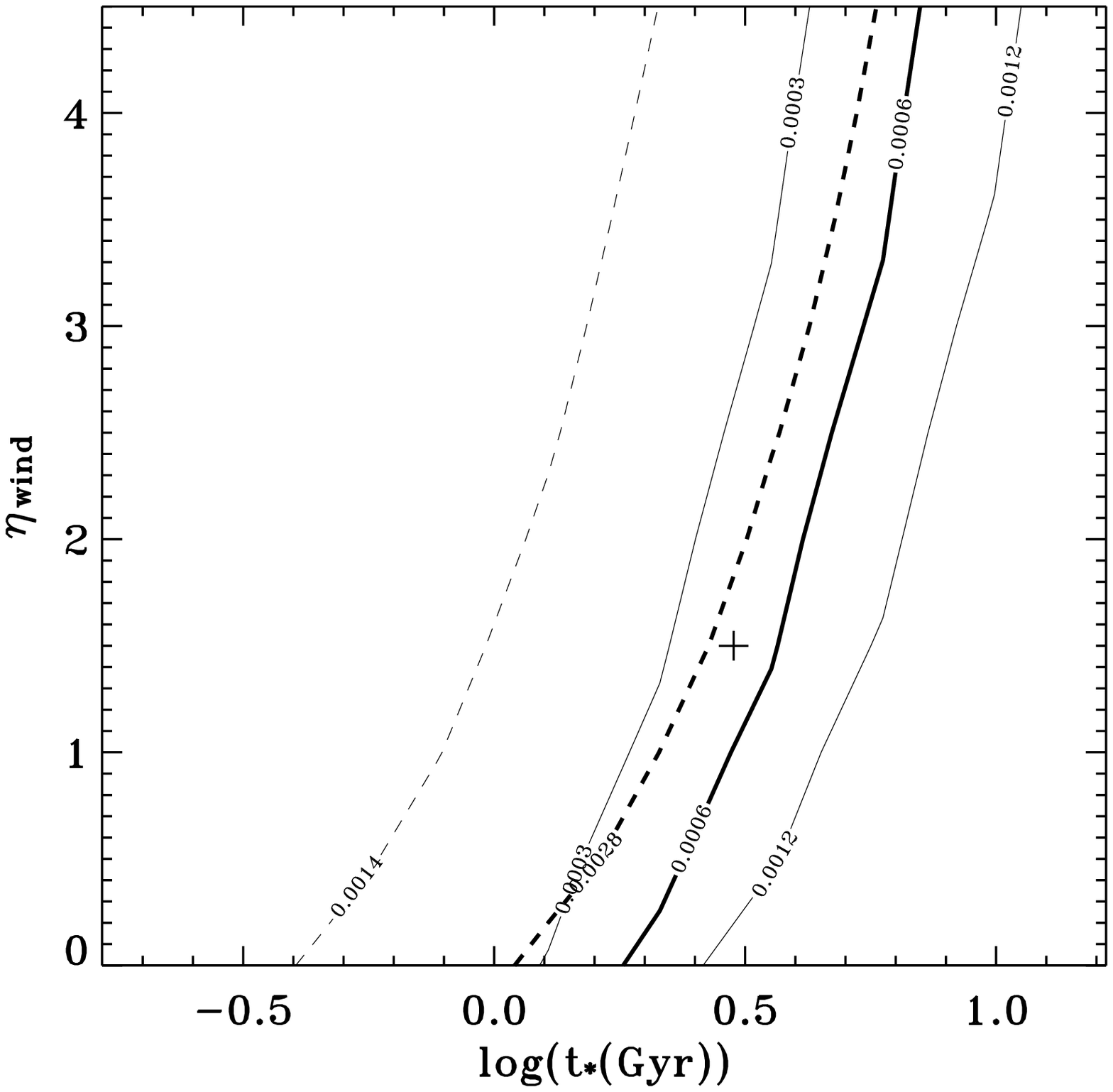}\\
    \end{tabular} \cpt{\emph{Upper left plot}: Cosmic Star Formation Rate
    as a  function of  redshift.   Data points, uniformally  corrected
    from various   observational  biases, come   from  \cite{Hughes98,
    Steidel99,      Flores99,   Glazebrook99,  Yan99,    Massarotti01,
    Giavalisco03}.  The  solid line  is  our canonical  model ($t_*=3$
    Gyr, $\eta_{\rm w}=1.5$).  The dashed line is our best fit model for the
    SFR ($t_*=1.5$ Gyr, $\eta_{\rm w}=1$).    \emph{Upper right plot}:  contour  of
    constant star  formation    rate (solid contours:  $z=0$,   dashed
    contours:  $z=3$)  as  a function  of  $t_*$  and  $\eta_{\rm w}$.  Bold
    contours are the  observed   value, while others correspond  to  a
    factor of 2 above and below the observed central value.  The cross
    is the position of our canonical model.   \emph{Middle left plot}: Cosmic
    Stellar    Density as      a   function    of     redshift,   from
    \cite{Dickinson03}. The   solid  line is   the  prediction  of our
    canonical  model, while  the dashed  line is   our best fit  model
    ($t_*=10$ Gyr,  $\eta_{\rm w}=0.5$).    \emph{Middle right  plot}:  contour  of
    constant stellar density  (solid contours: $z=0$, dashed contours:
    $z=3$).  \emph{Lower left plot}: cosmic gas density in damped Lyman alpha
    systems, from   \cite{Somerville01}. The  solid and   dashed lines
    corresponds to the mass fraction in cold  gas, as predicted by our
    canonical model. \emph{Lower  right plot}: contour  of constant  cold gas
    density  (solid   contours:    $z=0$,  dashed  contours:  $z=3$).}
    \label{sfrobs} \end{figure*}

  One of the main goal of this paper is  to compute the star formation
  history in a  hierarchical universe.  We will  now compare the model
  predictions to the observed star formation rate. Figure~\ref{sfrobs}
  shows  the observational  data  points, usually  referred to as  the
  `Madau plot', compiled  by \cite{Elbaz05}  and uniformally corrected
  from   cosmological  distances, incompletness  and  dust
  absorption.    Original data    points  came   from  \cite{Hughes98,
  Steidel99,    Flores99,   Glazebrook99,      Yan99,    Massarotti01,
  Giavalisco03}.   These data points indicates  an epoch  of peak star
  formation rate at  $z \simeq 2$, followed by  a rapid fall off (by a
  factor around ten) between $z=1$ and $z=0$.
 
  The  shape  and the  normalization  of   the  Madau plot  are   well
  reproduced by the parameter choice $t_*  \simeq 1.5$ Gyr and $\eta_{\rm w}
  \simeq 1$. It is worth noticing that  this star formation time scale
  corresponds roughly to the value one can  infer from the consumption
  timescale in local galaxies  \citep{ Kennicutt94, Kennicutt98}.  The
  wind efficiency we obtain from this  fitting exercise is close to the
  value  inferred  from $H_\alpha$  observations  of high-SFR galaxies
  \citep{Martin99}.  Galactic winds are required in order to reproduce
  the rapid fall off of the star formation rate at low redshift.

  The  observational  constraints  on  our  two  main  star  formation
  parameters   are   summarized   in   the   upper   right   plot   of
  Figure~\ref{sfrobs},  which  represents  contours of  constant  star
  formation rates  at $z=0$ and $z=3$  in the $\eta_{\rm w}$  - $t_*$ plane.
  The observed central value is shown as the bold lines. The two other
  contours correspond  to a SFR value  a factor of 2  higher and lower
  than the central value. As we  see, the main constraint given by the
  observed SFR is that $t_*$ must be lower than 5 Gyr. If not, the SFR
  will be underestimate at high redshift by a factor more than 2.  The
  contours also indicate that winds  are required to reproduce the low
  SFR observed  at low redshift.   More qualitatively, winds  are also
  required to  prevent from `the overcooling problem'  and to remove
  baryons from the cold gas phase.

  \begin{figure*}
    \begin{tabular}{cc}
      \includegraphics[width=0.48\hsize]{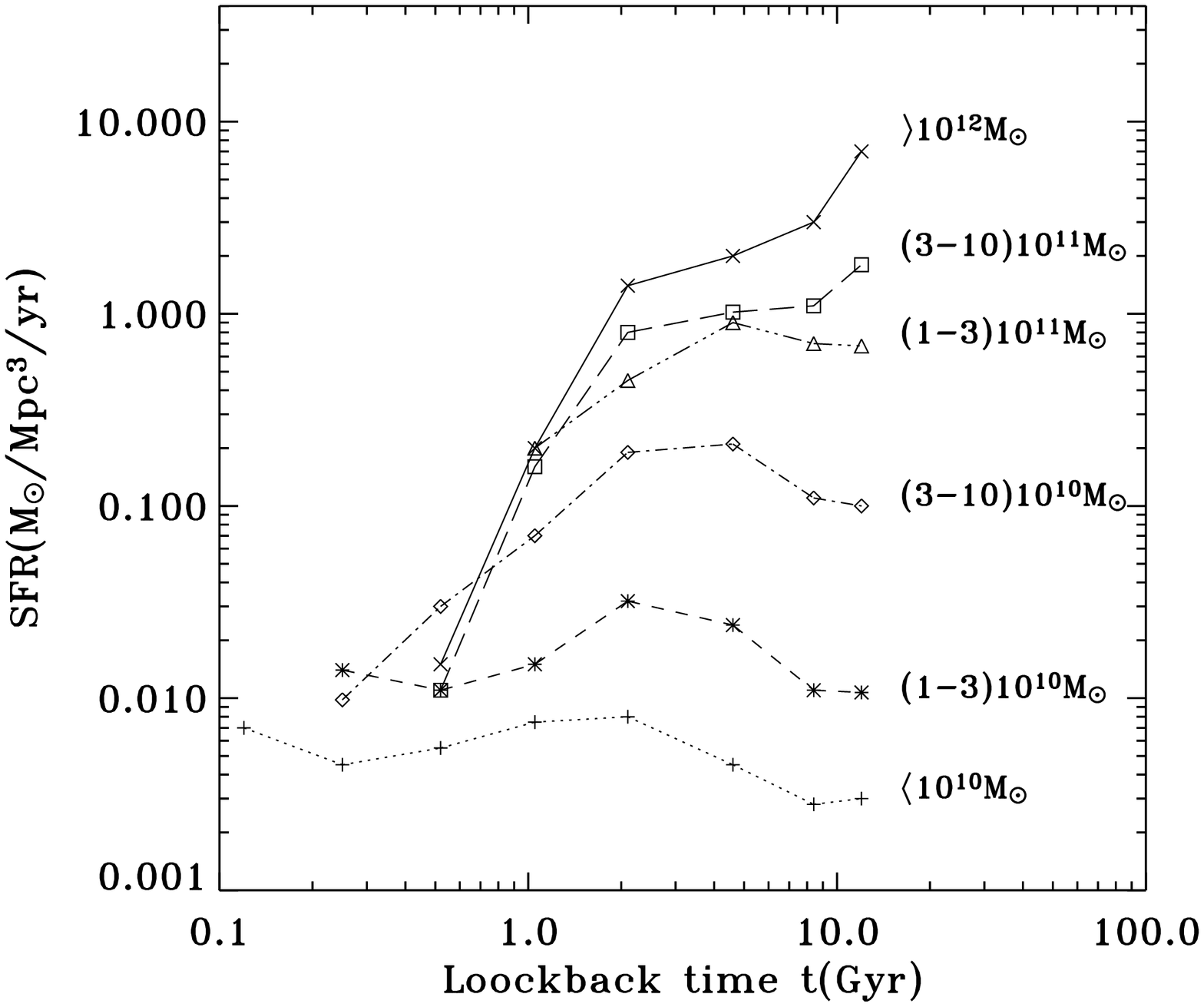}&
      \includegraphics[width=0.48\hsize]{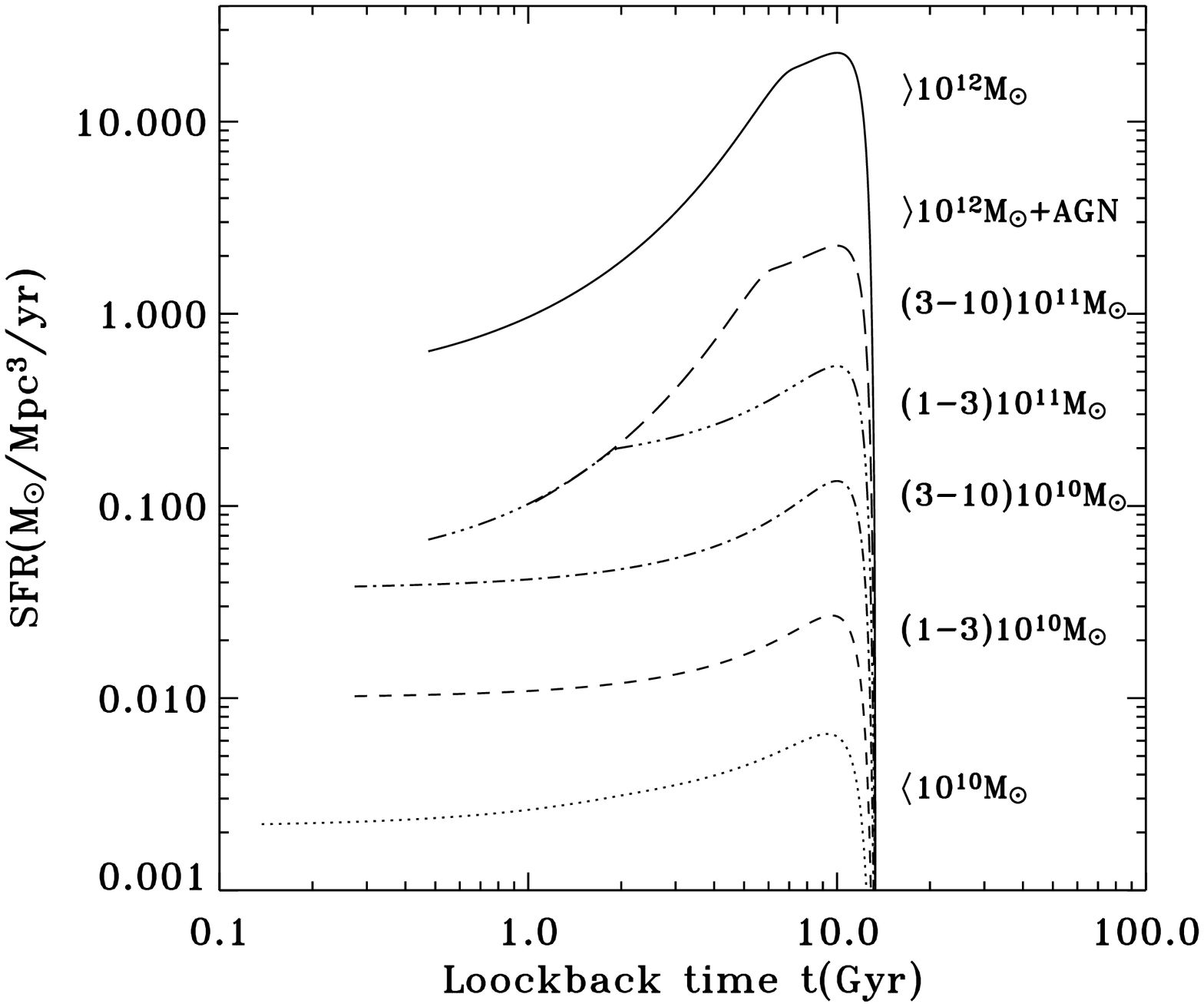}\\
      \includegraphics[width=0.48\hsize]{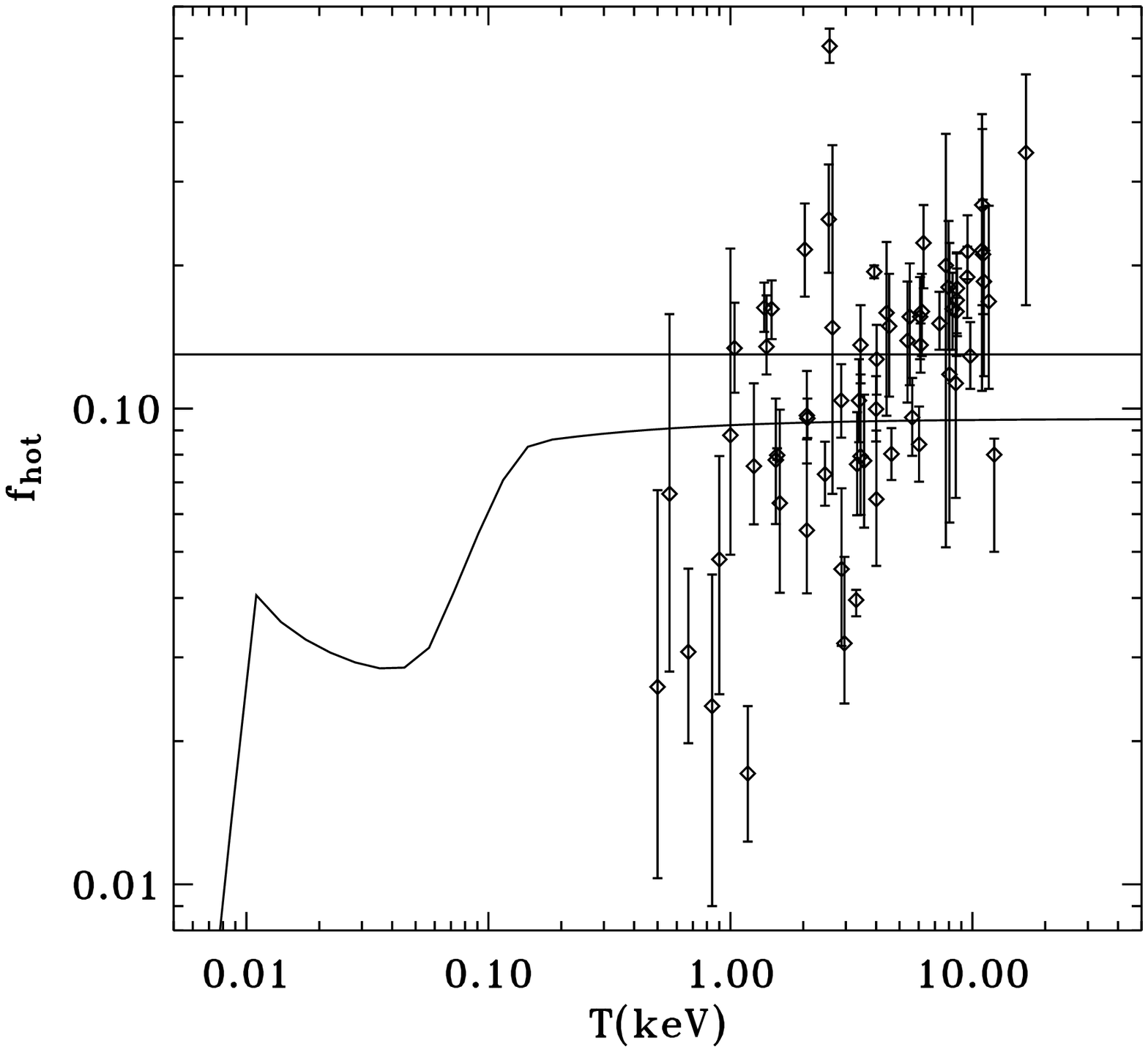}&
      \includegraphics[width=0.48\hsize]{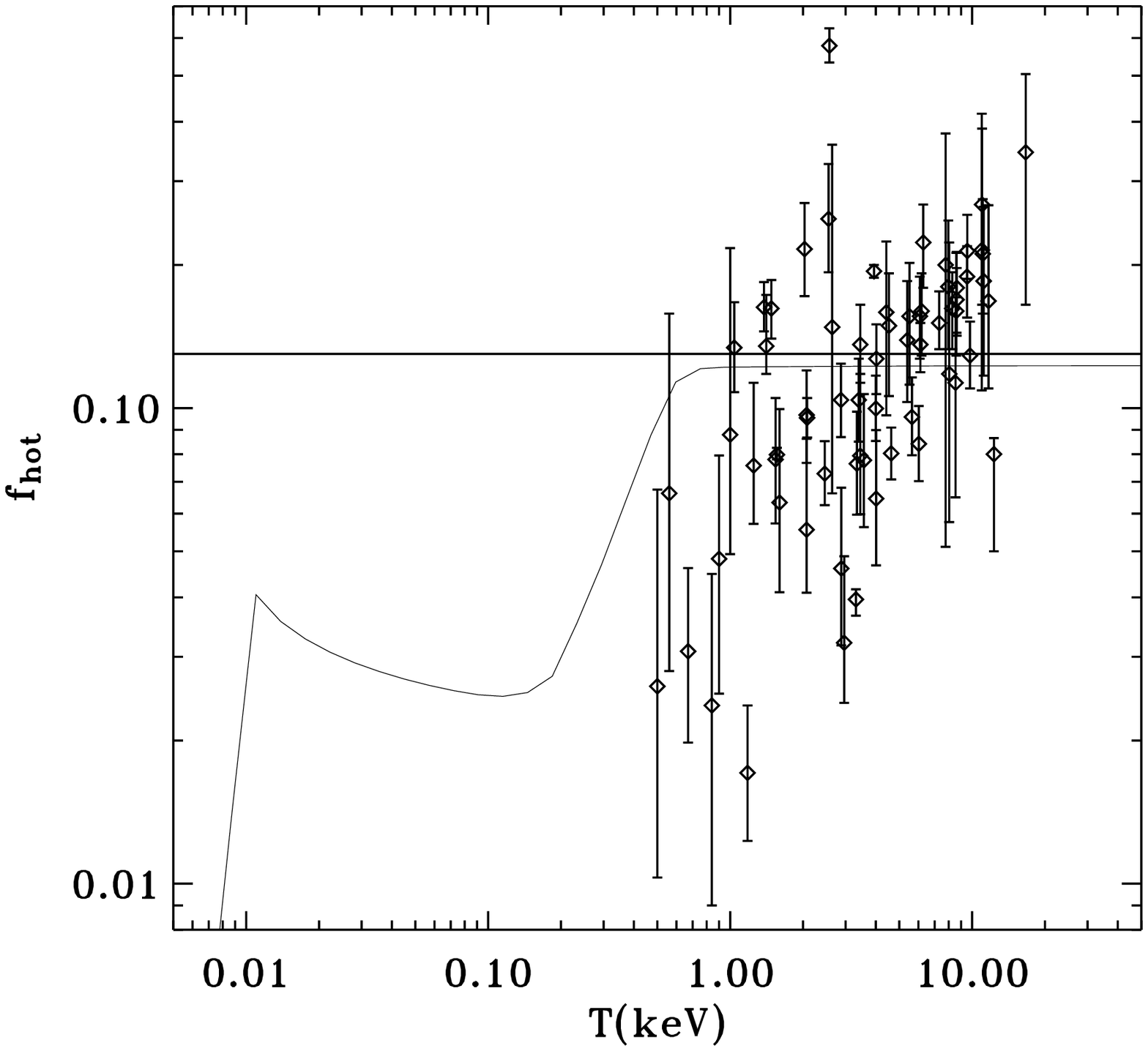}\\
    \end{tabular}
    \cpt{\emph{Top}:  average star  formation  history as  a function  of
      lookback time for different  final stellar masses.  The observed
      values from \cite{Heavens04} are  plotted on the top left panel.
      As in the article, the curves are offset vertically successively
      by 0.5  in log  except for the  most massive galaxies  which are
      offset  by an  additional 1.0.   The top  right panel  shows the
      average star  formation history of our  canonical model ($t_*=3\
      Gyr$  and $\eta_{\rm w}=1.5$).   For  the most  massive galaxies,  a
      second model with metal-rich  cooling and `superwinds' is also
      plotted,  in  better  agreement  with the  data.   \emph{Bottom}:  mass
      fraction  in the  hot  X-ray emitting  gas  in several  observed
      groups and  clusters \citep{Sanderson03a}. The  lower left solid
      line is the  prediction of our canonical model,  while the lower
      right solid line is the prediction of our `superwind' model.}
    \label{sfrdezdeM0obs}
  \end{figure*}
  
  \subsection{Cosmic Stellar Density}
  
  A  complementary method  to investigate  the history  of  the baryon
  assembly in the  universe is to observe the  evolution of the global
  stellar mass  density $\Omega_*(z)$.  This is an  independent way of
  constraining the  model, since  the SFR is  related to  young stars,
  while cosmic  stellar observations  focus on old,  red and  low mass
  stars.   The  middle  left  plot of  Figure~\ref{sfrobs}  shows  the
  observational    data    points    of   $\Omega_*$    compiled    by
  \cite{Dickinson03}   from    various   near-infrared   and   optical
  observations \citep{Cole01,Brinchmann00,Cohen02,Dickinson03}.

  In order  to perform  an accurate comparison,  we need to  take into
  account the  death of  short-lived stars during  the history  of the
  universe.  Since most of the stars are formed when the universe is a
  few Gyr young,  this effect is likely to be  important.  We use here
  the stellar Initial Mass  Function (IMF) proposed by \cite{Kroupa93}
  in order  to estimate  the amount  of stars still  alive at  a given
  redshift. The corresponding predicted  stellar density is plotted in
  the middle left plot of  Figure~\ref{sfrobs}. Our best fit model for
  this observations is now  $t_*=10$ Gyr and $\eta_{\rm w}=0.5$, in complete
  disagreement with our best-fit model for the Cosmic SFR. This rather
  surprising result is due  entirely to the high redshift observations
  of $\Omega_*$.   Low redshift observations,  on the other  hand, are
  compatible with the SFR constraints.

  This puzzle was identified  as a possible `missing galaxy problem'
  \citep{Nagamine04}.   This  indicates   that,  if  the  IMF  remains
  universal, the star formation rate  should fall off at high redshift
  more strongly than in  Figure~{\ref{sfrobs}}. In the model, this can
  be  obtained  for  $t_*=10$  Gyr,  but  then,  as  can  be  seen  in
  Figure~\ref{sfrdezparams}, the SFR is  too high at low redshift. The
  possible  inconsistency between  the observed  SFR and  the observed
  stellar density is discussed in great details in \cite{Dickinson03}.
  The  mass-luminosity  relation might  be  poorly  estimated at  high
  redshift  because  the  galaxy   luminosity  function  is  not  well
  constrained.  Another  solution is to  invoke an evolving  IMF, with
  more high mass  stars at high redshift.  Finally,  the last solution
  is  to invoke a  new physical  process, not  included in  the model,
  which inhibits star formation at high redshift.

  \subsection{Extragalactic Background Light}
  
  Another    strong  observational constraint   is  the  Extragalactic
  Background Light (EBL), integrated from the UV to the IR. The EBL is
  an estimate of the  total amount of energy emitted  by stars and AGN
  in the  history   of  the universe.  Consequently,    the cumulative
  luminosity of all stars created by the model cannot exceed the value
  of this  background.    \cite{Madau00}  compute the  value  of   the
  observed   EBL, integrating  from  0.2 to   2000 $\mu m$,  and found
  $I_{\rm  EBL} = 55$ nW/m$^2$/sr.  Following  the method presented in
  \cite{Madau00}, we compute  the  EBL corresponding to our  canonical
  model       \begin{equation} I_{\rm EBL}=\frac{c}{4\pi}    \int_{\rm
  0}^{+\infty}              \frac{\rho_{\rm              bol}(t)}{1+z}
  \left|\frac{dt}{dz}\right|  {\rm d}z,     \end{equation} with    the
  bolometric  emissivity   at    epoch  t \begin{equation}   \rho_{\rm
  bol}(t)=\int_{\rm      0}^t    L(\tau)  \dot{\rho}_*(t-\tau)    {\rm
  d}\tau.  \end{equation}  In   this   expression,   $L(\tau)$ is  the
  bolometric luminosity of a single stellar cluster of unit mass, as a
  function  of the  cluster  age  $\tau$.  We   use for $L(\tau)$  the
  analytical  approximation  given by    \cite{Madau00} for   a  solar
  metallicity  stellar population.   We  find for our canonical  model
  $I_{\rm EBL}  \simeq   100$ nW/m$^2$/sr, larger than    the expected
  value, but  within observational uncertainties (for  example optical
  data from  \cite{Bernstein99} and  IR  data from  \cite{Chary01} and
  \cite{Fixsen98} lead to $I_{\rm  EBL}  \simeq 80$ nW/m$^2$/sr).  Our
  best    fit model for  the  Cosmic  SFR ($t_*   \simeq  1.5$ Gyr and
  $\eta_{\rm w} \simeq 1$) gives $I_{\rm EBL} \simeq 150$ nW/m$^2$/sr,
  and  can  be   therefore considered   as  being  ruled   out  by the
  observational constraint.

  \subsection{Cosmic Baryon Budget}
  
  The   observed   Baryon Budget is   discussed   in  great details in
  \cite{Fukugita98}.  They  infer  from various observations  at $z=0$
  baryon mass fractions of  $\Omega_{\rm back} \simeq 0.002$, $\Omega_*  =
  0.0035$,  $\Omega_{\rm cold} = 0.00063$  and $\Omega_{\rm hot} = 0.017$.  We
  note   immediately  that, if  one   considers  the most  recent WMAP
  estimate \citep{Spergel03}  of $\Omega_{\rm b}  \simeq 0.04$, roughly 50\%
  of the baryons are missing.  The  most striking disagreement between
  these estimates and our model is the very low baryon fraction in the
  background (or Lyman alpha forest) at $z=0$.  As it is now admitted,
  Lyman alpha observations at low  redshift are very uncertain, so  we
  consider here  that most of the missing  baryons are  in fact in the
  diffuse  background. Recent observations  from \cite{Penton04} seems
  to confirm that at least  30\% of the baryons  lie in the Lyman alpha
  forest.

  Another  consequence of the  analysis performed by \cite{Fukugita98}
  is that baryons in the condensed  phase (cold gas  + stars) are only
  10\% of the total amount of baryons in the universe ($\Omega_{\rm b}
  \simeq   0.04$).  This strongly  supports the   presence of galactic
  winds, in order to  overcome the `overcooling  problem'.  Additional
  support is  given by the fact  that a large fraction ($\simeq$ 40\%)
  of baryons are today in the hot gas  phase that is  to say plasma in
  groups and clusters.    Our  canonical model (with   winds) predicts
  fractions (relative to $f_{\rm  b}$) of $f_{\rm back}  \simeq$ 50\%,
  $f_{\rm hot}   \simeq$  30\%, $f_{\rm  cold}  \simeq$ 1\%   and $f_*
  \simeq$ 20\% (10\% with   the  IMF  of \cite{Kroupa93}), in    rough
  agreement with present   day observations.  Without  winds, the same
  model is much more  difficult to reconcile with  observations, since
  we obtain  in this  case $f_{\rm back}  \simeq$  40\%,  $f_{\rm hot}
  \simeq$ 20\%, $f_{\rm cold} \simeq$ 4\% and $f_* \simeq$ 40\%.
  
  Another  strong observational constraint  come from the evolution of
  the cold gas density, deduced from  the observations of Damped Lyman
  Alpha   Systems  (DLAS)  in  distant    quasars spectra.   Following
  \cite{Somerville01},  we     use  observations    performed       by
  \cite{StorrieLombardi96}  and   \cite{Zwaan97}  to   estimate    the
  evolution of $\Omega_{\rm cold}$  as a function  of redshift.  These
  observations of  DLAS  are  lower  limit estimates  of  $\Omega_{\rm
  cold}$ (see \cite{Somerville01}  for a discussion).  Dust absorption
  is likely to have a  strong effect in biasing  these results. We use
  the method proposed by \cite{Pei99} to correct from these effects.

  Observational  data points  are  shown in   the  lower left  plot of
  Figure~\ref{sfrobs}.  The  curve reaches its peak value $\Omega_{\rm
  cold} \simeq  0.004$ at a redshift  $z \simeq  2$, in good agreement
  with our canonical model, which, in this case,  is also the best fit
  model.  It is worth noticing that  the  observed $\Omega_{\rm cold}$
  curve is proportional   to  the observed SFR   curve.   This can  be
  considered as a nice consistency check in the observational data and
  provides support to our simple star formation model.

  The lower  right   plot of   Figure~\ref{sfrobs} shows contours   of
  constant $\Omega_{\rm  cold}$   at  $z=0$  and $z=3$  in  our  model
  parameter  space.   Contours  appear   as   straight  lines in   the
  $t_*$-$\eta_{\rm  w}$ plane.  This is  consistent with the fact that
  the cold gas fraction depends mainly on the gas depletion time scale
  which  can be  estimated   to be  $t_*    / (\eta_{\rm  w}+1)$  (see
  Section~\ref{chain}).  The low observed value of $\Omega_{\rm cold}$
  favors a small depletion time scale with $1 < t_* / (\eta_{\rm w}+1)
  < 3$ Gyr, a rather tight constraint.

  \subsection{Halo Star Formation History}

  We  now  analyze recent  observations  of  individual galaxies  star
  formation history. Using Extended  Press \& Schechter theory, we can
  apply  our model  to predict  the baryon  history  within individual
  halos, in  an average  sense.  Using the  Sloan Digital  Sky Survey
  (SDSS),  \cite{Heavens04} infer  from $10^5$  nearby  galaxy optical
  spectra the age distribution  of their stellar population.  For each
  galaxy, they  deduce its individual star formation  history (using a
  Salpeter initial mass function).   Finally, they compute the average
  SFR  of all  galaxies in  a given  {\it stellar}  mass  range.  This
  quantity can be directly compared to the prediction of the model, if
  one converts the  halo Virial mass into a halo  star mass, using our
  model `star-to-mass ratio'.

  Interestingly, the observed  average star formation histories (upper
  left plot of Figure~\ref{sfrdezdeM0obs}) seem to indicate that large
  galaxies ($M_*>10^{12} M_{\odot}$)   form stars earlier   than small
  ones  ($M_*<10^{10}\    M_{\odot}$).   This behavior     was  called
  `anti-hierarchical'  by  \cite{Heavens04} and  was identified   as a
  potential  problem for    the  hierarchical scenario of    structure
  formation. The upper  right plot of Figure~\ref{sfrdezdeM0obs} shows
  our canonical model  predictions ($t_* = 3$  Gyr and $\eta_{\rm w} =
  1.5$). The exact quantity we plot is \begin{eqnarray} \dot{\rho}_* =
  \dot{f}_*(M_{\rm  0},z_{\rm 0})M_{\rm  0}   n(M_{\rm 0})  \Delta  M,
  \end{eqnarray}  where $n(M_{\rm 0})$ is  the Press \& Schechter halo
  mass  function.  We  see that the  same `anti-hierarchical' behavior
  (downsizing  effect)  is  reproduced   by   our model, within    the
  hierarchical collapse framework . There are two explanations. First,
  low mass halos have   a total mass  $M_{\rm 0}$  very close to   the
  Minimal  Mass $M_{\rm min}$,  so   that the mass  fraction in  `star
  forming  progenitors' is smaller than for  high mass galaxies.  This
  can  be seen in  Figure~\ref{figwacc},  where  the diffuse gas  mass
  accretion   rate vanishes as  $M_{\rm  0}  \rightarrow M_{\rm min}$.
  Second, for large mass  halos, the cooling  efficiency drops at  low
  redshifts,  as more and  more  progenitors reach $T_{\rm max}$,  the
  maximum cooling  temperature. At this  point, no  more  fresh gas is
  accreted onto the cold disc and star formation is slowing down.

  The  observed   individual star  formation  histories are  therefore
  successfully  reproduced by the  model {\it on a qualitative level.}
  If one looks  carefully  into Figure~\ref{sfrdezdeM0obs}, one   sees
  that for  large    galaxies, the predicted star   formation  history
  disagree with   the   observed one. The  sudden    drop in the  star
  formation rate  (interpreted as  the  end of gas  cooling  into cold
  discs) happens earlier in the model (look-back time  between 7 and 8
  Gyr) than in the data (look-back time around 2 Gyr). Moreover, after
  gas cooling ends,  the decrease in the  star formation  rate is much
  faster in the data than in the model.

  The first feature can  be accounted  for if  one takes into  account
  metal enrichment   into the cooling  gas.  For  a metallicity of one
  tenth or one third solar,  cooling can be significantly higher  than
  for a primordial H and He plasma.  We have tried a new model with an
  increased Maximum  Cooling temperature $T_{\rm  max}=2 \times  10^6$
  K. In this case, the position of the knee in the star formation rate
  curve agrees perfectly with  observational data.  On the other hand,
  the  strong decrease  after the  knee is  difficult to explain. Note
  however that the measure  of the SFH  from the optical spectrum is a
  very  complex operation  and  the uncertainties  are  important. For
  example, a large part of the star formation is enshrouded and may be
  missed.   Moreover,  the IMF  is     probably not a  Salpeter   one.
  Consequently, the     decrease might    be  less accentuated    than
  presented. Nevertheless, if the  decrease is confirmed, one solution
  is to invoke  very strong winds  that remove most   of the cold  gas
  accumulated before the  end of disc  accretion.  To illustrate this,
  we further modify the canonical model, introducing what we call here
  `superwinds',    with parameters  $T_{\rm   w}  \simeq  10^7$ K  and
  $\eta_{\rm w} = 15$.  This last model (see Fig.~\ref{sfrdezdeM0obs})
  is finally able to reproduce the  observed star formation history in
  large galaxies.

  This  `superwinds' scenario is completely ruled  out by the observed
  global baryon  history (see previous  section). We have therefore to
  consider   2  coexisting  wind  models   of  very  different nature:
  `galactic winds', driven by supernovae bubbles, for normal and dwarf
  galaxies,  and `superwinds', possibly  driven  by a massive  central
  black hole or AGN  for massive galaxies.  Building a self-consistent
  model  along  those lines is beyond   the scope  of  this paper, but
  several attempts of `superwinds' models  can already be found in the
  literature \citep{Springel05}.

  \subsection{Hot Gas Fraction in X-ray Clusters}

  The observed amount of hot gas  in groups and clusters ($\Omega_{\rm
  hot} = 0.017$  in \cite{Fukugita98}) put  a strong constraint on the
  hierarchical scenario, mostly by requiring  galactic winds to  solve
  the    `overcooling   problem'.   \cite{Sanderson03a}  have analyzed
  various X-ray  observations of groups  and clusters and computed the
  fraction of  hot gas ($f_{\rm hot}$) as  a function of  the observed
  X-ray, emission   weighted, temperature.     These data points   are
  plotted        with    their     corresponding      error    bars in
  Figure~\ref{sfrdezdeM0obs}.

  First, we    learn from these   observations that   most  baryons in
  clusters have to be in the hot phase, unless  we are ready to face a
  serious  crisis with the  WMAP  constraint $\Omega_{\rm b} =  0.04$.
  Second, there is a clear correlation between the fraction of hot gas
  in each halo  and the X-ray  temperature. Moreover, this correlation
  $f_{\rm hot}  \propto   T_{\rm X}$  is  often  used  as  a  possible
  explanation  for the  observed $L_{\rm  X}  - T_{\rm X}$ relation in
  clusters \citep{Neumann01}.

  We  have plotted on the same  figure our  canonical model prediction
  for the hot gas fraction as a function of the Virial temperature. In
  small galaxies,   where both  galactic winds   and  gas  cooling are
  important, the hot gas fraction have its minimum around $f_{\rm hot}
  \simeq 3\%$. For  large halos, cooling  is less and  less efficient,
  and  in the   same time,  winds can   not escape  the halo potential
  well. This double mechanism  (cooling   + winds) has the   important
  consequence of  refilling  with   hot gas  the  parent   halo.  This
  qualitative picture is interesting,    but  when one  compares   our
  canonical model predictions with the X-ray data, the result is quite
  disappointing.   The  refilling  mechanism  we  have  just explained
  occurs  at too low  temperature  ($T \simeq  0.1$  keV), while  data
  suggests a significantly higher transition temperature. Note however
  that the observations  concern the  center of  the  clusters and the
  extrapolation to larger   radii (using $\beta$  model)   is not safe
  \citep{Neuman05}.

  One solution might  come again from  our `superwinds'  scenario.  We
  modify our canonical model  by first increasing the  Maximum Cooling
  temperature up to $T_{\rm max}=2 \times 10^6$ K, as it should be for
  the case of a  realistic,  metal-rich, plasma.   We then modify  our
  wind  parameters to $T_{\rm  w} \simeq  10^7$ K and  $\eta_{\rm w} =
  15$, as the  `superwinds' model we use in  the last section. We plot
  the    hot gas  fraction     we obtain   for  this   new   model  in
  Figure~\ref{sfrdezdeM0obs}.  The transition  from `gas poor' to `gas
  rich' regime  occurs now  at a much  more realistic  temperature ($T
  \simeq  1$ keV).  This transition  is however still  much sharper in
  our model than it is in the data.  As suggested in the last section,
  we could therefore improve our  wind model by explicitly introduce 2
  different feedback scenarios: `supernovae driven' and `AGN driven'.

  \cite{Kay04}  have   performed hydrodynamical simulations  of galaxy
  clusters with a   kind of  feedback  which  is  very   close to  our
  superwind  scenario.  Indeed, they  heat the  dense  and cold gas of
  their  clusters at a temperature of   $17$ keV.  This corresponds in
  our notations to   $T_{\rm w} \simeq  2\times 10^7$   K (compared to
  $T_{\rm w}  \simeq 10^7$   K in  our superwind model).    Using this
  strong feedback,   they  reproduce   the   observed cluster  $L_{\rm
  X}-T_{\rm X}$ relation with the correct level of  entropy in the ICM
  core. Such a strong feedback seems  therefore essential to reproduce
  both global fraction of hot gas and entropy profile.

\subsection{Stellar and HI mass functions}

\begin{figure}                                      
\centering
\includegraphics[width=\hsize]{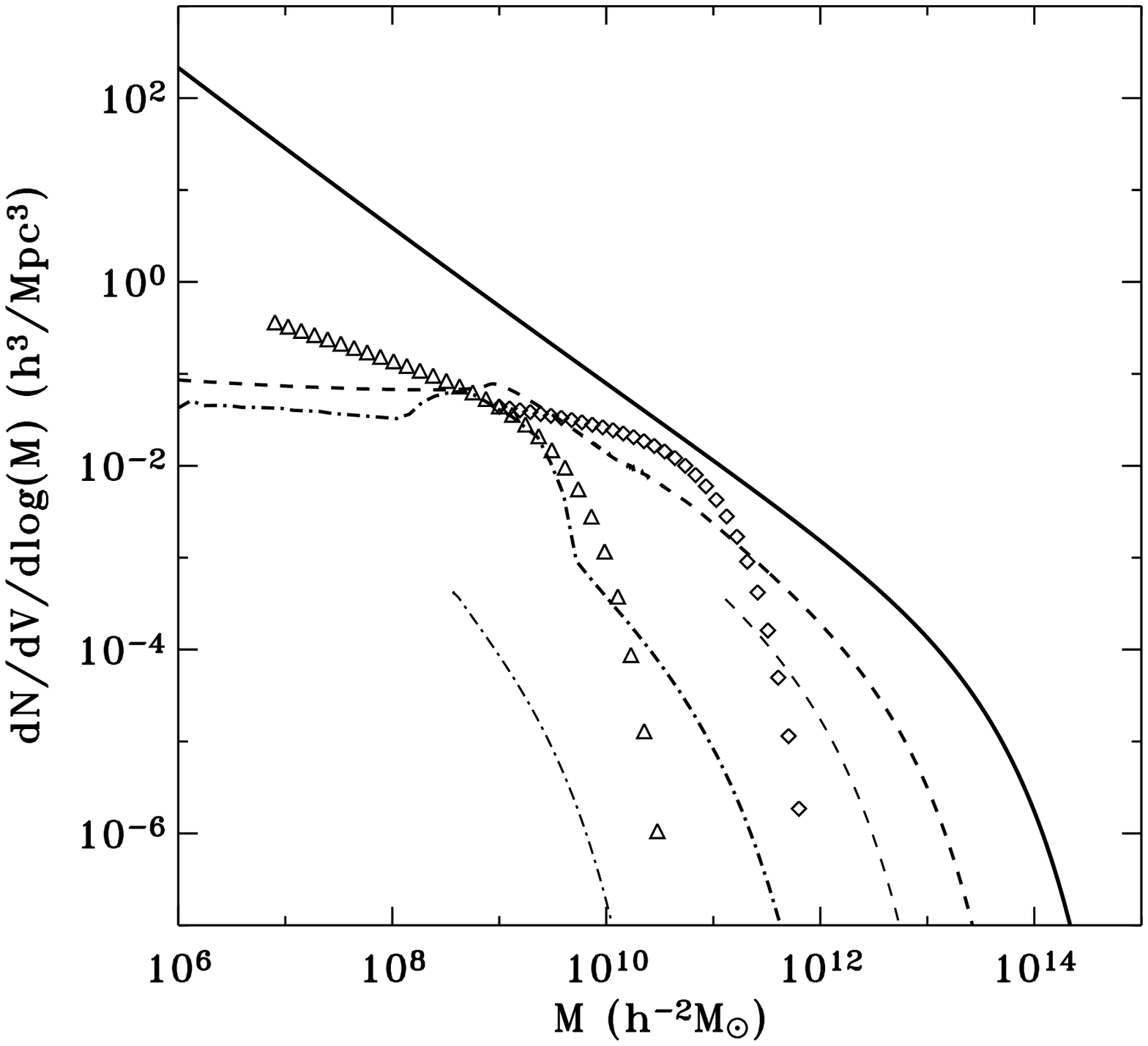}
\cpt{Predicted and  observed  stellar and  HI  mass function.  The
    	thick dashed line is  the stellar  mass function predicted  by
    	our standard model, whereas  the thin dashed  line is  for our
    	superwind model. Shown as diamonds is the Schechter fit to the
    	observed  stellar mass function \citep{Cole01}. Similarly, the
    	thick dot-dashed line is the  cold gas mass function predicted
    	by our standard model whereas  the thin dot-dashed line is for
    	our superwind   model.   The Schechter  fit   to the   HI mass
    	function   from \cite{Zwaan05}   is  shown  as triangles.  For
    	comparison, the thick  continuous line is the  Press-Schechter
    	mass    function,  assuming    a universal   baryon   fraction
    	everywhere.}
\label{ndembar} 
\end{figure}

In this  final part,  we  investigate the stellar  and cold   gas mass
function at z=0.   We reach here the limit   of our model  because, as
mentionned  before,  our analytical  predictions are  for  baryon mass
fractions  as a function of halo  masses, whereas observed baryon mass
fractions are   usually   given as  a  function  of  individual galaxy
masses. The conversion between galaxy and halo masses can be performed
using  the  'halo occupation number' \citep{Kravtsov04b},  which gives
the  average number  of galaxy satellites  within  $R_{\rm 200}$  as a
function of the parent halo mass.  We  intend to apply this correction
to our model predictions in a  future paper.  Nevertheless, we compare
our predicted stellar and cold gas halo  mass function to the observed
stellar \citep{Cole01} and HI galaxy mass function \citep{Zwaan05}.

As presented in figure~\ref{ndembar}  the global normalisation of both
stellar  and cold  gas mass  function  is in  good  agreement with the
observations. Indeed the integral of  this mass function multiplied by
the mass is nothing else that the z=0 cosmic baryon  budget for this 2
phases.  As for the observations,  the profile shape shows a  fall-off
for high mass  and  a flatening  for  lower  masses.  This  is  easily
explained by our predicted stellar and cold gas mass fraction. For the
high  mass   tail, the  fall   off  is due   to  the fast   decline of
Press-Schechter mass function, where as  for low mass the flatening is
due to the minimal halo mass $M_{\rm min}$ above which the fraction of
baryons become weaker and weaker.

If   we look in more   details, we see   an  encouraging agreement for
intermediate mass.    The decline of   the high  mass tail  is however
steeper  in the observations, both  for the HI mass function $n(M_{\rm
HI})$ and the stellar one $n(M_*)$.  We  have identified 2 reasons for
that.  First, as highlighted before, the halo occupation number should
be taken into account, because it would  decompose each large halos in
a  collection of  lower mass  halos.  As  a consequence, the high-mass
decline may be more steep.  Second, superwinds may also lower the cold
and  stellar mass fraction in high  mass halos \citep{Springel05}.  To
illustrate this  point, we have plotted  the effects of  our superwind
scenario on the mass function.  Note that  a more complex model with 2
different winds would result in  more realistic predictions.  For  low
mass  galaxies,  the model tends   to underestimate the  cold gas mass
function.  Here again the halo occupation number would explain part of
this discrepancy by  increasing the number  of low mass  halos. In our
model the  transition between the  star forming  halos and the diffuse
background   occurs abruptly  at   $M_{\rm  min}$.    As seen in   our
simulations, this transition is in fact  much smoother: this is likely
to improve  the agreement between the  predicted and the observed mass
function.

  \section{Conclusion}

  We have studied the baryon budget  evolution in the framework of the
  hierarchical  scenario of  structure formation,  using  4  different
  phases: diffuse background (Ly$\alpha$  Forest), hot gas  (plasma in
  the halo of galaxies, groups and cluster), cold discs and stars.  We
  have paid particular attention to the star formation  rate, as it is
  a key observational constraint for our current cosmological model.
  
  We have analyzed the baryon history for the universe as a whole, but
  also on a halo-by-halo basis.  For that purpose, we have developed a
  fully self-consistent (though simple)  analytical model.  These last
  two point are the most original aspects of  our work. Our analytical
  model has proven to be an efficient tool to quickly compute accurate
  predictions  for the  baryon    budget  history.  It  is   currently
  available  as  a set  of IDL  routines, and  can be  provided by the
  authors upon request.

  In  order   to validate this   model,  we  have  performed numerical
  simulations  of galaxy formation  using the  AMR  code  RAMSES.  Our
  highest resolution run reach $512^3$  dark matter particles and half
  a  billion  AMR cells, which is  among  the largest galaxy formation
  simulations  performed so far. We  have also analyzed the simulation
  results of \cite{Springel03b} based on the SPH code GADGET. We found
  in  all cases a  good agreement between   simulations and our model.
  This  cross-validation has allowed us   to use our  model to analyze
  observational data.
    
  We  have explored  our physical parameter   space $t_*-\eta_{\rm w}$
  (star  formation time scale   and wind efficiency) and  compared the
  model  results  to the  cosmic observations   of the  comoving  star
  formation rate, the  evolution of the comoving  density of stars and
  cold gas,   and  the intensity   of   the  integrated  extragalactic
  background.   The  conclusion is  that   the parameters  $t_*=M_{\rm
  cold}/\dot{M}_*=3\   {\rm Gyr}$  and  $\eta_{\rm     w}=\dot{M}_{\rm
  wind}/\dot{M}_*=1.5$ are favored.  It means that winds with ejection
  rates around 1 or 2  times the star  formation rate are required  to
  prevent the overcooling problem.
  
  Comparisons  with  individual  halo  properties,  such  as   the age
  distribution of  stars  in  galaxies and the   hot  gas fraction  in
  clusters  seems  to indicate that high   velocity and high intensity
  outflows  (`superwinds') are  required  in  massive galaxies.  The
  origin of these  violent  outflows  could  come from a   central AGN
  \citep{Springel05}.  The modeling of such winds and their exact role
  in    the metal  enrichment   of    the  IGM  are currently    under
  investigation.

  \begin{acknowledgements}
    
    The simulations presented  here   were performed at    `Centre de
    Calcul  pour  la Recherche et  la Technologie'  (CCRT).  We thank
    Alexandre  Refregier for providing  us IDL routines to compute the
    cosmological evolution of the halo model (ICOSMO package \cite{Refregier05}). We thank Volker Springel
    for providing us his simulations results. We thank David Elbaz for
    his help on analyzing observational data points of the Cosmic Star
    Formation Rate.  We thank Alastair  Sanderson for providing us hot
    gas  fraction in  clusters.We   also  thank Pierre-Alain Duc   for
    stimulating discussions.

  \end{acknowledgements}
  
  \newpage

\appendix

  \section{Numerical methods}
\label{method_num}

  \subsection{Hybrid scheme for high-Mach flow}
    
  Our hydrodynamical solver is   built using a  multidimensional
  unsplit Godunov scheme, with second order accuracy both in space and
  in time.  In order to capture correctly shock waves, the total fluid
  energy  is  used as  independent  variable.    As soon  as  velocity
  gradients (velocity undivided  differences in each direction) in the
  flow have a comparable magnitude to the fluid sound speed, the
  fluid temperature can be computed accurately.  In the opposite case,
  which happens   routinely in cosmological  applications,  the  fluid
  temperature is dominated by truncation errors in the velocity field.
  Since the velocity field is  mainly determined by the  gravitational
  acceleration, this arises  when  the mesh resolution $\Delta   x$ is
  greater than the Jeans length.  In order  to deal with this problem,
  particularly acute     in  our case,   since  we  need   an accurate
  temperature determination for  the cooling and heating functions, we
  are left with two possibilities

  \begin{itemize}
    \item we refine our mesh sufficiently in order to resolve the Jeans
      length. This solution is actually adopted in most star formation 
      numerical studies \citep{Truelove97,Boss00}. Except for our smallest boxes,
      this strategy is not applicable here.
    \item we switch off the total energy conservative update in regions
      where truncation errors are too large, and use instead the
      internal fluid energy as independent variable. The problem is
      now to design the correct switching strategy.
  \end{itemize}

  Note  that  this problem, rather    specific to cosmology, has  been
  identified  by  several   authors,  both in   the  Godunov framework
  \citep{Bryan95}   and in the SPH   framework \citep{Shapiro96}.  The
  solution  adopted by these various   authors was to carefully detect
  shock waves formation,  in order to  turn  off spurious  heating (or
  cooling) before  shock heating actually  occurs.  In RAMSES, we have
  adopted the following approach.  At  the end of  the hydro step,  we
  update the  fluid internal energy $e  = \rho \epsilon$ by solving in
  parallel two different sets of equations.  \begin{eqnarray} \partial
  _{\rm t} E + \partial _{\rm  x} u (E+P) =  0 ~~~\mbox{and}~~~ e = E-
  \frac{1}{2} \rho  u^2 \label{econs}, \end{eqnarray} \begin{eqnarray}
  \partial _{\rm t} e + \partial _{\rm x} u e = -P \partial _{\rm x} u
  = - (\gamma -1) e \partial _{\rm x} u.  \label{eprim} \end{eqnarray}
  The  first one is the standard  conservative update, followed by the
  subtraction of the fluid  kinetic energy, while the second one
  is a direct, non  conservative update of the  internal energy.  This
  hybrid approach is a classical  trick in fluid dynamics, to overcome
  certain  problems  associated to   the conservative   formulation in
  equation~\ref{econs}    \citep{Toro97}.  In   order   to solve   for
  equation~\ref{eprim}, we  need to compute  (and  store) the velocity
  divergence $\partial _{\rm  x} u$ for  each fluid element.   We then
  decide which update between $e_{\rm prim}$ and $e_{\rm cons}$ is the
  valid one by comparing the conservative sound  speed $c_s^2 = \gamma
  (  \gamma  -1) e_{\rm cons}$ to   an estimate of  the local velocity
  truncation  errors $\Delta x \partial  _{\rm  x} u$ \begin{eqnarray}
  \label{switch} e _{\rm final}= e_{\rm cons}  & & \mbox{if}~~~ e_{\rm
  cons}  \ge \alpha  (\Delta x \partial   _{\rm x} u)^2,\\ \nonumber e
  _{\rm final}=  e_{\rm  prim} &  & \mbox{otherwise}.   \end{eqnarray}
  $\alpha$ is a free  dimensionless parameter of  order unity. We take
  as  phase value $\alpha=0.5$  in all the simulations presented here.
  Note that  our     approach  differs  from  the  one   proposed   in
  \cite{Bryan95} in several   aspects, the most  important one  to our
  opinion is that criterion~\ref{switch} is Galilean invariant as
  for the Euler equations.

  The consequence of our hybrid energy update is that shock heating is
  turned off  completely in small mass halos,  where truncation errors
  dominate the  fluid internal  energy, due to  a lack  of resolution.
  This introduces an effective low mass cut off in the simulated halos
  distribution, below which the temperature evolution is governed only
  by adiabatic compressional heating  and UV heating: small mass halos
  are  therefore considered  as  sitting in  the  smooth, Lyman  alpha
  background.  Careful  numerical experiments with  RAMSES, under the
  refinement parameters used in this paper, indicate that this minimal
  mass sits around 100-200 dark matter particles.

  \subsection{Refinement strategy}
  \label{sectrefine}

  We  adopt  here  a   refinement  strategy  based  on  the  classical
  `quasi-lagrangian' approach. Our goal is to describe as accurately
  as possible  the collapse of individual Fourier  modes. We therefore
  refine cells when  the mass contained in that  cell exceeds a certain
  threshold. When  only one fluid  is present (dark matter  say), this
  threshold translates  into a  critical particle number,  above which
  new  refinements are  triggered.   This approach  has  been used  by
  several  authors performing  AMR  N body   simulations in  the
  literature \citep{Kravtsov97, Abel00, Teyssier02}.

  In the  present case, we are  dealing with 3  different fluids: dark
  matter,  gas  and  stars.   Moreover,  2 fluids  are  treated  using
  particles (dark matter  and stars), and one is  described using grid
  cells. When cooling is included  in the gas energy equation, baryons
  sink  very  efficiently  in  the  center of  their  host  halos,  and
  ultimately dominates the total mass distribution. We have adopted the
  following approach in all the simulations performed here: we define
  the total number of `particles' as
  \begin{eqnarray}
    N_{\rm tot} =\Delta x^3 (\rho_{\rm dm}/m_{\rm dm}+\rho_{\rm g}/m_{\rm g}+\rho_{*}/m_{*}),
  \end{eqnarray}
  where $m_{\rm dm}$ is the mass of  each dark matter particle, $m_{\rm g} = f_{\rm b}
  m_{\rm dm}$ is the initial gas mass  of each cell in our base grid ($f_{\rm b}
  =  \Omega_{\rm b}  /  \Omega_m$), and  $m_*$  is  the  mass of  each  star
  particle.  We  set $m_*=m_{\rm g}$,  to  prevent  spurious refinements  or
  de-refinement in  star forming  regions. In this  way, the  grid, as
  well as the  source term for the Poisson equation,  are not aware of
  the underlying star formation process.

  AMR cells are refined, at  each level of refinement, if the effective
  number exceed $N_{\rm tot}=40$.  This  threshold is rather high compared
  to  previous N  body AMR  simulations, where  $N_{\rm tot}  \simeq 5-10$
  \citep{Kravtsov97, Abel00, Teyssier02}.  In our current gas and dark
  matter runs, we want to  minimize the appearance of gas cells devoid
  of  dark  matter particles.   This  is  known  to lead  to  spurious
  oscillations in the gas density \citep{Teyssier02}.

  We  set the  maximum level  of refinement  to  $\ell_{\rm max}=5$.  This
  number is  rather low compared  to previous N body  AMR simulations,
  where    $\ell_{\rm max}   \simeq   7-8$    \citep{Kravtsov97,   Abel00,
  Teyssier02}.   Since we  are  dealing with  strong  cooling in  high
  density regions, there is nothing to prevent the gas from collapsing
  into dense cold  knots. The AMR grid will  be automatically refined,
  leading to very small cells.  If our maximum refinement level is too
  high, dark  matter particles sitting in gas  dominated, high density
  regions  will  suffer  from  two body  relaxation  effects.  Setting
  $\ell_{\rm max}=5$ gives us a good compromise between spatial resolution
  and cautiousness.

  As  one can see, our   refinement  strategy is rather cautious,  and
  based on a   very conservative  choice  of  parameters,  compared to
  previous, more aggressive strategies proposed in the AMR literature.
  For our largest runs,  starting with a base  grid of  $512^3$ coarse
  cells, this  allows us to reach a  formal resolution of $16384^3$ in
  the densest regions  of   the flows. This  configuration  was
  computed with 256 processors and 350h wall-clock time on a massively
  parallel computer.

  \section{Filtering Mass}
 \label{Filtering}
  Thoul \& Weinberg (97) and \cite{Gnedin00} were the first authors to
  carefully examine how baryons fall  (or not) into dark matter halos
  potential  wells  in presence  of  a  UV  background.  These  authors
  realized  that  it  is  only  above  a  critical  mass,  called  the
  `Filtering Mass'  $M_{\rm F}$, that baryons and dark  matter assemble in
  comparable amount inside virialized  halos. In halos of mass lower
  than $M_{\rm F}$, the  baryon fraction was found to  be (on average) lower
  than the universal value.  This  means that below this mass, baryons
  can be considered as a quasi-homogeneous, diffuse component.

  \cite{Gnedin00} noticed that the  instantaneous Jeans mass $M_{\rm J}$ was
  not the  correct way  of computing $M_{\rm F}$.   The correct  estimate of
  $M_{\rm F}$ at  a given epoch  is based on  the average of the  Jeans mass
  over the  past history of the  universe.  We therefore  need a model
  for  the thermal  history  of  the background,  in  other words  the
  average temperature $\bar{T}$. 

  At very early epochs, due to residual electrons, the gas temperature
  is tightly coupled to the cosmic black body.   After $z \simeq 200$,
  the gas temperature finally decouples and evolves  as $(1+z)^2$.  At
  a  given redshift (called here  the reionization redshift) $z_{\rm r}$, we
  assume that baryons are promptly  heated to a fixed temperature $T_{\rm r}
  \simeq 10^4$ K, and remains roughly isothermal afterwards. Note that
  a more  realistic computation of the  ionization fraction would lead
  to  a   more   complex    behavior    for the  temperature     after
  reionization. Our model  for the background temperature is therefore
  given   by   the   following  equations    \begin{eqnarray} \bar{T}=
  \left\{\begin{array}{llll} 2.73(1+z)  & & z>200,\\  0.0136(1+z)^2 & &
  200>z>z_{\rm r},\\  T_{\rm r} & &  z<z_{\rm r}.\\ \end{array}\right.  \end{eqnarray} The
  two parameters $z_{\rm r}$ and $T_{\rm r}$ must  be carefully chosen in order to
  recover the proper thermal history of the universe.  Our simulations
  are based on the \cite{Haardt96} model  for the UV background.  This
  rather complex model can be  roughly  reproduced with $T_{\rm r} \simeq  6
  \times  10^3$~K and $z_{\rm r}  \simeq 6$.  For a  different  model with a
  harder UV  spectrum, the  corresponding reionization temperature has
  to be  set to  a  higher  value.  Recent  observations  by  the WMAP
  satellite suggest  a larger value  of the reionization redshift $z_{\rm r}
  \simeq 20$. This latter value will  be used as a reference parameter
  for comparison to observational data.
  
  As soon as  the background temperature is known,  we can compute the
  Jeans mass
  \begin{eqnarray}
    M_{\rm J}=\frac{4 \pi}{3} \bar{\rho}\left(\frac{2\pi a}{k_{\rm J}}\right)^3,
  \end{eqnarray}
  where the Jeans scale is given by
  \begin{eqnarray}
    k_{\rm J}=a \sqrt{4 \pi G \bar{\rho}\frac{3 \mu m_{\rm H}}{5 k_{\rm B} \bar{T}} },
  \end{eqnarray}
  and  $\bar{\rho}$  is the  average  mass  density  in the  universe.
  Finally,  the  Filtering  Mass  is  simply given  by  the  following
  integral over the expansion factor \citep{Gnedin00}
  \begin{eqnarray}
    M_{\rm F}^{\frac{2}{3}}=\frac{3}{a} \int_{\rm 0} ^a da'
    M_{\rm J}^{\frac{2}{3}}(a')(1-\sqrt{\frac{a'}{a}}).
  \end{eqnarray}
  Thus,  dark   matter  halos  with  masses  lower   than  $M_{\rm F}$  (or
  equivalently with  Virial temperature lower than  $T_{\rm F}$) are tracers
  of  the diffuse baryon  background, while  dark matter  halos with
  masses  greater  than $M_{\rm F}$  contain  an  hot  baryon component  in
  hydrostatic  equilibrium,  whose  mass  fraction  is  close  to  the
  universal one $f_{\rm b} = \Omega_{\rm b} / \Omega_{\rm m} \simeq 0.13$.
  
  \section{Cooling model}
  \label{coolingmodel}

  The Filtering Mass is the first important mass  scale. It is however
  not the only  one needed to properly  define star forming halos.  We
  need  to  determine a  second mass  scale   above which  hot gas  in
  hydrostatic equilibrium is  able to cool and  fall in the center  of
  the host dark matter halo.

  If  cooling is ignored,   this   hydrostatic plasma  can indeed   be
  considered as   another  diffuse component.   Numerical  simulations
  \citep{Eke98},   theoretical arguments \citep{Suto98,     Komatsu01,
  Ascasibar03} and observations    of X-ray emission in  large   X-ray
  clusters \citep{Neumann99} all suggest  that the maximum overdensity
  during adiabatic collapse lies around $10^5$,  which is a rather low
  value  compared to typical   overdensities of  galactic  discs.  The
  diffuse  background  is therefore defined as   all dark matter halos
  with  masses lower  than the Filtering   Mass, plus all  dark matter
  halos whose hot plasma is not able to cool and condense as a disc.

  \begin{figure}
    \centering \includegraphics[width=\hsize]{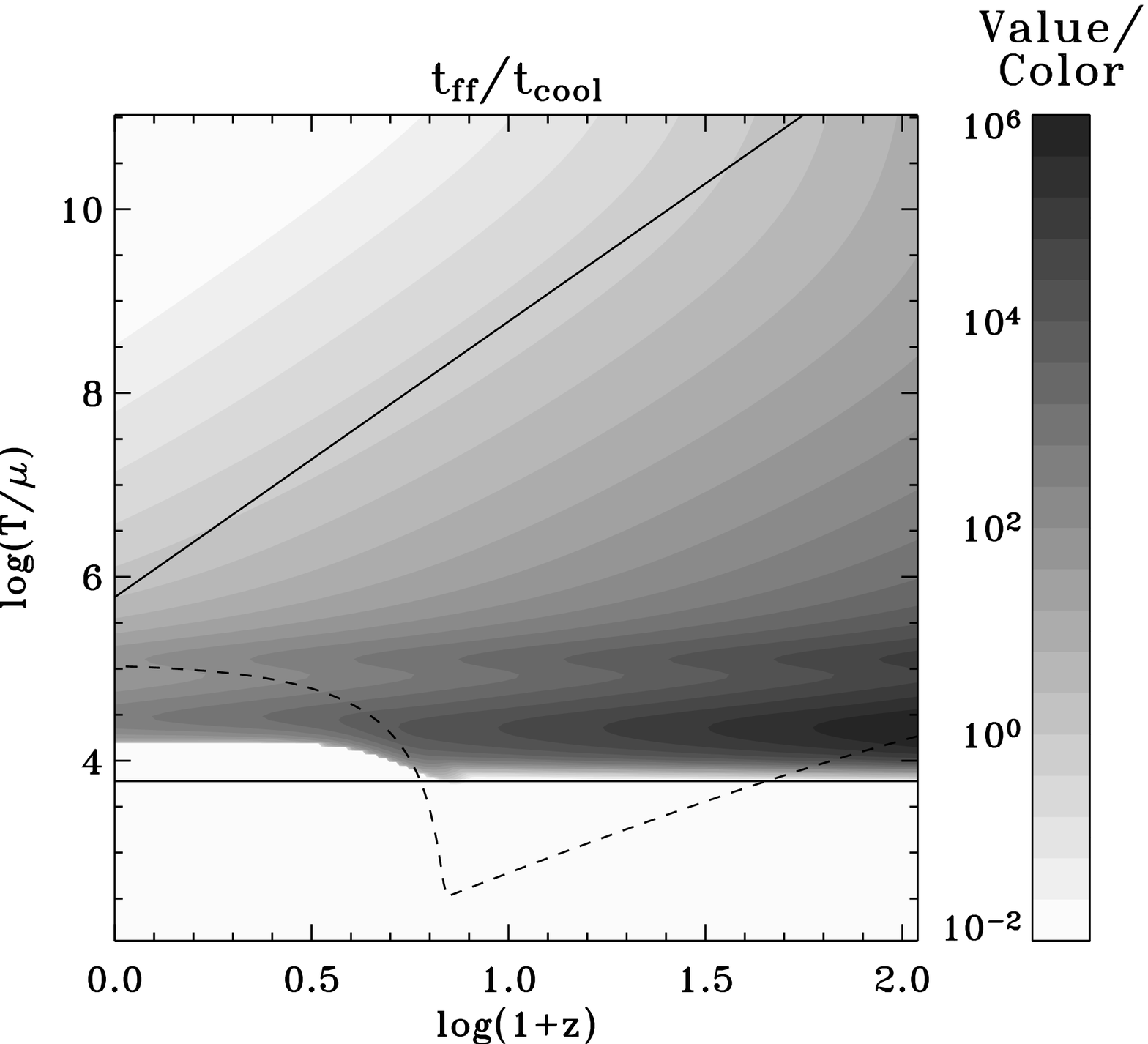}
    \cpt{Ratio of  free-fall time over cooling time  for a typical
      core gas density in the  temperature and redshift plane, for the
      \cite{Haardt96} reionization model  ($z_{\rm r} \simeq 6$).  The upper
      solid line  shows the redshift evolution of  the maximum cooling
      temperature $T_{\rm max}$, while the lower solid line corresponds to
      the minimum cooling temperature $T_{\rm cool}$.  The region enclosed
      by these 2 solid lines has $t_{\rm cool} \le t_{\rm ff}$.  Also shown as
      the dashed line is the evolution of the Filtering Mass, in units
      of Virial temperature.}
    \label{figtcool}
  \end{figure}
  
  In this paper, we only  consider atomic cooling processes. We ignore
  molecular   cooling  processes.  There  are   two reasons  for that.
  First,  we   do not  consider     Pop III  star   formation, as   in
  \cite{Abel02} and \cite{Sokasian04},  since it is currently a matter
  of debate  to  determine if  this first generation   of  stars has a
  negative or a  positive  feedback on molecular  hydrogen  formation.
  Second, in this paper, we perform numerical  simulations that do not
  have enough  resolution to describe  properly molecular  cooling and
  fragmentation  inside HI  discs, as  opposed  to the  recent work of
  \cite{Kravtsov04b}.

  Semi-analytical models  of galaxy formation  have developed various
  methods to  compute the atomic  cooling efficiency as a  function of
  halo  mass and merging  history. The  cooling model  we use  here is
  based on  a major  simplification of such  models, in order  to deal
  with analytical  formulae. We take  the typical core density  in the
  halo  gas distribution  to be  roughly constant  with $\delta_{\rm core}
  \simeq 10^5$.  Using cooling and heating  functions corresponding to
  the \cite{Haardt96} UV background, we compute the  ratio of the net
  cooling  time over  the local  free-fall time  $t_{\rm cool}  / t_{\rm ff}$,
  assuming isothermality at the Virial temperature of the halo.

  We  plot  in  Figure~\ref{figtcool}  this  ratio as  a  function  of
  redshift and  halo temperature.  We  also plot the evolution  of the
  Filtering Mass (in units of  Virial temperature) as the dashed line.
  Darker contours are  for halos with very fast  cooling rates, while
  lighter  ones are  for  halos  with cooling  time  slower than  the
  dynamical time. 

  The transition between fast and slow  cooling occurs in this plot at
  the curve defined by $t_{\rm cool} = t_{\rm ff}$.  This curve can be
  approximated  roughly  by 2 straight lines,   with a maximum cooling
  temperature  given by  \begin{eqnarray} T_{\rm max}  \simeq 6 \times
  10^5  (1+z)^3 \mbox{K}, \label{eqtmax}  \end{eqnarray} and a minimum
  cooling temperature given by  \begin{eqnarray} T_{\rm cool} \simeq 6
  \times  10^3  \mbox{K}.  \end{eqnarray}   We   finally compute   the
  accretion  rate of hot gas  into   the central, cold,  centrifugally
  supported disc, for these  2 different regimes \begin{itemize} \item
  Fast cooling: cooling  is assumed    to fast  enough so that    mass
  accretion onto  the disc is purely limited  by the orbital  decay of
  infalling  gas  clumps.  For purely radial   orbits, the  time scale
  associated to  this orbital decay is  equal to $t_{\rm orb} = R_{\rm
  200}/V_{\rm 200}$.   For  more complex orbits,  the  orbital path is
  larger, and so is the associated time scale. We introduce a new free
  parameter, the  `mean  orbital length', and  compute  the associated
  orbital decay time scale as $t_{\rm orb} = R_{\rm orb}/V_{\rm 200}$.
  The cooling rate is  finally given by  $\omega_{\rm cool} \simeq 1 /
  t_{\rm  orb}$.  \item Slow cooling: cooling  is the limiting process
  that controls  mass accretion onto  the disc.  We simplify even more
  by discarding  completely the  cooling flow description,  and assume
  $\omega_{\rm cool} \simeq 0$.    \end{itemize} In  this   analytical
  model, we have introduced    the free parameter $R_{\rm orb}$   that
  allows us to  vary the orbital  time scale relative to purely radial
  accretion.  In the latter  case, the time  scale reduces to  $R_{\rm
  200}/V_{\rm 200}$, {\it which is  independent of halo mass}.  In the
  following, we assume that  $R_{\rm orb}$ is proportional to  $R_{\rm
  200}$,  so  that  $t_{\rm orb}$   can  also be  considered as  being
  independent of halo mass.  This unknown (but constant) ratio $R_{\rm
  orb}/R_{\rm 200}$ will be determined using our numerical simulations
  as  calibration tools.  Note that  this simple model can be improved
  in many  ways.   Semi-analytical work are good   examples  of a more
  accurate modeling.  However, our simple model offers the possibility
  to  perform analytical calculations and   turns out to be reasonably
  accurate.

\newpage

  \nocite{*}
  
  \bibliography{yann}

\begin{thebibliography}{105}
\expandafter\ifx\csname natexlab\endcsname\relax\def\natexlab#1{#1}\fi

\bibitem[{{Abel} {et~al.}(2000){Abel}, {Bryan}, \& {Norman}}]{Abel00}
{Abel}, T., {Bryan}, G.~L., \& {Norman}, M.~L. 2000, \apj, 540, 39

\bibitem[{{Abel} {et~al.}(2002){Abel}, {Bryan}, \& {Norman}}]{Abel02}
{Abel}, T., {Bryan}, G.~L., \& {Norman}, M.~L. 2002, Science, 295, 93

\bibitem[{{Ascasibar} {et~al.}(2003){Ascasibar}, {Yepes}, {M{\" u}ller}, \&
  {Gottl{\" o}ber}}]{Ascasibar03}
{Ascasibar}, Y., {Yepes}, G., {M{\" u}ller}, V., \& {Gottl{\" o}ber}, S. 2003,
  \mnras, 346, 731

\bibitem[{{Benson} {et~al.}(2002){Benson}, {Lacey}, {Baugh}, {Cole}, \&
  {Frenk}}]{Benson02}
{Benson}, A.~J., {Lacey}, C.~G., {Baugh}, C.~M., {Cole}, S., \& {Frenk}, C.~S.
  2002, \mnras, 333, 156

\bibitem[{{Bernstein}(1999)}]{Bernstein99}
{Bernstein}, R.~A. 1999, in Astronomical Society of the Pacific Conference
  Series, 487--+

\bibitem[{{Blanchard} {et~al.}(1992){Blanchard}, {Valls-Gabaud}, \&
  {Mamon}}]{Blanchard92}
{Blanchard}, A., {Valls-Gabaud}, D., \& {Mamon}, G.~A. 1992, \aap, 264, 365

\bibitem[{{Bond} {et~al.}(1991){Bond}, {Cole}, {Efstathiou}, \&
  {Kaiser}}]{Bond91}
{Bond}, J.~R., {Cole}, S., {Efstathiou}, G., \& {Kaiser}, N. 1991, \apj, 379,
  440

\bibitem[{{Boss} {et~al.}(2000){Boss}, {Fisher}, {Klein}, \& {McKee}}]{Boss00}
{Boss}, A.~P., {Fisher}, R.~T., {Klein}, R.~I., \& {McKee}, C.~F. 2000, \apj,
  528, 325

\bibitem[{{Brinchmann} \& {Ellis}(2000)}]{Brinchmann00}
{Brinchmann}, J. \& {Ellis}, R.~S. 2000, \apjl, 536, L77

\bibitem[{{Bryan} {et~al.}(1995){Bryan}, {Norman}, {Stone}, {Cen}, \&
  {Ostriker}}]{Bryan95}
{Bryan}, G.~L., {Norman}, M.~L., {Stone}, J.~M., {Cen}, R., \& {Ostriker},
  J.~P. 1995, {Comput. Phys. Comm.}, 89, 149

\bibitem[{{Cen} \& {Ostriker}(1992)}]{Cen92}
{Cen}, R. \& {Ostriker}, J.~P. 1992, \apjl, 399, L113

\bibitem[{{Chary} \& {Elbaz}(2001)}]{Chary01}
{Chary}, R. \& {Elbaz}, D. 2001, \apj, 556, 562

\bibitem[{{Cohen}(2002)}]{Cohen02}
{Cohen}, J.~G. 2002, \apj, 567, 672

\bibitem[{{Cole} {et~al.}(1994){Cole}, {Aragon-Salamanca}, {Frenk}, {Navarro},
  \& {Zepf}}]{Cole94}
{Cole}, S., {Aragon-Salamanca}, A., {Frenk}, C.~S., {Navarro}, J.~F., \&
  {Zepf}, S.~E. 1994, \mnras, 271, 781

\bibitem[{{Cole} {et~al.}(2000){Cole}, {Lacey}, {Baugh}, \& {Frenk}}]{Cole00}
{Cole}, S., {Lacey}, C.~G., {Baugh}, C.~M., \& {Frenk}, C.~S. 2000, \mnras,
  319, 168

\bibitem[{{Cole} {et~al.}(2001){Cole}, {Norberg}, {Baugh}, {Frenk},
  {Bland-Hawthorn}, {Bridges}, {Cannon}, {Colless}, {Collins}, {Couch},
  {Cross}, {Dalton}, {De Propris}, {Driver}, {Efstathiou}, {Ellis},
  {Glazebrook}, {Jackson}, {Lahav}, {Lewis}, {Lumsden}, {Maddox}, {Madgwick},
  {Peacock}, {Peterson}, {Sutherland}, \& {Taylor}}]{Cole01}
{Cole}, S., {Norberg}, P., {Baugh}, C.~M., {et~al.} 2001, \mnras, 326, 255

\bibitem[{{Cooray}(2000)}]{Cooray00}
{Cooray}, A. 2000, \prd, 62, 103506

\bibitem[{{Cooray} \& {Sheth}(2002)}]{Cooray02}
{Cooray}, A. \& {Sheth}, R. 2002, \physrep, 372, 1

\bibitem[{{de Avillez} \& {Berry}(2001)}]{DeAvillez01}
{de Avillez}, M.~A. \& {Berry}, D.~L. 2001, \mnras, 328, 708

\bibitem[{{Dickinson} {et~al.}(2003){Dickinson}, {Papovich}, {Ferguson}, \&
  {Budav{\' a}ri}}]{Dickinson03}
{Dickinson}, M., {Papovich}, C., {Ferguson}, H.~C., \& {Budav{\' a}ri}, T.
  2003, \apj, 587, 25

\bibitem[{{Eke} {et~al.}(1998){Eke}, {Navarro}, \& {Frenk}}]{Eke98}
{Eke}, V.~R., {Navarro}, J.~F., \& {Frenk}, C.~S. 1998, \apj, 503, 569

\bibitem[{{Elbaz}(2005)}]{Elbaz05}
{Elbaz}, D. 2005, Habilitation \`a diriger les recherches

\bibitem[{{Elmegreen}(2002)}]{Elmegreen02}
{Elmegreen}, B.~G. 2002, \apj, 577, 206

\bibitem[{{Fixsen} {et~al.}(1998){Fixsen}, {Dwek}, {Mather}, {Bennett}, \&
  {Shafer}}]{Fixsen98}
{Fixsen}, D.~J., {Dwek}, E., {Mather}, J.~C., {Bennett}, C.~L., \& {Shafer},
  R.~A. 1998, \apj, 508, 123

\bibitem[{{Flores} {et~al.}(1999){Flores}, {Hammer}, {Thuan}, {C{\' e}sarsky},
  {Desert}, {Omont}, {Lilly}, {Eales}, {Crampton}, \& {Le F{\`
  e}vre}}]{Flores99}
{Flores}, H., {Hammer}, F., {Thuan}, T.~X., {et~al.} 1999, \apj, 517, 148

\bibitem[{{Fontana} {et~al.}(2003){Fontana}, {Donnarumma}, {Vanzella},
  {Giallongo}, {Menci}, {Nonino}, {Saracco}, {Cristiani}, {D'Odorico}, \&
  {Poli}}]{Fontana03}
{Fontana}, A., {Donnarumma}, I., {Vanzella}, E., {et~al.} 2003, \apjl, 594, L9

\bibitem[{{Fukugita} {et~al.}(1998){Fukugita}, {Hogan}, \&
  {Peebles}}]{Fukugita98}
{Fukugita}, M., {Hogan}, C.~J., \& {Peebles}, P.~J.~E. 1998, \apj, 503, 518

\bibitem[{{Giavalisco} {et~al.}(2004){Giavalisco}, {Dickinson}, {Ferguson},
  {Ravindranath}, {Kretchmer}, {Moustakas}, {Madau}, {Fall}, {Gardner},
  {Livio}, {Papovich}, {Renzini}, {Spinrad}, {Stern}, \&
  {Riess}}]{Giavalisco03}
{Giavalisco}, M., {Dickinson}, M., {Ferguson}, H.~C., {et~al.} 2004, \apjl,
  600, L103

\bibitem[{{Glazebrook} {et~al.}(1999){Glazebrook}, {Blake}, {Economou},
  {Lilly}, \& {Colless}}]{Glazebrook99}
{Glazebrook}, K., {Blake}, C., {Economou}, F., {Lilly}, S., \& {Colless}, M.
  1999, \mnras, 306, 843

\bibitem[{{Gnedin}(1998)}]{Gnedin98}
{Gnedin}, N.~Y. 1998, \mnras, 299, 392

\bibitem[{{Gnedin}(2000)}]{Gnedin00}
{Gnedin}, N.~Y. 2000, \apj, 542, 535

\bibitem[{{Gnedin}(2004)}]{Gnedin04}
{Gnedin}, N.~Y. 2004, \apj, 610, 9

\bibitem[{{Gnedin} \& {Ostriker}(1997)}]{Gnedin97}
{Gnedin}, N.~Y. \& {Ostriker}, J.~P. 1997, \apj, 486, 581

\bibitem[{{Haardt} \& {Madau}(1996)}]{Haardt96}
{Haardt}, F. \& {Madau}, P. 1996, \apj, 461, 20

\bibitem[{{Haarsma} {et~al.}(2000){Haarsma}, {Partridge}, {Windhorst}, \&
  {Richards}}]{Haarsma00}
{Haarsma}, D.~B., {Partridge}, R.~B., {Windhorst}, R.~A., \& {Richards}, E.~A.
  2000, \apj, 544, 641

\bibitem[{{Hatton} {et~al.}(2003{\natexlab{a}}){Hatton}, {Devriendt}, {Ninin},
  {Bouchet}, {Guiderdoni}, \& {Vibert}}]{Hatton03}
{Hatton}, S., {Devriendt}, J.~E.~G., {Ninin}, S., {et~al.} 2003{\natexlab{a}},
  \mnras, 343, 75

\bibitem[{{Hatton} {et~al.}(2003{\natexlab{b}}){Hatton}, {Devriendt}, {Ninin},
  {Bouchet}, {Guiderdoni}, \& {Vibert}}]{Somerville03}
{Hatton}, S., {Devriendt}, J.~E.~G., {Ninin}, S., {et~al.} 2003{\natexlab{b}},
  \mnras, 343, 75

\bibitem[{{Heavens} {et~al.}(2004){Heavens}, {Panter}, {Jimenez}, \&
  {Dunlop}}]{Heavens04}
{Heavens}, A., {Panter}, B., {Jimenez}, R., \& {Dunlop}, J. 2004, \nat, 428,
  625

\bibitem[{{Hernquist} \& {Springel}(2003)}]{Springel03c}
{Hernquist}, L. \& {Springel}, V. 2003, \mnras, 341, 1253

\bibitem[{{Hughes} {et~al.}(1998){Hughes}, {Serjeant}, {Dunlop},
  {Rowan-Robinson}, {Blain}, {Mann}, {Ivison}, {Peacock}, {Efstathiou}, {Gear},
  {Oliver}, {Lawrence}, {Longair}, {Goldschmidt}, \& {Jenness}}]{Hughes98}
{Hughes}, D.~H., {Serjeant}, S., {Dunlop}, J., {et~al.} 1998, \nat, 394, 241

\bibitem[{{Jenkins} {et~al.}(2001){Jenkins}, {Frenk}, {White}, {Colberg},
  {Cole}, {Evrard}, {Couchman}, \& {Yoshida}}]{Jenkins01}
{Jenkins}, A., {Frenk}, C.~S., {White}, S.~D.~M., {et~al.} 2001, \mnras, 321,
  372

\bibitem[{{Katz}(1992)}]{katz92}
{Katz}, N. 1992, \apj, 391, 502

\bibitem[{{Katz} {et~al.}(1996){Katz}, {Weinberg}, \& {Hernquist}}]{Katz96}
{Katz}, N., {Weinberg}, D.~H., \& {Hernquist}, L. 1996, \apjs, 105, 19

\bibitem[{{Kauffmann} {et~al.}(1999){Kauffmann}, {Colberg}, {Diaferio}, \&
  {White}}]{Kaufmann99}
{Kauffmann}, G., {Colberg}, J.~M., {Diaferio}, A., \& {White}, S.~D.~M. 1999,
  \mnras, 303, 188

\bibitem[{{Kauffmann} {et~al.}(1993){Kauffmann}, {White}, \&
  {Guiderdoni}}]{Kaufmann93}
{Kauffmann}, G., {White}, S.~D.~M., \& {Guiderdoni}, B. 1993, \mnras, 264, 201

\bibitem[{{Kay} {et~al.}(2002){Kay}, {Pearce}, {Frenk}, \& {Jenkins}}]{Kay02}
{Kay}, S.~T., {Pearce}, F.~R., {Frenk}, C.~S., \& {Jenkins}, A. 2002, \mnras,
  330, 113

\bibitem[{{Kay} {et~al.}(2004){Kay}, {Thomas}, {Jenkins}, \& {Pearce}}]{Kay04}
{Kay}, S.~T., {Thomas}, P.~A., {Jenkins}, A., \& {Pearce}, F.~R. 2004, \mnras,
  355, 1091

\bibitem[{{Kennicutt}(1998)}]{Kennicutt98}
{Kennicutt}, R.~C. 1998, \apj, 498, 541

\bibitem[{{Kennicutt} {et~al.}(1994){Kennicutt}, {Tamblyn}, \&
  {Congdon}}]{Kennicutt94}
{Kennicutt}, R.~C., {Tamblyn}, P., \& {Congdon}, C.~E. 1994, \apj, 435, 22

\bibitem[{{Komatsu} \& {Seljak}(2001)}]{Komatsu01}
{Komatsu}, E. \& {Seljak}, U. 2001, \mnras, 327, 1353

\bibitem[{{Kravtsov}(2003)}]{Kravtsov03}
{Kravtsov}, A.~V. 2003, \apjl, 590, L1

\bibitem[{{Kravtsov} {et~al.}(2004{\natexlab{a}}){Kravtsov}, {Berlind},
  {Wechsler}, {Klypin}, {Gottl{\" o}ber}, {Allgood}, \&
  {Primack}}]{Kravtsov04b}
{Kravtsov}, A.~V., {Berlind}, A.~A., {Wechsler}, R.~H., {et~al.}
  2004{\natexlab{a}}, \apj, 609, 35

\bibitem[{{Kravtsov} {et~al.}(2004{\natexlab{b}}){Kravtsov}, {Gnedin}, \&
  {Klypin}}]{Kravtsov04}
{Kravtsov}, A.~V., {Gnedin}, O.~Y., \& {Klypin}, A.~A. 2004{\natexlab{b}},
  \apj, 609, 482

\bibitem[{{Kravtsov} {et~al.}(1997){Kravtsov}, {Klypin}, \&
  {Khokhlov}}]{Kravtsov97}
{Kravtsov}, A.~V., {Klypin}, A.~A., \& {Khokhlov}, A.~M. 1997, \apjs, 111, 73

\bibitem[{{Kroupa} {et~al.}(1993){Kroupa}, {Tout}, \& {Gilmore}}]{Kroupa93}
{Kroupa}, P., {Tout}, C.~A., \& {Gilmore}, G. 1993, \mnras, 262, 545

\bibitem[{{Lacey} \& {Cole}(1993)}]{Lacey93}
{Lacey}, C. \& {Cole}, S. 1993, \mnras, 262, 627

\bibitem[{{Ma} \& {Fry}(2000)}]{Ma00}
{Ma}, C.~P. \& {Fry}, J.~N. 2000, \apj, 543, 503

\bibitem[{{Madau} \& {Pozzetti}(2000)}]{Madau00}
{Madau}, P. \& {Pozzetti}, L. 2000, \mnras, 312, L9

\bibitem[{{Marinoni} \& {Hudson}(2002)}]{Marinoni02}
{Marinoni}, C. \& {Hudson}, M.~J. 2002, \apj, 569, 101

\bibitem[{{Martin}(1999)}]{Martin99}
{Martin}, C.~L. 1999, \apj, 513, 156

\bibitem[{{Martin} \& {Kennicutt}(2001)}]{Martin01}
{Martin}, C.~L. \& {Kennicutt}, R.~C. 2001, \apj, 555, 301

\bibitem[{{Massarotti} {et~al.}(2001){Massarotti}, {Iovino}, \&
  {Buzzoni}}]{Massarotti01}
{Massarotti}, M., {Iovino}, A., \& {Buzzoni}, A. 2001, \apjl, 559, L105

\bibitem[{{Mayer} \& {Moore}(2004)}]{Mayer04}
{Mayer}, L. \& {Moore}, B. 2004, \mnras, 354, 477

\bibitem[{{McKee} \& {Ostriker}(1977)}]{McKee77}
{McKee}, C.~F. \& {Ostriker}, J.~P. 1977, \apj, 218, 148

\bibitem[{{Mihos} \& {Hernquist}(1994)}]{Mihos94}
{Mihos}, J.~C. \& {Hernquist}, L. 1994, \apj, 437, 611

\bibitem[{{Muanwong} {et~al.}(2002){Muanwong}, {Thomas}, {Kay}, \&
  {Pearce}}]{Muanwong02}
{Muanwong}, O., {Thomas}, P.~A., {Kay}, S.~T., \& {Pearce}, F.~R. 2002, \mnras,
  336, 527

\bibitem[{{Nagamine} {et~al.}(2004){Nagamine}, {Cen}, {Hernquist}, {Ostriker},
  \& {Springel}}]{Nagamine04}
{Nagamine}, K., {Cen}, R., {Hernquist}, L., {Ostriker}, J.~P., \& {Springel},
  V. 2004, \apj, 610, 45

\bibitem[{{Navarro} \& {White}(1993)}]{Navarro93}
{Navarro}, J.~F. \& {White}, S.~D.~M. 1993, \mnras, 265, 271

\bibitem[{{Neumann}(2005)}]{Neuman05}
{Neumann}. 2005, submitted

\bibitem[{{Neumann} \& {Arnaud}(1999)}]{Neumann99}
{Neumann}, D.~M. \& {Arnaud}, M. 1999, \aap, 348, 711

\bibitem[{{Neumann} \& {Arnaud}(2001)}]{Neumann01}
{Neumann}, D.~M. \& {Arnaud}, M. 2001, \aap, 373, L33

\bibitem[{{Pei} {et~al.}(1999){Pei}, {Fall}, \& {Hauser}}]{Pei99}
{Pei}, Y.~C., {Fall}, S.~M., \& {Hauser}, M.~G. 1999, \apj, 522, 604

\bibitem[{{Penton} {et~al.}(2004){Penton}, {Stocke}, \& {Shull}}]{Penton04}
{Penton}, S.~V., {Stocke}, J.~T., \& {Shull}, J.~M. 2004, \apjs, 152, 29

\bibitem[{{Press} \& {Schechter}(1974)}]{Press74}
{Press}, W.~H. \& {Schechter}, P. 1974, \apj, 187, 425

\bibitem[{{Refregier}(2005)}]{Refregier05}
{Refregier}. 2005, in prep

\bibitem[{{Refregier} \& {Teyssier}(2002)}]{Refregier02}
{Refregier}, A. \& {Teyssier}, R. 2002, \prd, 66, 043002

\bibitem[{{Sanderson} \& {Ponman}(2003)}]{Sanderson03b}
{Sanderson}, A.~J.~R. \& {Ponman}, T.~J. 2003, \mnras, 345, 1241

\bibitem[{{Sanderson} {et~al.}(2003){Sanderson}, {Ponman}, {Finoguenov},
  {Lloyd-Davies}, \& {Markevitch}}]{Sanderson03a}
{Sanderson}, A.~J.~R., {Ponman}, T.~J., {Finoguenov}, A., {Lloyd-Davies},
  E.~J., \& {Markevitch}, M. 2003, \mnras, 340, 989

\bibitem[{{Seljak}(2000)}]{Seljak00}
{Seljak}, U. 2000, \mnras, 318, 203

\bibitem[{{Shapiro} {et~al.}(1996){Shapiro}, {Martel}, {Villumsen}, \&
  {Owen}}]{Shapiro96}
{Shapiro}, P.~R., {Martel}, H., {Villumsen}, J.~V., \& {Owen}, J.~M. 1996,
  \apjs, 103, 269

\bibitem[{{Sokasian} {et~al.}(2004){Sokasian}, {Yoshida}, {Abel}, {Hernquist},
  \& {Springel}}]{Sokasian04}
{Sokasian}, A., {Yoshida}, N., {Abel}, T., {Hernquist}, L., \& {Springel}, V.
  2004, \mnras, 350, 47

\bibitem[{{Somerville} \& {Primack}(1999)}]{Somerville99}
{Somerville}, R.~S. \& {Primack}, J.~R. 1999, \mnras, 310, 1087

\bibitem[{{Somerville} {et~al.}(2001){Somerville}, {Primack}, \&
  {Faber}}]{Somerville01}
{Somerville}, R.~S., {Primack}, J.~R., \& {Faber}, S.~M. 2001, \mnras, 320, 504

\bibitem[{{Spergel} {et~al.}(2003){Spergel}, {Verde}, {Peiris}, {Komatsu},
  {Nolta}, {Bennett}, {Halpern}, {Hinshaw}, {Jarosik}, {Kogut}, {Limon},
  {Meyer}, {Page}, {Tucker}, {Weiland}, {Wollack}, \& {Wright}}]{Spergel03}
{Spergel}, D.~N., {Verde}, L., {Peiris}, H.~V., {et~al.} 2003, \apjs, 148, 175

\bibitem[{{Springel} {et~al.}(2005){Springel}, {Di Matteo}, \&
  {Hernquist}}]{Springel05}
{Springel}, V., {Di Matteo}, T., \& {Hernquist}, L. 2005, \apjl, 620, L79

\bibitem[{{Springel} \& {Hernquist}(2003{\natexlab{a}})}]{Springel03a}
{Springel}, V. \& {Hernquist}, L. 2003{\natexlab{a}}, \mnras, 339, 289

\bibitem[{{Springel} \& {Hernquist}(2003{\natexlab{b}})}]{Springel03b}
{Springel}, V. \& {Hernquist}, L. 2003{\natexlab{b}}, \mnras, 339, 312

\bibitem[{{Steidel} {et~al.}(1999){Steidel}, {Adelberger}, {Giavalisco},
  {Dickinson}, \& {Pettini}}]{Steidel99}
{Steidel}, C.~C., {Adelberger}, K.~L., {Giavalisco}, M., {Dickinson}, M., \&
  {Pettini}, M. 1999, \apj, 519, 1

\bibitem[{{Storrie-Lombardi} {et~al.}(1996){Storrie-Lombardi}, {McMahon}, \&
  {Irwin}}]{StorrieLombardi96}
{Storrie-Lombardi}, L.~J., {McMahon}, R.~G., \& {Irwin}, M.~J. 1996, \mnras,
  283, L79

\bibitem[{{Sugiyama}(1995)}]{sugiyama95}
{Sugiyama}, N. 1995, \apjs, 100, 281

\bibitem[{{Suto} {et~al.}(1998){Suto}, {Sasaki}, \& {Makino}}]{Suto98}
{Suto}, Y., {Sasaki}, S., \& {Makino}, N. 1998, \apj, 509, 544

\bibitem[{{Teyssier}(2002)}]{Teyssier02}
{Teyssier}, R. 2002, \aap, 385, 337

\bibitem[{{Thacker} \& {Couchman}(2000)}]{Thacker00}
{Thacker}, R.~J. \& {Couchman}, H.~M.~P. 2000, \apj, 545, 728

\bibitem[{{Theuns} {et~al.}(1998){Theuns}, {Leonard}, {Efstathiou}, {Pearce},
  \& {Thomas}}]{Theuns98}
{Theuns}, T., {Leonard}, A., {Efstathiou}, G., {Pearce}, F.~R., \& {Thomas},
  P.~A. 1998, \mnras, 301, 478

\bibitem[{{Thomas} {et~al.}(2002){Thomas}, {Muanwong}, {Kay}, \&
  {Liddle}}]{Thomas01}
{Thomas}, P.~A., {Muanwong}, O., {Kay}, S.~T., \& {Liddle}, A.~R. 2002, \mnras,
  330, L48

\bibitem[{{Tinsley}(1980)}]{Tinsley80}
{Tinsley}, B.~M. 1980, Fundamentals of Cosmic Physics, 5, 287

\bibitem[{{Toro}(1997)}]{Toro97}
{Toro}, E. 1997, Book

\bibitem[{{Truelove} {et~al.}(1997){Truelove}, {Klein}, {McKee}, {Holliman},
  {Howell}, \& {Greenough}}]{Truelove97}
{Truelove}, J.~K., {Klein}, R.~I., {McKee}, C.~F., {et~al.} 1997, \apjl, 489,
  L179+

\bibitem[{{van den Bosch}(2002)}]{VanDenBosch02}
{van den Bosch}, F.~C. 2002, \mnras, 331, 98

\bibitem[{{White}(2002)}]{White02}
{White}, M. 2002, \apj, 143, 241

\bibitem[{{White} \& {Frenk}(1991)}]{White91}
{White}, S.~D.~M. \& {Frenk}, C.~S. 1991, \apj, 379, 52

\bibitem[{{Yan} {et~al.}(1999){Yan}, {McCarthy}, {Freudling}, {Teplitz},
  {Malumuth}, {Weymann}, \& {Malkan}}]{Yan99}
{Yan}, L., {McCarthy}, P.~J., {Freudling}, W., {et~al.} 1999, \apjl, 519, L47

\bibitem[{{Yepes} {et~al.}(1997){Yepes}, {Kates}, {Khokhlov}, \&
  {Klypin}}]{Yepes97}
{Yepes}, G., {Kates}, R., {Khokhlov}, A., \& {Klypin}, A. 1997, \mnras, 284,
  235

\bibitem[{{Zwaan} {et~al.}(1997){Zwaan}, {Briggs}, {Sprayberry}, \&
  {Sorar}}]{Zwaan97}
{Zwaan}, M.~A., {Briggs}, F.~H., {Sprayberry}, D., \& {Sorar}, E. 1997, \apj,
  490, 173

\bibitem[{{Zwaan} {et~al.}(2005){Zwaan}, {Meyer}, {Staveley-Smith}, \&
  {Webster}}]{Zwaan05}
{Zwaan}, M.~A., {Meyer}, M.~J., {Staveley-Smith}, L., \& {Webster}, R.~L. 2005,
  \mnras, 359, L30

\end{thebibliography}

  
  
\end{document}